\documentclass[10pt,twocolumn,aps,prx,floatfix,superscriptaddress,nofootinbib]{revtex4-2}
\usepackage{amsmath}
\usepackage{amsfonts}
\usepackage{amssymb}
\usepackage{graphicx}
\usepackage{color}
\usepackage{bm}
\usepackage{dsfont}
\usepackage{soul}
\usepackage{mathtools}
\usepackage[normalem]{ulem}

\DeclarePairedDelimiterX\braket[2]{\langle}{\rangle}{#1 \delimsize\vert #2}
\usepackage[final]{hyperref} 
\hypersetup{
	colorlinks=true,       
	linkcolor=blue,        
	citecolor=blue,        
	filecolor=magenta,     
	urlcolor=blue         
}

\newcommand{\beq}{\begin{equation}}
\newcommand{\eneq}{\end{equation}}

\begin{document}

\title{Information transport and transport-induced entanglement in open fermion chains}

\author{Andrea Nava}
\affiliation{Institut f\"ur Theoretische Physik, Heinrich-Heine-Universit\"at, 40225 D\"usseldorf, Germany}

\author{Claudia Artiaco}
\affiliation{Department of Physics, KTH Royal Institute of Technology, Stockholm 106 91, Sweden}
\affiliation{\mbox{Institut f\"ur Theoretische Physik, Universit\"at zu K\"oln, Z\"ulpicher Stra{\ss}e 77, D-50937 Cologne, Germany}}

\author{Yuval Gefen}
\affiliation{Department of Condensed Matter Physics, Weizmann Institute, 7610001 Rehovot, Israel}

\author{Igor Gornyi}
\affiliation{\mbox{Institute for Quantum Materials and Technologies, Karlsruhe Institute of Technology, 76131 Karlsruhe, Germany}} 
\affiliation{\mbox{Institut f\"ur Theorie der Kondensierten Materie, Karlsruhe Institute of Technology, 76131 Karlsruhe, Germany}}

\author{Mikheil Tsitsishvili}
\affiliation{Institut f\"ur Theoretische Physik, Heinrich-Heine-Universit\"at, 40225  D\"usseldorf, Germany}

\author{Alex Zazunov} 
\affiliation{Institut f\"ur Theoretische Physik, Heinrich-Heine-Universit\"at, 40225  D\"usseldorf, Germany}

\author{Reinhold Egger}
\affiliation{Institut f\"ur Theoretische Physik, Heinrich-Heine-Universit\"at, 40225 D\"usseldorf, Germany}

\begin{abstract} 
Understanding the entanglement dynamics in quantum many-body systems under steady-state transport conditions is an actively pursued challenging topic.
Hydrodynamic equations, akin to transport equations for charge or heat, would be of great interest but face severe challenges because of the inherent nonlocality of entanglement and the difficulty of identifying conservation laws. We show that progress is facilitated by using information as key quantity related to---but distinct from---entanglement. Employing the recently developed  ``information lattice'' framework, we 
formulate general continuity equations governing the flow of spatially and scale-resolved information currents in nonequilibrium open quantum systems. 
To illustrate our theory, we consider noninteracting fermion chains coupled to dissipative reservoirs, using Lindblad master equations.
By relating the information lattice to a noise lattice constructed from particle-number fluctuations, we show that information is experimentally accessible via noise measurements. Similarly, local information currents can be obtained by measuring particle currents, onsite occupations, and covariances of particle numbers and/or particle currents. Using the fermionic negativity to quantify bipartite entanglement, we also study transport-induced entanglement and its relation to information currents. For a clean particle-hole symmetric chain, we find that information currents are shielded from entering the information lattice.
Impurities or particle-hole asymmetry break this effect, causing information current flow and entanglement between end segments of the chain.  Our work opens the door to systematic investigations of information transport and entanglement generation in driven open quantum systems far from equilibrium.
 \end{abstract}
\maketitle

\section{Introduction}\label{sec1}

Entanglement is a key resource in quantum information processing, underlying both potential advantages of quantum computation and enhanced capabilities of quantum communication~\cite{bennett2000quantum, horodecki2009quantum, nielsen2010quantum, wilde2013quantum}. In many-body systems, states can be classified in terms of the scaling of the von Neumann entropy, which quantifies entanglement in pure states, with subsystem size and time \cite{Amico_2008, Cirac_2012, laflorencie2016quantum, zeng2019quantum, eisert2010colloquium, srednicki1993entropy, hastings2007area, wolf2008area, vidal2003entanglement, calabrese2005evolution, Kobayashi_2024, Cheraghi2025}. 
Furthermore, long-range patterns of entanglement are the defining features of topological phases of matter, for both pure and mixed states.
These long-range entanglement structures determine how quantum information is stored, protected, and manipulated, and they strongly constrain the possible dynamical pathways the system can undergo while preserving its topological character \cite{hamma2005ground, kitaev2006topological, levin2006detecting,  hastings2011topological, Savary_2017, fan2024diagnostics, Sang2024mixed, sang2025mixed, masmendoza2025graphical}. 

A natural and fundamental question that then arises in the context of many-body physics is \textit{how entanglement is transported}.
How do quantum correlations behind entanglement propagate in a system governed by local Hamiltonian dynamics and local decoherence or dissipation?
Which mechanisms determine the flow of quantum information in nonequilibrium settings?
A complementary perspective concerns the influence of transport processes involving other physical characteristics of the system---such as charge or energy---on the system's information content and entanglement properties: Can transport, in itself, generate, enhance, or redistribute quantum entanglement?  

While various aspects of the entanglement dynamics have been 
addressed in the literature~\cite{calabrese2005evolution, Kraus_2008, Galve_2009, Zueco_2009, Sarkar_2009, Verstraete_2009, Vedral_2009, Scholak_2010, Swingle2012, Nozaki2013, Lin_2013, Shankar_2013, Aolita_2015, mahajan_2016, Brange_2017, Nahum2017,  Keyserlingk2018,  Lee_2019, Cao2019a, Alba2019, Panda_2020, Sancho2020, Vovcenko_2021,  Alberton_2021, Fraenkel2021, Klein2022, Ruggiero_2022, Bertini2022, Carollo2022, Alba_2023, Fraenkel2023,  Muzzi_2025, Tan2025, Tirrito2025, Poboiko2025, Shekhar2025}, a systematic and physically transparent description of quantum-information 
transport and transport-induced entanglement in diverse dynamical settings, in particular, for quantum many-body systems operated under transport conditions,
remains a challenging area of research.
Our general aim is to establish a unifying picture describing the transport of information and entanglement in quantum many-body systems under nonequilibrium steady-state (NESS) conditions.
Here a key issue is the local quantification of information and entanglement, particularly for systems that support charge transport through them and are therefore inherently open. In this context, the corresponding currents should be specified for mixed quantum states. 
Adopting such a local-dynamics perspective plays a crucial conceptual role in understanding the mutual relations between quantum correlations, entanglement, and nonequilibrium transport. In addition, it resonates with the spirit of the entanglement quasiparticle picture \cite{calabrese2005evolution}, see also, e.g., Refs.~\cite{Cao2019a, Alba2019, Bertini2022, Carollo2022, Cheraghi2025}.
This description was originally introduced to analyze the post-quench time evolution of entanglement in closed systems and typically relies on the availability of exact solutions or on the knowledge of asymptotic states. Although it provides a powerful and intuitive tool to reconstruct entropy contributions, it remains to some extent system-specific and approximate for generic systems. Other examples are the membrane picture~\cite{Nahum2017, Jonay2018, zhou2020entanglement, Mezei2018, Mezei2020} and the entanglement tsunami~\cite{liu2014entanglement}. This shows that developing genuinely local perspectives for the dynamics of quantum information in many-body systems can, in principle, disclose fundamental questions in quantum matter. Throughout the paper, we carefully distinguish between entanglement and quantum information, where the latter is quantified in terms of subsystem von Neumann entropies.  However, we will also comment on 
consequences of our results for transport-induced entanglement.

Hydrodynamics affords a remarkably universal effective macroscopic description of transport in terms of only a few slowly varying local fields. In this spirit, it would be highly insightful to construct a hydrodynamic theory for the entanglement dynamics of quantum many-body systems in a NESS by means of, e.g., a suitably adapted version of the Martin–Siggia–Rose formalism \cite{Martin_1973}, recently applied for describing monitored classical stochastic processes \cite{Gopalakrishnan_2026} and a classical hydrodynamic
framework \cite{Hauser2026}.
However, the construction of such a theory faces two major hurdles.  First, entanglement refers to nonlocal quantum correlations where it is difficult to identify local entanglement currents.  Second, even for a closed system, it is unclear how to obtain conservation laws based on standard quantifiers for bipartite entanglement, such as the fermionic negativity (FN) for mixed states \cite{shapourian2017partial,shapourian2017many, Shapourian2019a, shapourian2019entanglement, eisert2018entanglement, Alba_2023}. 
Fortunately, by considering \emph{information} as the key quantity of interest, which is defined in terms of subsystem von Neumann entropies~\cite{nielsen2010quantum}, progress is possible. 
Indeed, as we demonstrate below, the associated information currents are locally conserved and easily accessible experimental quantities under NESS conditions. They offer a natural macroscopic language for correlation spreading in open systems, with entanglement transport emerging as a distinctive---but not unique---manifestation of the underlying information flow.

In this paper, we address a minimal model to identify the fundamental physical mechanisms behind information transport in nonequilibrium open systems.  The resulting insights allow for systematic extensions to more complex settings and generalized transport equations. We formulate general continuity equations which govern the flow of local and scale-resolved information currents in nonequilibrium open quantum many-body systems.   
Results obtained from these fundamental equations are illustrated for the exactly solvable model of a noninteracting fermion chain driven into a current-carrying NESS via couplings to external reservoirs.
We analyze how the flow of quantum information can be manipulated on different spatial scales by the change of external control parameters, e.g., applied bias voltages and gate voltages. Our theory is constructed by starting from the recently proposed information-lattice framework~\cite{Klein2022, Artiaco2025}. 
Within the information lattice, total information (i.e., the average number of binary questions that one can answer by knowing the system state) decomposes into local contributions that evolve according to local unitary dynamics, in contrast to global measures such as the subsystem von Neumann entropy, which capture total correlations between a subsystem and its environment.
Specifically, this approach assigns to every spatial region of a one-dimensional (1D) chain the quantum information accessible at each scale (\emph{aka} layer) $\ell$, and arranges these spatially and scale-resolved quantities into a lattice.  The resulting information lattice captures how information is distributed throughout the system. While the nodes of the information lattice represent local information, its links are associated with local information currents. In fact, on the information lattice, local information behaves like a hydrodynamic quantity obeying continuity-like local conservation laws~\cite{Bauer2025}. In particular, under a two-site unitary gate, local information can only flow between adjacent layers $\ell$ and $\ell \pm 1$. 

The information-lattice framework is currently under active development~\cite{Flor2025}, and applications are being explored along various directions. In particular, this concept has allowed for the development of novel approximate methods for the efficient simulation of the time evolution of large-scale interacting quantum systems via the ``Local-Information Time Evolution'' approach~\cite{Klein2022, Artiaco2024, Harkins2025}. Moreover, the information lattice provides a universal characterization of quantum many-body states~\cite{Artiaco2025} in different phases (e.g., localized, critical, ergodic, or topological) and of quantum quench dynamics~\cite{Bauer2025, Artiaco2025out, barata2025hadronic}, as well as a witness for long-range magic~\cite{bilinskaya2025witnessing}. 

While previous work on information lattices has established the foundations of the local-information perspective, it has predominantly been focused on closed systems and pure states. A comprehensive description of local information currents that incorporates the interplay between unitary dynamics and dissipation---and thus captures the full scope of information transport in open quantum many-body systems---has so far remained elusive. The present paper closes this gap by addressing information flows in open quantum systems, based on systematically exploring local information currents in nonequilibrium settings with mixed states.

Below, we specify general continuity equations for the flow of local and scale-resolved information currents  in open fermion systems under transport conditions (thereby also determining transport-induced entanglement). We illustrate our theory with the exactly solvable limit of noninteracting lattice fermions, where the unitary dynamics is governed by a short-range quadratic Hamiltonian. The chain is coupled to external environments through local Lindblad jump operators acting on specific sites. These jump operators are linear in the fermionic operators and model the processes of injection and/or removal of particles, thus allowing us to study NESS configurations carrying a finite current. For such open many-fermion systems, Gaussian states---where the density matrix is the exponential of a quadratic form in fermionic creation and annihilation operators---remain Gaussian under Lindbladian time evolution, and hence we speak of \textit{quasi-free} fermions,  even though particles may still be scattered by inhomogeneities. The full many-body density matrix is then completely determined in terms of the two-body correlation matrix, whose dynamics obey a closed set of equations of motion \cite{Zunkovic_2014, Barthel_2022, Fazio2025}. This feature drastically simplifies the characterization of information transport and makes it possible to evaluate local information currents for large system sizes. 

Building on this solvable setting, we investigate current-carrying NESSs and analyze how nonequilibrium conditions in open systems shape the flow of local information and entanglement transport properties quantified by the FN \cite{shapourian2017partial, shapourian2017many, Shapourian2019a, shapourian2019entanglement, eisert2018entanglement, Alba_2023}.
We demonstrate that long-range quantum correlations and entanglement between spatially distant subsystems may be induced under nonequilibrium conditions that concurrently enable charge transport through the fermionic system. In particular, we show how correlations absent at equilibrium are generated under transport conditions when defects are present in the system, see also Refs.~\cite{Fraenkel2021, Fraenkel2023}. 
In addition, we highlight how deviations from particle-hole symmetry strongly affect the information currents flowing in the system.

One of the key results of our work is that the local information lattice and the associated information currents are not only deep theoretical constructs but can also be
measured experimentally without need for quantum state tomography.   
Indeed, for the class of open quasi-free fermionic systems studied here, local information can be probed through a ``noise lattice” quantifying the corresponding local particle number fluctuations (variances) defined in analogy to the information lattice. 
Similarly, by evaluating the time derivative of the noise lattice, we find that local information currents can be accessed by measuring
 particle currents, onsite occupation numbers, and covariances of particle numbers and/or particle currents.  Alternatively, one can measure local information currents by exploiting the fact that they are fully determined by the correlation matrix defined in Eq.~\eqref{corrdef} below.  For quasi-free fermions,  the correlation matrix can be expressed by means of Wick's theorem in terms of particle number and/or particle current correlations. Such correlation functions are experimentally accessible, e.g., through interferometric approaches  \cite{Knap_2013,Pedernales_2014,DelRe_2024,Liu_2025,Yu_2011,Kastner_2020,Wang_2025}.
 Our ideas therefore suggest experimental routes for measuring the predicted patterns of information flow in platforms such as quantum dot arrays or ultracold fermionic atoms, and can enable controlled studies of transport-induced entanglement in realistic open many-body quantum systems.

\begin{figure}
\includegraphics[width=0.48\textwidth]{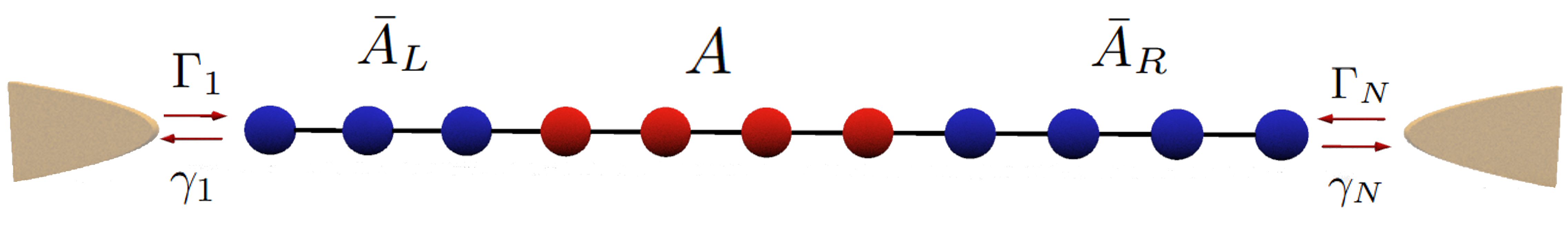}
\caption{Schematic of a spinless fermion chain with $N$ sites, connected at its ends to fermionic reservoirs with single-particle injection (removal) rates $\Gamma_1$ and $\Gamma_N$ ($\gamma_1$ and $\gamma_N$), respectively. A possible partition into a subsystem $A$ (red sites) with complement $\bar A={\bar A}_L \cup {\bar A}_R$ (blue sites) is indicated. }  \label{fig1}
\end{figure}

The remainder of this paper is structured as follows. In Sec.~\ref{sec2}, we develop our theory for general (potentially interacting) open fermionic systems under transport conditions, formulating a continuity equation for local and scale-resolved information currents using the information lattice.  The concepts established in Sec.~\ref{sec2} are then
implemented for the exactly solvable case of quasi-free fermions in Sec.~\ref{sec22}, where we also introduce the noise lattice.
In Sec.~\ref{sec3}, we employ this exact solution in order to reveal characteristic patterns of local information currents in various parameter regimes.
Specifically, we focus on a case where reservoirs are attached only to the ends of the fermionic chain, see Fig.~\ref{fig1}. We show that the presence of a defect and/or particle-hole asymmetry have dramatic consequences on local information currents. For the clean and particle-hole symmetric case, we reveal a ``shielding effect'' where information currents injected from the environment are prohibited from entering the bulk of the information lattice.  By breaking particle-hole symmetry, the shielding effect is also broken, and information current can flow from one end of the chain to the other.  In the absence of defects, these information currents are characterized by short-range correlations.  If defects are present in the chain, however, long-range quantum correlations, corresponding to information currents existing on length scales of the order of the system size, emerge in NESS configurations but are absent at equilibrium. Particle-hole asymmetry then leads to asymmetric information current patterns.  We contrast our results for the large-bias regime with those obtained for small applied bias voltage (linear response regime).  For clarity, we focus on the properties of the current-carrying NESS reached at long times in Sec.~\ref{sec3}, but the theoretical framework allows one to capture the full time evolution of the system toward the NESS as well.
In Sec.~\ref{sec4}, we then discuss transport-induced entanglement as quantified by the FN, and show that the resulting picture is fully consistent with the reported flow of local information in the information lattice.
The paper concludes with a summary and an outlook in Sec.~\ref{sec5}.  Analytical arguments supporting our conclusions in Secs.~\ref{sec3} and \ref{sec4} can be found in Appendix \ref{appendix:exact for N=3}, where we provide the NESS solution for an open fermionic system with $N=3$ sites.  
In Appendix \ref{appendix:unraveling}, we comment on unraveled solutions of the Lindblad equation, which describe, e.g., the case of measurement devices as the environment. In addition, since for the unraveled pure-state trajectories, the von Neumann entropy quantifies bipartite entanglement, this approach establishes a direct connection between information and entanglement.
Below, we use units with $\hbar=k_B=1$.

\section{Information lattice and information currents}\label{sec2}

We consider the von Neumann information flow inside the information lattice \cite{Klein2022,Artiaco2024,Artiaco2025,Bauer2025,Artiaco2025out} for an arbitrary  1D fermion chain with $N$ sites, allowing for local particle exchange with fermionic reservoirs, see Fig.~\ref{fig1}. The Lindblad equation governing the system dynamics is presented in Sec.~\ref{sec2a}.  In Sec.~\ref{sec2b}, we derive a continuity equation under transport conditions for the local and scale-resolved information currents by employing the information lattice.  
In Sec.~\ref{sec2c}, we consider the simple cases $N=2$ and $N=3$, respectively, before turning to the case of arbitrary $N$ in Sec.~\ref{sec2d}.  Finally, in Sec.~\ref{sec2f}, we comment on standard particle currents.

\subsection{Model and Lindblad equation}\label{sec2a}

We study a spinless lattice fermion chain with $N$ sites. Apart from a Hamiltonian ($H$) contribution, which determines the unitary time evolution of the system density matrix $\rho(t)$, we allow for the presence of fermionic environments exchanging single particles at local sites along the chain, see Fig.~\ref{fig1} for an example.  As a consequence, NESS currents flowing through the chain can be modeled. For weak system-environment couplings (Born approximation) and assuming very short environmental memory times (Markov approximation), the time evolution of $\rho(t)$ is governed by the Lindblad master equation (LME)  \cite{Breuer2007},  
\begin{equation}\label{eq:rho-equ} \dot{\rho}(t) = -i\left[H,\rho(t) \right]+\sum_k {\cal D}[L_{k}]\rho(t),
\end{equation}
with the dissipator 
\begin{equation}
{\cal D}[L]\rho = L \rho L^{\dagger}-\frac{1}{2}\{ L^{\dagger}L,\rho\}
\label{eq:dissipator}
\end{equation} 
and the anticommutator $\{\cdot,\cdot\}$.
The environment-induced injection (or removal) of fermions into (from) the chain is encoded by a set of jump operators $\{ L_{k}\}$, leading to a non-unitary time evolution of $\rho(t)$.  

In order to model local particle exchange with fermionic reservoirs, we choose the jump operators $L_k\in \{L_{j,{\rm inj}}, L_{j,{\rm rem}}\}$ as onsite operators injecting or removing a single fermion,
\begin{equation}\label{jumpop}
    L_{j,\mathrm{inj}} = \sqrt{\Gamma_j} \, c_j^\dagger ,\quad 
    L_{j,\mathrm{rem}} = \sqrt{\gamma_j} \, c_j^{},
\end{equation} 
linear in fermionic annihilation (creation) operators $c_j$ ($c^\dagger_j$) for sites $j=1,\ldots,N$, with $\Gamma_j$ ($\gamma_j$) is the corresponding injection (removal) rate.
For more general jump operators extending over several sites, our approach still applies as long as all jump operators remain linear in  fermionic operators, i.e.,
$L_k$ is connected to $\{c_j^{\dagger}\}$ or $\{c_j^{}\}$ by real-valued  $N\times N$
matrices $\Gamma$ and $\gamma$, respectively, which are no longer diagonal  \cite{Guimaraes_2016,Cinnirella2025}. For the onsite case in Eq.~\eqref{jumpop},
\begin{equation}\label{Gamdiag}
\Gamma={\rm diag}(\Gamma_1,\ldots,\Gamma_N),\quad
\gamma={\rm diag}(\gamma_1,\ldots,\gamma_N).
\end{equation}
Moreover, for the setup in Fig.~\ref{fig1}, all rates in Eq.~\eqref{Gamdiag} vanish except for $j=1$ (left end) and $j=N$ (right chain end). 
For a microscopic model of the system-environment coupling, these rates can be computed from Fermi's golden rule \cite{Breuer2007,Benenti_2009}.

Assuming that the environments correspond to metallic leads, $\Gamma_{1 (N)}$ and $\gamma_{1 (N)}$ are thereby expressed in terms of Fermi distribution factors $f_{L/R}$ and hybridization energies $g_{L/R}= 2\pi |\tau_{L/R}|^2 \nu_{L/R}$ for the left and right ($L/R$) lead \cite{Benenti_2009},
 \begin{equation}\label{ratesF}
    \Gamma_{1\, (N)} = g_{L(R)} f_{L(R)},\quad   
    \gamma_{1\, (N)} = g_{L(R)} \left(1-f_{L(R)}\right),
\end{equation}
where the lead density of states $\nu_{L/R}$ is taken at the respective chemical potential $\mu_{L/R}$ and $\tau_{L/R}$ is a hopping amplitude for 
eigenmodes injected from the left or right side.  The hybridization scales $g_{L/R}$ do not depend on energy under
the standard wide-band approximation for the leads.  
The Fermi factors in Eq.~\eqref{ratesF} are 
\begin{equation}\label{fermidist}
f_L= \frac{1}{e^{(\epsilon_{1}-\bar\mu-eV/2)/T_{L}}+1}, \quad 
f_R= \frac{1}{e^{(\epsilon_{N}-\bar \mu+eV/2)/T_{R}}+1},
\end{equation}
with the applied bias voltage $eV=\mu_L-\mu_R$ (we assume $V\ge 0$), the mean chemical potential $\bar\mu=(\mu_L+\mu_R)/2$, and the lead temperatures $T_{L/R}$.   

Two paradigmatic limits investigated below are the large-bias regime ($eV\gg T_{L/R}$) and the linear-response regime ($eV\ll T_{L/R}$). (i) For large voltage bias, we may approximate  Eq.~\eqref{fermidist} by $f_L=1$ and $f_R=0$,  where Eq.~\eqref{ratesF} simplifies to
\begin{equation}\label{largebias}
    \Gamma_{1}\!=\!g_{L}\!\equiv\! g(1+\delta), \quad
    \Gamma_{N}\!=\!\gamma_{1}= 0,\quad
    \gamma_{N}\!=\!g_{R}\!\equiv\!g(1-\delta),
\end{equation}
with an overall system-environment coupling strength $g$.
For $\delta=0$, the driven chain preserves particle-hole symmetry. The parameter $\delta$ then quantifies particle-hole asymmetry.
(ii) In the linear-response regime, for simplicity, we assume $g_L=g_R=g$, $T_L=T_R=T$, and $\epsilon_1=\epsilon_N=\epsilon$.
From Eq.~\eqref{fermidist}, we get 
\begin{equation}
f_{L/R}\simeq \frac12[1+\delta \pm V/V_b]+{\cal O}(V^2),
\end{equation}
with ${\delta=-\tanh[(\epsilon-\bar\mu)/2T]}$ and 
$eV_b=2T\cosh^2[(\epsilon-\bar\mu)/2T]$.   Equation \eqref{ratesF} then yields the rates
\begin{eqnarray}
    \Gamma_{1} &=& \frac{g}{2}(1+\delta+\phi),\quad    \gamma_{1} = \frac{g}{2}(1-\delta-\phi), \nonumber \\ \label{smallbias}
    \Gamma_{N} &=& \frac{g}{2}(1+\delta-\phi),\quad    \gamma_{N} = \frac{g}{2}(1-\delta+\phi),
\end{eqnarray}
with the drive parameter $\phi \equiv V/V_b\ll 1$. Particle-hole asymmetry again corresponds to $\delta\ne 0$. 

\subsection{Information currents}\label{sec2b}

We next consider the total von Neumann information $I(\rho)$ for the general fermion chain
in Sec.~\ref{sec2a}, where $\rho=\rho(t)$. 
With the von Neumann entropy $S(\rho)$, it is defined as  
\begin{equation}\label{totalinfo}
I(\rho) = N-S(\rho) = N+\mathrm{Tr}\left(\rho\log_{2}\rho\right).
\end{equation}
The von Neumann information (or total information) $I(\rho)$ is the average number of bits that can be predicted about measurement outcomes in the asymptotic limit of infinitely many measurements from the quantum state $\rho$~\cite{Shannon1948, Klein2022}. Operationally, one considers an infinite number of identical copies of the state $\rho$, performs infinitely many optimal projective measurements---one per copy---and counts how many binary outcomes are predetermined rather than random. For example, a pure qubit state allows one such binary outcome to be predicted with certainty when measured in its eigenbasis, yielding $I(\rho) =1$ bit, whereas a maximally mixed state allows none. Hence, $I(\rho)$ quantifies the predictive advantage of knowing $\rho$ over complete ignorance.

From the time evolution of $I(\rho)$, we can extract
local information currents flowing between the system and the environment.
In fact, using $\mathrm{Tr} \dot{\rho}=0$, which follows from the LME,  we get 
\begin{equation} \label{I-derivative}
    \frac{d}{dt}I(\rho)= \mathrm{Tr}\left(\dot{\rho}\log_{2}\rho\right).
\end{equation}
Inserting Eq.~\eqref{eq:rho-equ} for $\dot{\rho}$, the contribution from the unitary term vanishes because of $\left[\rho,\log_{2}\rho\right]=0$ and the cyclic property of the trace. 
Indeed, for a closed system (all $L_k=0$), $I(\rho)$ must be conserved.  
However, for open systems, 
\begin{align}\label{eq:I-der-rho}
\frac{d}{dt}I(\rho)&=-\mathcal{I}_{\rm tot}=-\!\sum_{k}\mathcal{I}_{k},\\
\mathcal{I}_{k} &= -\mathrm{Tr}\left[\left( {\cal D}[L_k]\rho \right)\log_{2}\rho\right],\label{eqIk}
\end{align}
where $\mathcal{I}_{k}$ is the information current flowing between
the system and the environment due to the jump operator $L_k$ and $\mathcal{I}_{\rm tot}$ is the total information current.

We now turn to the von Neumann information $I_A \equiv I(\rho_A)$ for an arbitrary 
subsystem $A$ of the fermion chain with size $N_{A}<N$, where the time-dependent reduced density matrix is $\rho_{A}=\mathrm{Tr}_{\bar{A}}\rho$ with 
the complement $\bar{A}$ of $A$, see Fig.~\ref{fig1}.
In analogy to Eq.~\eqref{totalinfo}, 
\begin{equation} \label{eq:info_subsystem} 
I_{A}=N_{A}+\mathrm{Tr}_A\left(\rho_{A}\log_{2}\rho_{A}\right),
\end{equation}
with $S_A = S(\rho_A) = - \mathrm{Tr}_A\left(\rho_{A}\log_{2}\rho_{A}\right)$.
By differentiating with respect to time,  similarly to Eq.~\eqref{I-derivative}, we find
\begin{equation} \label{IA-derivative}
    \frac{d}{dt}I_A(\rho)= \mathrm{Tr}_A \left(\dot{\rho}_A\log_{2}\rho_A\right),
\end{equation}
with
\begin{align} \label{rhoA-lindblad}
    \dot{\rho}_A(t) =& -i[H_A,\rho_A(t)] - i \mathrm{Tr}_{\bar{A}}  \left[H_{A,A_L},\rho(t)\right] \\\nonumber
    & - i \mathrm{Tr}_{\bar{A}} \left[H_{A,A_R},\rho(t)\right] +\sum_k \mathrm{Tr}_{\bar{A}} \left( {\cal D}[L_k]\rho(t) \right),
\end{align}
where $H_A$, $H_{A,A_L}$, and $H_{A,A_R}$ are Hamiltonian contributions involving only sites in $A$, in $(A,\bar{A}_L)$, and in $(A, \bar{A}_R)$, respectively. Inserting Eq.~\eqref{rhoA-lindblad} into Eq.~\eqref{IA-derivative} and noting that the first contribution in Eq.~\eqref{rhoA-lindblad} vanishes,
we find
\begin{equation}\label{IAdeco}
\frac{dI_{A}}{dt} =-\mathcal{I}_{{L}}-\mathcal{I}_{{R}}-\mathcal{I}_{E},
\end{equation}
with information currents originating from the left (${\cal I}_L)$ or right (${\cal I}_R)$ subsystem $\bar{A}_{L/R}$ and from the environment (${\cal I}_E$).

By considering Eqs.~\eqref{eq:I-der-rho}, \eqref{eq:info_subsystem}, and \eqref{IAdeco} for every possible subsystem $A$ forming a connected segment, 
we can construct the information current lattice.
To that end, we label arbitrary connected subsystems by $A=(\ell,n)$, where $\ell=N_{A}-1$ (with $\ell=0,\ldots,N-1$) determines the layer of the information lattice and $n=j_{L}+\ell/2$ specifies the midpoint of the segment $A$.    
We refer to $I_{(\ell,n)}$ ($i_{(\ell,n)}$) as information pertaining to a ``triangle''  (``site'') of the information lattice.
Using $I_{A=(\ell,n)}$ in Eq.~\eqref{eq:von_neumann_entropy}, we define local (scale-resolved) information as~\cite{Klein2022, Artiaco2025}
\begin{equation}\label{eq:triangles-to-sites}
i_{(\ell,n)}=I_{(\ell,n)}-I_{(\ell-1,n-\frac{1}{2})}-I_{(\ell-1,n+\frac{1}{2})}+I_{(\ell-2,n)} ,
\end{equation} 
with the convention $I_{(\ell<0,n)}= 0$. As a consequence, we have $i_{(0,n)}=I_{(0,n)}$ for
the bottom layer $\ell=0$ of the information lattice, and $i_{(1,n)}=I_{(1,n)}-I_{(0,n-\frac{1}{2})}-I_{(0,n+\frac{1}{2})}$ for layer $\ell=1$. 
Moreover, the total information in Eq.~\eqref{totalinfo} is represented by $I(\rho)=I_{(N-1,(N+1)/2)}$. 
The local information $i_{(\ell,n)}$ measures the additional predictive power gained by having access to the reduced density matrix of the subsystem $A=(\ell,n) $, beyond what can already be inferred from the reduced density matrices of all smaller subsystems contained within $A$. By construction, this quantity is non-negative, $i_{(\ell,n)} \ge 0$. As an example, consider the two-site singlet state, $\lvert \psi_{\text{singlet}} \rangle = \frac{1}{\sqrt{2}}\bigl(\lvert 10 \rangle - \lvert 01 \rangle\bigr) $. The reduced density matrix of each individual site is maximally mixed, implying $i_{(0,1)} = i_{(0,2)} = 0$. Consequently, knowledge of the singlet state does not enhance predictability for single-site measurements, since both outcomes for any measurement occur with equal probability $1/2$. By contrast, the value $i_{(1,\frac{3}{2})} = 2$ indicates that the full information content of $\lvert \psi_{\text{singlet}} \rangle$---amounting to two bits---becomes available only through measurements acting jointly on both sites.

Below, we use unitary information currents flowing from $A=(\ell,n)$ to the left ($L$) or right ($R$) site on the subsequent layer $\ell+1$ of the information lattice,  
\begin{equation}\label{unit}
\mathcal{I}_{(\ell,n),R/L} \equiv \mathcal{I}_{(\ell,n)\to (\ell+1,n\pm 1/2)} ,  
\end{equation}
with the back-currents  given by  
\begin{equation}
\mathcal{I}_{(\ell+1,n\pm 1/2)\to (\ell,n)}= -\mathcal{I}_{(\ell,n),R/L}.
\end{equation}
The information currents \eqref{unit} are caused by unitary contributions to the LME~\cite{Klein2022, Artiaco2024}.
In addition, information currents $\mathcal{I}_{(\ell,n),E}$ may flow from segment $A=(\ell,n)$ to the environment, with a respective back-current of opposite sign. 
We note that, in more general models with long-range hopping, information currents can flow from layer $\ell$ to any layer up to $\ell+r$, where $r$ is the range of the hopping or interaction terms in the Hamiltonian. In the following, we focus on nearest-neighbor Hamiltonians resulting in currents of the form in Eq.~\eqref{unit}. 

In order to acquire intuition, we first study these information currents for the
simple cases $N=2$ and $N=3$ in Sec.~\ref{sec2c}. In Sec.~\ref{sec2d}, we then turn to arbitrary $N$. In what follows,  
we take diagonal matrices $\Gamma$ and $\gamma$, see Eq.~\eqref{Gamdiag}.  

\subsection{Information currents for \texorpdfstring{$N=2,3$}{N=2,3}}\label{sec2c}

\begin{figure}
\center
\includegraphics[width=0.9\linewidth]{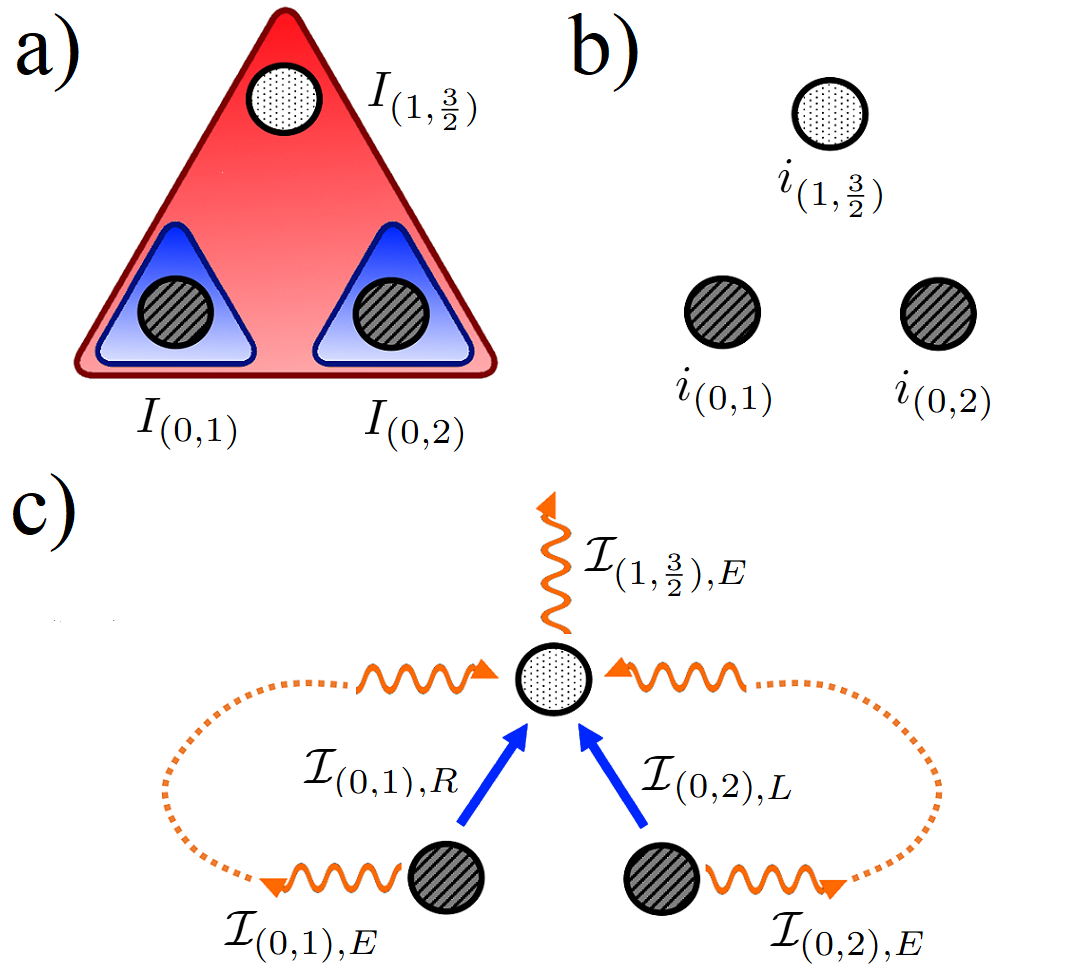}
\caption{Information lattice and information currents for system size $N=2$.  Circles indicate the possible subsystems $A=(\ell, n)$. Filled circles on the bottom layer $\ell=0$ correspond to physical sites $(n=1,2)$ representing information in the single-site density matrices. a) Triangle representation of the information lattice using $I_{(\ell,n)}$ in  Eq.~\eqref{eq:von_neumann_entropy}. b) Site representation using $i_{(\ell,n)}$ in Eq.~\eqref{eq:triangles-to-sites}. c) Information currents in the site representation, see Eq.~\eqref{eq:info-currents}, with the unitary terms \eqref{unitarycurr2} shown as blue arrows and the dissipative terms \eqref{disscurr2} as wavy orange arrows. Dotted lines refer to the interpretations 
discussed in the text.}
\label{fig2}
\end{figure}

Let us start with the case  $N=2$. The corresponding information lattice is 
depicted in Fig.~\ref{fig2}, where the three possible subsystems 
  $A=(\ell,n)$  
are   $\{(0,1)$, $(0,2),(1,\frac{3}{2})\}$. The triangular representation of the 
information lattice in terms of $I_{(\ell,n)}$ is shown in Fig.~\ref{fig2}(a), and the alternative site representation employing $i_{(\ell,n)}$ in Fig.~\ref{fig2}(b). 
From Eq.~\eqref{IAdeco}, the information currents 
in the triangular representation follow by computing $\frac{d}{dt}I_A$ for all choices of $A=(\ell,n)$.  

Using  the notation in Eq.~\eqref{unit}, we obtain
\begin{eqnarray}\nonumber
\frac{dI_{(0,1)}}{dt} & =& -\mathcal{I}_{(0,1),R}- \mathcal{I}_{(0,1),E},  \\
\label{infocurN2}
\frac{dI_{(0,2)}}{dt} & =& -\mathcal{I}_{(0,2),L}- \mathcal{I}_{(0,2),E}, \\ \nonumber
\frac{dI_{(1,\frac{3}{2})}}{dt} &=& - \mathcal{I}_{(1,\frac{3}{2}),E}.
\end{eqnarray}
To switch to the site representation in Fig.~\ref{fig2}(b), we use Eq.~\eqref{eq:triangles-to-sites} and obtain the local information currents for an open fermionic chain with $N=2$,
\begin{eqnarray}
\frac{di_{(0,1)}}{dt} & =& -\mathcal{I}_{(0,1),R} -\mathcal{I}_{(0,1),E},\nonumber \\
\frac{di_{(0,2)}}{dt} & =& -\mathcal{I}_{(0,2),L} -\mathcal{I}_{(0,2),E},
\label{eq:info-currents}\\ 
\nonumber \frac{di_{(1,\frac{3}{2})}}{dt} & =& - \mathcal{I}_{(1,\frac{3}{2}),E} + 
\mathcal{I}_{(0,1),R} +  \mathcal{I}_{(0,2),L} + \sum_n
\mathcal{I}_{(0,n),E}.
\end{eqnarray}
These information currents are illustrated in Fig.~\ref{fig2}(c).

We note that all unitary information currents sum up to zero since in a closed system total information is conserved. In fact, for nearest-neighbor Hamiltonians, unitary
information currents only connect neighboring layers of the information lattice~\cite{Klein2022, Artiaco2024}. In particular, they do not flow horizontally within a given layer.  
It is tempting to give a similar interpretation to the dissipative information currents, claiming that they flow between neighboring layers as well. For instance, $\mathcal{I}_{(0,n),E}$ flows out from information lattice site $i_{(0,n)}$ and flows back into $i_{(1,\frac{3}{2})}$, see Fig.~\ref{fig2}(c). In effect, the only information current truly flowing out of the system is then given by $\mathcal{I}_{(1,\frac{3}{2}),E}$. In principle, this is a valid interpretation for $N=2$. However, as we discuss below, a more physical viewpoint emerges by demanding that dissipative information currents connect each subsystem $A$ directly with the environment. 

\begin{figure}
\center
\includegraphics[width=0.9\linewidth]{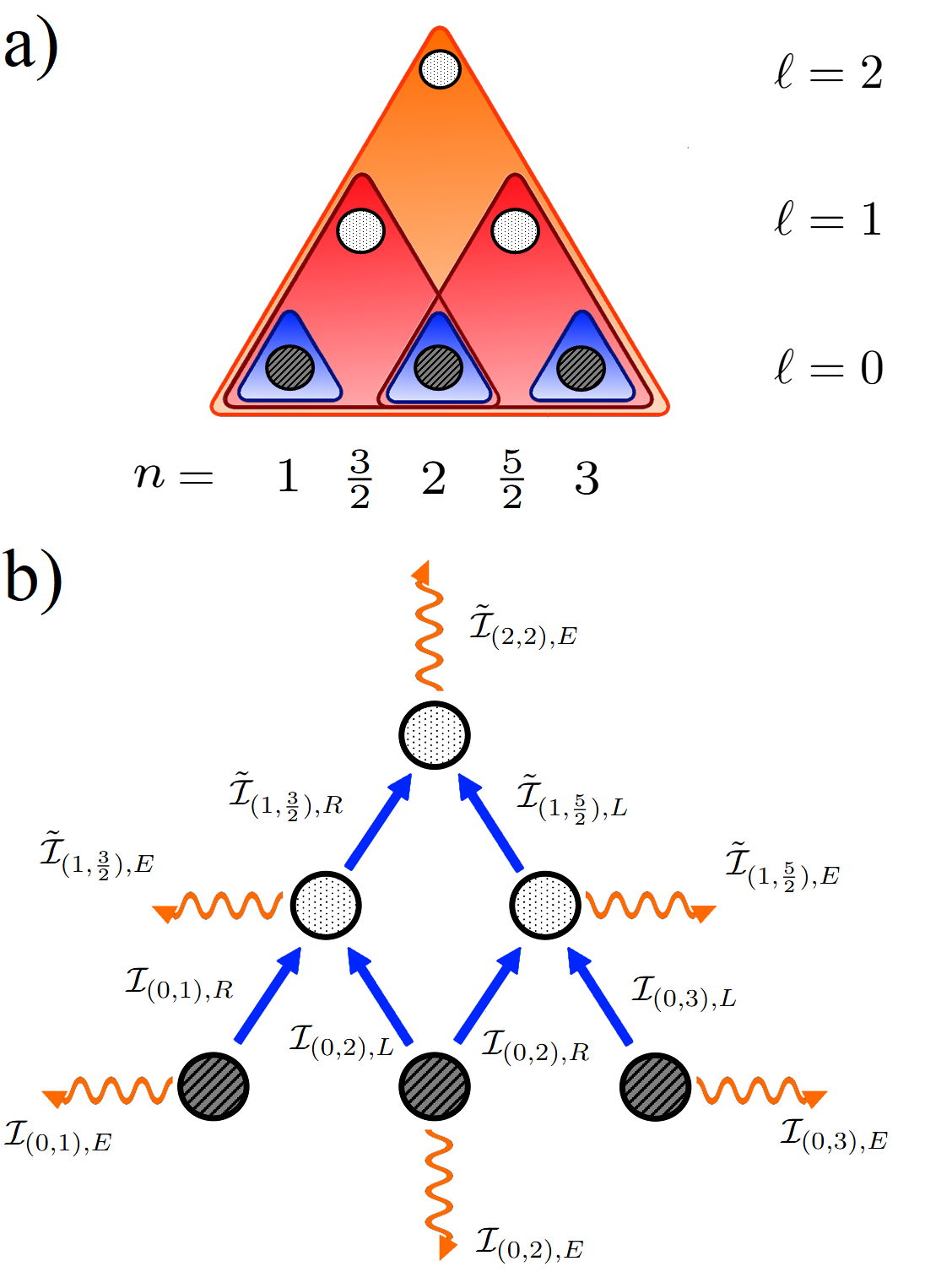}
\caption{Information lattice and information currents for $N=3$. 
Circles indicate all possible subsystems $A=(\ell, n)$. Filled circles on the bottom layer  correspond to physical sites.  a) Triangular representation of the information lattice.  b) Effective information currents $\tilde{\cal I}_{(\ell,n),R/L/E}$ in site representation from Eqs.~\eqref{conservN3} and \eqref{effcurrent}, distinguishing unitary (blue arrows) and dissipative terms (wavy orange arrows). For the bottom layer $\ell=0$, one finds $\tilde{\cal I}_{(0,n),R/L/E}= {\cal I}_{(0,n),R/L/E}$. }
\label{fig3}
\end{figure}

We now turn to the case $N=3$, see also Appendix \ref{appendix:exact for N=3}.  From this case, the information currents for arbitrary $N$ can be deduced rather straightforwardly, see Sec.~\ref{sec2d}. 
The triangular representation in Fig.~\ref{fig3}(a) then consists of six  terms,
$\{ I_{(0,1)},I_{(0,2)},I_{(0,3)},I_{(1,\frac{3}{2})},I_{(1,\frac{5}{2})},I_{(2,2)}\}$.
  Using Eq.~\eqref{IAdeco}, we obtain the corresponding conservation laws for information currents, 
\begin{eqnarray}
\frac{dI_{(0,1)}}{dt} & =&-\mathcal{I}_{(0,1),R}-\mathcal{I}_{(0,1),E},\nonumber \\
\frac{dI_{(0,2)}}{dt} & =&-\mathcal{I}_{(0,2),L}-\mathcal{I}_{(0,2),R}-\mathcal{I}_{(0,2),E},\nonumber \\
\frac{dI_{(0,3)}}{dt} & =&-\mathcal{I}_{(0,3),L}-\mathcal{I}_{(0,3),E}, \nonumber\\
\frac{dI_{(1,\frac{3}{2})}}{dt} & =&-\mathcal{I}_{(1,\frac{3}{2}),R}-\mathcal{I}_{(1,\frac{3}{2}),E},\nonumber \\
\nonumber \frac{dI_{(1,\frac{5}{2})}}{dt} & =&-\mathcal{I}_{(1,\frac{5}{2}),L}-\mathcal{I}_{(1,\frac{5}{2}),E} ,\\
\label{curr3}
\frac{dI_{(2,2)}}{dt} & = & -\mathcal{I}_{(2,2),E}. 
\end{eqnarray}
As for $N=2$, explicit expressions for the information currents in Eq.~\eqref{curr3} follow from Eq.~\eqref{eq:curr_block}. 
However, we here do not specify the resulting lengthy expressions since they are not particularly instructive.

From Eq.~\eqref{eq:triangles-to-sites}, the information current conservation laws in site representation then follow as, see Fig.~\ref{fig3}(b) for an illustration, 
\begin{eqnarray}
\frac{di_{(0,1)}}{dt} & =&-\mathcal{I}_{(0,1),R}-\mathcal{I}_{(0,1),E},\nonumber \\
\frac{di_{(0,2)}}{dt} & =&-\mathcal{I}_{(0,2),L}-\mathcal{I}_{(0,2),R}-\mathcal{I}_{(0,2),E},\nonumber \\
\frac{di_{(0,3)}}{dt} & =&-\mathcal{I}_{(0,3),L}-\mathcal{I}_{(0,3),E},\nonumber  \\
\nonumber
\frac{di_{(1,\frac{3}{2})}}{dt} & =&\mathcal{I}_{(0,1),R}+\mathcal{I}_{(0,2),L}-\tilde{\mathcal{I}}_{(1,\frac{3}{2}),R}-\tilde{\mathcal{I}}_{(1,\frac{3}{2}),E},\\ \nonumber 
\frac{di_{(1,\frac{5}{2})}}{dt} & =&\mathcal{I}_{(0,2),R}+\mathcal{I}_{(0,3),L}-\tilde{\mathcal{I}}_{(1,\frac{5}{2}),L}-\tilde{\mathcal{I}}_{(1,\frac{5}{2}),E},\\ 
\frac{di_{(2,2)}}{dt} & = &\tilde{\mathcal{I}}_{(1,\frac{3}{2}),R}+\tilde{\mathcal{I}}_{(1,\frac{5}{2}),L}-\tilde{\mathcal{I}}_{(2,2),E} ,\label{conservN3}
\end{eqnarray}
where we introduce effective information currents (denoted by the tilde in $\tilde{\cal I}_{(\ell,n),R/L/E}$):
\begin{eqnarray}
\tilde{\mathcal{I}}_{(1,\frac{3}{2}),R} & =&\mathcal{I}_{(1,\frac{3}{2}),R}-\mathcal{I}_{(0,2),R},\nonumber \\
\tilde{\mathcal{I}}_{(1,\frac{5}{2}),L} & =&\mathcal{I}_{(1,\frac{5}{2}),L}-\mathcal{I}_{(0,2),L},\nonumber \\
\tilde{\mathcal{I}}_{(1,\frac{3}{2}),E} & =&\mathcal{I}_{(1,\frac{3}{2}),E}-\mathcal{I}_{(0,1),E}-\mathcal{I}_{(0,2),E},  \nonumber \\
\tilde{\mathcal{I}}_{(1,\frac{5}{2}),E} & =&\mathcal{I}_{(1,\frac{5}{2}),E}-\mathcal{I}_{(0,2),E}-\mathcal{I}_{(0,3),E},\label{effcurrent}\\ \nonumber
\tilde{\mathcal{I}}_{(2,2),E}  & =& \mathcal{I}_{(2,2),E}-\tilde{\mathcal{I}}_{(1,\frac{3}{2}),E}-\tilde{\mathcal{I}}_{(1,\frac{5}{2}),E} -\sum_{n=1}^3\mathcal{I}_{(0,n),E} .
\end{eqnarray}

We emphasize that the passage from $\{I_{(\ell,n)}\}$ to $\{i_{(\ell,n)}\}$ is necessary to truly define a ``local'' information lattice. To illustrate this point, let us consider the case $N=3$ in Fig.~\ref{fig3}. While information flowing away from the ``triangle" $I_{(1,\frac{3}{2})}$ is absorbed by $I_{(2,2)}$, we are not able to decompose this information into a contribution that delocalizes on all three sites and a contribution where information flows from $I_{(0,2)}$ into $I_{(1,\frac{5}{2})}$, thus affecting only two sites of the chain. 
The decomposition in terms of $\{i_{(\ell,n)}\}$ resolves this question since $i_{(\ell,n)}$ is the information pertaining to subsystem $A=(\ell,n)$ that is not already contained in any sub-subsystem.  Nonetheless, the triangular decomposition is also of interest, e.g., for developing a hydrodynamic approach.  We therefore discuss both decompositions in this paper.

The  effective information currents $\tilde{\cal I}_{(\ell,n),R/L/E}$ appearing in Eq.~\eqref{effcurrent} have the following properties. First, the unitary information currents  $\tilde{\cal I}_{(\ell,n), R/L}$ only flow in the vertical direction between $(\ell,n)$ and $(\ell+1,n\pm 1/2)$. Second, all dissipative information currents $\tilde{\cal I}_{(\ell,n),E}$ are directly exchanged with the environment for each subsystem $A=(\ell,n)$. 
For instance,  the unitary term $\tilde{\mathcal{I}}_{(1,\frac{3}{2}),R}$ flowing away from the site $i_{(1,\frac{3}{2})}$ is obtained by subtracting from $\mathcal{I}_{(1,\frac{3}{2}),R}$ the current $\mathcal{I}_{(0,2),R}$ flowing out from $i_{(0,2)}$. For the bottom layer of the information lattice, we then define $\tilde{\cal I}_{(0,n),R/L/E}={\cal I}_{(0,n),R/L/E}$. 
The effective information currents $\tilde{\cal I}_{(\ell,n),R/L/E}$ are our key quantities of interest, and we simply refer to them as ``information currents'' below.  
Finally, we note that all unitary (dissipative) information currents sum up to zero $\left(\mathcal{I}_{(2,2),E} \right)$.

\begin{figure}
\center
\includegraphics*[width=0.9\linewidth]{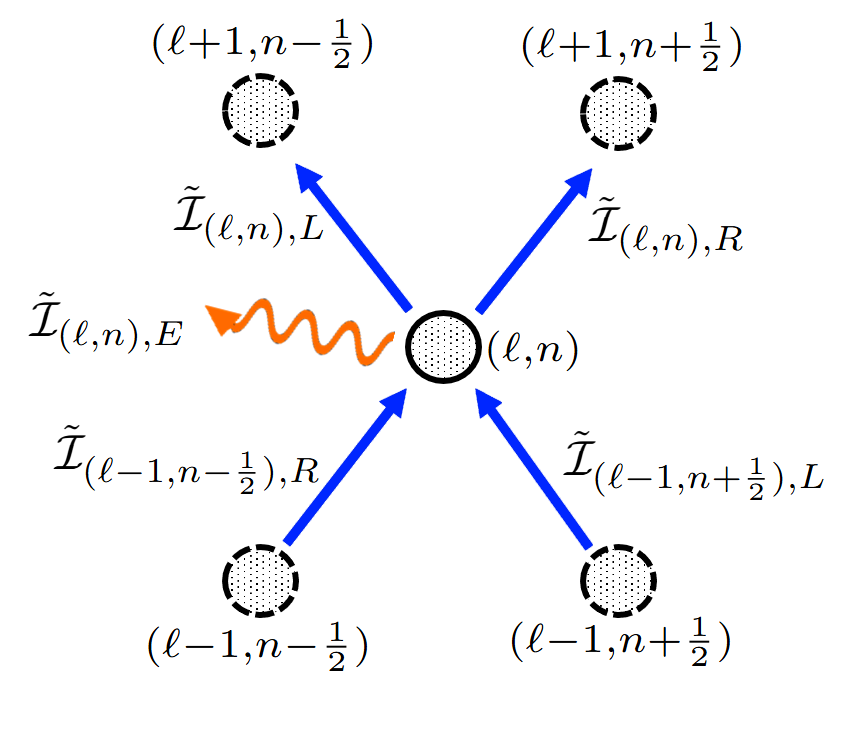}
\caption{Information currents for arbitrary $N$ at a generic site $(\ell,n)$ of the information lattice, see Eq.~\eqref{fig4formula}. Unitary (dissipative) terms are shown as blue  (wavy orange) arrows. }
\label{fig4}
\end{figure}

\subsection{Information currents for arbitrary $N$}\label{sec2d}

We have now collected all ingredients needed for specifying the information currents for an arbitrary system size $N$.  In the triangular representation, they follow again from Eq.~\eqref{IAdeco}.
Using Eq.~\eqref{eq:triangles-to-sites}, we then find the general expression
\begin{eqnarray}\label{fig4formula}
\frac{di_{(\ell,n)}}{dt} &=& -\tilde{\mathcal{I}}_{(\ell,n),L} -\tilde{\mathcal{I}}_{(\ell,n),R} \label{eq:lattice_currents} \\ \nonumber
&+&\tilde{\mathcal{I}}_{(\ell-1,n-\frac12),R} +\tilde{\mathcal{I}}_{(\ell-1,n+\frac12),L}  -\tilde{\mathcal{I}}_{(\ell,n),E},
\end{eqnarray}
where we have generalized  Eq.~\eqref{effcurrent} for the local information currents to arbitrary $N$,
\begin{eqnarray}
\tilde{\mathcal{I}}_{(\ell,n),L} & = & \mathcal{I}_{(\ell,n),L}-\mathcal{I}_{(\ell-1,n-\frac12),L}, \nonumber \\
\tilde{\mathcal{I}}_{(\ell,n),R} & = & \mathcal{I}_{(\ell,n),R}-\mathcal{I}_{(\ell-1,n+\frac12),R},  \label{eq:effective_currents}\\
\tilde{\mathcal{I}}_{(\ell,n),E} & = & \mathcal{I}_{(\ell,n),E}-\mathcal{I}_{(\ell-1,n-\frac12),E} \nonumber \\\nonumber
&-& \mathcal{I}_{(\ell-1,n+\frac12),E}+\mathcal{I}_{(\ell-2,n),E},
\end{eqnarray}
with the convention ${\cal I}_{(\ell<0,n),R/L/E}=0$.
The conservation law \eqref{eq:lattice_currents} is illustrated in Fig.~\ref{fig4}. We note that in the presence of long-range hopping terms of range $r$ in the system Hamiltonian, nonlocal information currents connecting site $(\ell,n)$ and $(\ell \pm x,n \pm \frac{x}{2})$ with $x=1,\ldots,r$ can emerge as a consequence of the nonlocal charge currents flowing between lattice sites at distances up to the maximum hopping range $r$.
Concerning the dissipative information currents, we note that environmental couplings acting at physical site $j$ can only affect the $i_{(\ell,n)}$ such that $j \in A=(\ell,n)$. For environments coupled only to the chain ends, see Fig.~\ref{fig1}, we conclude that only the outer two diagonals of the information lattice come with dissipative information currents. 
Moreover, the information current flowing out of the total system is $\mathcal{I}_{(N-1,\frac{N+1}{2}),E}$.

In order to characterize the information flow inside the information lattice in terms of  ``global'' currents, it is convenient to also define ``horizontal" information currents, 
\begin{eqnarray}
\mathcal{J}_{n^+}^{\mathrm{h}} &=& \sum_{\mathrm{even}\, \ell} \tilde{\mathcal{I}}_{(\ell,n),R} - \sum_{\mathrm{odd}\, \ell} \tilde{\mathcal{I}}_{(\ell,n+\frac12),L}, \nonumber \\
\mathcal{J}_{n^-}^{\mathrm{h}} &=& \sum_{\mathrm{even}\, \ell} \tilde{\mathcal{I}}_{(\ell,n),L} - \sum_{\mathrm{odd}\, \ell} \tilde{\mathcal{I}}_{(\ell,n-\frac12),R},
\label{eq:summed_currents_horizontal}
\end{eqnarray}
where summations extend over the possible values $0<\ell<N-1$ and $1+\frac{\ell}{2}<n<N-\frac{\ell}{2}$. Hence $\mathcal{J}_{n^\pm}^{\mathrm{h}}$ gives the total information current crossing a vertical line (strictly) located between $n$ and $n\pm \frac12$, respectively. Note that $\mathcal{J}_{n^+}^{\mathrm{h}} \neq \mathcal{J}_{(n+1)^-}^{\mathrm{h}}$ due to the dissipative currents at $n+\frac12$.  
Equation \eqref{eq:summed_currents_horizontal} can be simplified by switching from the effective information currents $\tilde{\cal  I}_{(\ell,n),L/R}$ given by Eq.~\eqref{eq:effective_currents} back to the information currents ${\cal I}_{(\ell,n),L/R}$ in triangular representation. 
With $\mathcal{I}_{(\ell,n)}=\mathcal{I}_{(\ell,n),L}+\mathcal{I}_{(\ell,n),R}$, we find
\begin{equation}\label{eq:summed_currents2}
\mathcal{J}_{n^\pm}^{\mathrm{h}} = \sum_{\mathrm{even}\, \ell}
\mathcal{I}_{(\ell,n)} - \sum_{ \mathrm{odd}\, \ell } \mathcal{I}_{(\ell,n\pm\frac12)} .
\end{equation}
We conclude that $\mathcal{J}_{n^\pm}^{\mathrm{h}}$ represents the difference between all outgoing information currents from information triangles centered at site $n$ and those at $n\pm \frac12$.
Similarly, we define ``vertical" information currents,
\begin{equation}
\mathcal{J}_{\ell}^{\mathrm{v}} = \sum_{n} \left[ \tilde{\mathcal{I}}_{(\ell,n),L} + \tilde{\mathcal{I}}_{(\ell,n),R} \right] ,
\label{eq:summed_currents_vertical}
\end{equation}
representing the total information current flowing from layer $\ell\to \ell+1$ of the information lattice (summed over all sites). Note that 
$\mathcal{J}_{n^\pm}^{\mathrm{h}}$ and $\mathcal{J}_\ell^{\rm v}$ are due to unitary terms.

\subsection{Particle vs. information currents}\label{sec2f}

One may wonder whether the above information currents can be related to conventional
particle currents. 
For a generic system, we split the Hamiltonian, $H=H^{({\rm sp})}+H^{({\rm int})}$, into a single-particle term $H^{({\rm sp})}$, with onsite energies $\epsilon_j$ and hopping strengths $J_{j,k}$ between sites $j$ and $k$, and an interacting part $H^{({\rm int})}$. Assuming only nearest-neighbor hopping
(similar relations holds for longer-range hopping),  $H^{({\rm sp})}_{j,j+1}=J_{j,j+1} \left(c^\dagger_{j} c^{}_{j+1} + c_{j+1}^\dagger c_{j}^{} \right)$ gives rise to a particle current from site $j$ to site $j+1$ described by the operator
\begin{equation}\label{opcurr}
    \hat {\cal I}^{(p)}_{j\to j+1}= iJ_{j,j+1} \left(c^\dagger_{j} c^{}_{j+1} - c_{j+1}^\dagger c_{j}^{} \right).
\end{equation}
The expectation value ${\cal I}^{(p)}_{j\to j+1}=\langle\hat {\cal I}^{(p)}_{j\to j+1}\rangle$, as well as the particle currents flowing from the system into the environment,
can again be expressed by using subsystem density matrices $\rho_A$, in analogy to Eqs.~\eqref{eq:I-der-rho} and~\eqref{eq:info_subsystem} for the information currents. Indeed, with $A$ corresponding to the sites $(j,j+1)$, we have
\begin{equation}\label{particlecurrentgeneral1}
    {\cal I}^{(p)}_{j\to j+1} = \mathrm{Tr}_{A}\left
    (\rho_A \hat {\cal I}^{(p)}_{j\to j+1}\right).
\end{equation}
Similarly, the particle current flowing from site $j$ into the environment is
described by ($\hat n_j=c_j^\dagger c_j^{}$)
\begin{equation}\label{particlecurrentgeneral2}
{\cal I}^{(p)}_{j\to E}=\mathrm{Tr}_{A}\left
    ( [\gamma_j \hat n_j-\Gamma_j(1-\hat n_j) ] \rho_A\right),
\end{equation}
with $A$ now corresponding to only site $j$.
Since the particle currents \eqref{particlecurrentgeneral1} and \eqref{particlecurrentgeneral2}, as well as the information currents 
appearing in Eq.~\eqref{IAdeco}, are determined by the same set of density operators $\rho_A$, one expects that both types of currents must be related to each other.  We derive this relation in detail for quasi-free fermions in Sec.~\ref{sec22b}. In general terms, steady-state particle currents defining the dc transport may then be considered as a source for information currents.  
In view of the  continuity equation in Eqs.~\eqref{fig4formula} and \eqref{eq:effective_currents} for local and scale-resolved information currents under nonequilibrium transport conditions, this intimate connection between information and particle currents is expected to shape  information patterns in concrete settings.
 
\section{Implementation for quasi-free fermions}\label{sec22}

In this section, we implement the theoretical framework developed in Sec.~\ref{sec2} to the exactly solvable model of a quasi-free fermion chain under transport conditions.
In Sec.~\ref{sec22a}, we introduce the model and the correlation matrix, and thereby adapt the general theory to this solvable limit.   We also express the information currents in terms of the correlation matrix.
In Sec.~\ref{sec22b}, we show how  local information currents and standard particle currents are related.
This relation opens up experimental routes to measure information currents.  
In particular, the information currents connecting layers 
$\ell=0$ and $\ell=1$ of the information lattice are fully determined by standard particle currents and onsite occupation numbers.  
Finally, in Sec.~\ref{sec22c}, by connecting the information lattice to a noise lattice, defined in terms of the particle-number fluctuations (variances) for all possible subsystems,  we show that the information lattice can be experimentally measured even for large system sizes.  In a similar fashion, we show that one can also measure the associated information currents in terms of particle currents, onsite occupations, and covariances of particle currents and/or particle numbers.

\subsection{Model and information currents}\label{sec22a}

The noninteracting Hamiltonian reads as
\begin{equation}
    H=\sum_{i,j=1}^N c_i^{\dagger} \mathcal{H}_{ij} c_j^{},
\end{equation}
with the Hermitian $N\times N$ matrix ${\cal H}$. Assuming real-valued nearest-neighbor hopping amplitudes $J_{j,j+1}$ and onsite energies $\epsilon_j$, 
${\cal H}$ has the tridiagonal form
\begin{equation}\label{HN3}
\mathcal{H}=\left(\begin{array}{ccccc}
\epsilon_{1} & J_{12}       & 0            & \cdots & 0 \\
J_{12}       & \epsilon_{2} & J_{23}       & \cdots & 0 \\
0            & J_{23}       & \epsilon_3   & \cdots & 0 \\
\vdots       &              &              &        & \vdots \\
0            & 0            &  \cdots      &  J_{N-1,N}   & \epsilon_N
\end{array}\right).
\end{equation}
The LME for a noninteracting Hamiltonian with jump operators linear in the fermionic operators describes the time evolution of quasi-free fermions. 

To proceed, we note that for quasi-free fermions with Gaussian initial states $\rho(0)$, the LME \eqref{eq:rho-equ} preserves Gaussianity of $\rho(t)$ at all times.  
The dynamics of the density matrix can thus be equivalently expressed in terms of the Hermitian $N\times N$ correlation matrix $C(t)$ with matrix elements \cite{Fazio2025}
\begin{equation}\label{corrdef}
C_{ij}(t) =\mathrm{Tr}\left[ \rho(t) c_{i}^{\dagger} c_{j}^{}\right].
\end{equation}
We note that $C$ is positive semidefinite ($C \succeq 0$), where diagonal elements
correspond to occupation numbers and ${\rm Tr}[C]$ is the expectation value of the total particle number.  
The Pauli exclusion principle also imposes $\mathbb{I}-C \succeq 0$, such that all eigenvalues $\nu_i$ of $C$ must satisfy $0 \le \nu_i \le 1$. 
From Eqs.~\eqref{eq:rho-equ} and \eqref{jumpop}, the equation of motion for $C(t)$ follows as \cite{Guimaraes_2016,Nava_2021,Berger2025} 
\begin{equation}
\dot{C}(t) =i\left [\mathcal{H}^{T} , C(t) \right] +\Gamma-\frac12 \{\Gamma+\gamma,C(t) \}.\label{eq:C-equ}
\end{equation}
Instead of dealing with the dynamics of the full $2^{N}\times2^{N}$ density matrix, we then only need to determine the time-dependent $N\times N$ correlation matrix.

For quasi-free fermions, the total von Neumann information $I(\rho)$, Eq.~\eqref{totalinfo}, can equivalently be expressed in terms of the correlation matrix $C=C(t)$ in Eq.~\eqref{corrdef} \cite{Peschel_2009}.
With the identity matrix $\mathbb{I}$, 
\begin{equation}\label{IC}
I(\rho) =N+\mathrm{Tr}\left[C\log_{2}C+(\mathbb{I}-C)\log_{2}(\mathbb{I}-C)\right].
\end{equation}
In terms of $C=C(t)$, the total information current (\ref{eq:I-der-rho}) is written as
\begin{eqnarray}\nonumber
\mathcal{I}_{\rm tot} &=& -\mathrm{Tr}\left(\dot{C}\,\log_{2}\frac{C}{\mathbb{I}-C}\right)
\\ \label{Itot2}
&=&-\mathrm{Tr}\left\{\bigl[ \Gamma - (\Gamma+\gamma) C\bigr] \log_{2}\frac{C}{\mathbb{I}-C}\right\},
\end{eqnarray}
where the second line follows from Eq.~\eqref{eq:C-equ}. 

The von Neumann information $I_A$ of a subsystem $A$, Eq.~\eqref{eq:info_subsystem}, can be fully expressed in terms of the reduced correlation matrix $C_{A}$ of dimension $N_A\times N_A$, obtained from $C$ by deleting all rows and columns related to $\bar{A}$. Similar to Eq.~\eqref{IC}, we find
\begin{align}
I_A\!&=\!N_A+\mathrm{Tr}_A\left[C_{A}\,\log_{2}C_{A}+(\mathbb{I}-C_{A})\,\log_{2}(\mathbb{I}-C_{A})\right]\nonumber \\
&\!=\!N_A+\sum_{i=1}^{N_A}\left[\,\mu_{i}\log_{2}\mu_{i}+(1-\mu_{i})\log_{2}(1-\mu_{i})\,\right],
 \label{eq:von_neumann_entropy}
\end{align}
 with the eigenvalues $\mu_i$ of $C_A$. 
As in Eq.~\eqref{Itot2}, we find 
\begin{equation}\label{eq:IA-der-CA} 
\frac{dI_{A}}{dt}=\mathrm{Tr}_A\left(\dot{C}_{A}\log_{2}\frac{C_{A}}{\mathbb{I}-C_{A}}\right).
\end{equation}
However, the dynamical equation for $C_{A}$ now depends on the full correlation matrix $C(t)$. In particular, in addition to information currents flowing between $A$ and the environment, information currents may now also connect $A$ and $\bar A$. 

To proceed, we express all matrices in block form with respect to subsystems $A$ and $\bar A$.  For concreteness, as sketched in Fig.~\ref{fig1}, we assume
$A$ to be a chain segment extending between sites $j_{L}$ and
$j_{R}$, i.e., $N_A=j_R-j_L+1$.  For the complement $\bar{A}=\bar{A}_L\cup \bar{A}_R$, the left subsystem $\bar{A}_{L}$ extends between sites $j=1$ and $j_{L}-1$, i.e., $N_{{\bar A}_L}=j_L-1$.
Similarly, $\bar{A}_{R}$ extends from $j=j_{R}+1$ to $N$, with $N_{{\bar A}_R}=N-j_R$. The full correlation matrix is written as 
\begin{equation}\label{Cblock}
C=\left(\begin{array}{ccc}
C_{\bar{A}_{L}} & C_{\bar{A}_{L},A} & C_{\bar{A}_{L},\bar{A}_{R}}\\
C_{\bar{A}_{L},A}^{\dagger} & C_{A} & C_{A,\bar{A}_{R}}\\
C_{\bar{A}_{L},\bar{A}_{R}}^{\dagger} & C_{A,\bar{A}_{R}}^{\dagger} & C_{\bar{A}_{R}} 
\end{array}\right).
\end{equation}
Similarly, we write the Hamiltonian matrix in the form
\begin{equation}\label{Hblock}
\mathcal{H}=\left(\begin{array}{ccc}
\mathcal{H}_{\bar{A}_{L}} & \mathcal{H}_{\bar{A}_{L},A} & \mathcal{H}_{\bar{A}_{L},\bar{A}_{R}}\\
\mathcal{H}_{\bar{A}_{L},A}^{\dagger} & \mathcal{H}_{A} & \mathcal{H}_{A,\bar{A}_{R}}\\
\mathcal{H}_{\bar{A}_{L},\bar{A}_{R}}^{\dagger} & \mathcal{H}_{A,\bar{A}_{R}}^{\dagger} & \mathcal{H}_{\bar{A}_{R}}
\end{array}\right) .
\end{equation}
In general, we have $N_{\bar{A}_{L}}\neq N_{A}\neq N_{\bar{A}_{R}}$, such that 
 off-diagonal blocks in Eqs.~\eqref{Cblock} and \eqref{Hblock} are not necessarily represented by square matrices.  
From Eq.~\eqref{eq:C-equ}, we can then extract the dynamical equation for $C_{A}$.

To simplify the derivation of this dynamical equation, we assume diagonal coupling 
matrices $\Gamma$ and $\gamma$, see Eq.~\eqref{Gamdiag}. In addition, we use ${\cal H}_{{\bar A}_L,{\bar A}_R}=0$, which  holds for the tridiagonal Hamiltonian ${\cal H}$ in Eq.~\eqref{HN3}.
It is convenient to define the auxiliary $N_A\times N_A$ matrices (keeping the $A$ dependence implicit)
\begin{eqnarray}
F_{\bar{A}_{L}} & =& \mathcal{H}_{\bar{A}_{L},A}^{T}C_{\bar{A}_{L},A}-C_{\bar{A}_{L},A}^{\dagger}\mathcal{H}_{\bar{A}_{L},A}^{*},\nonumber \\
F_{\bar{A}_{R}} & =&\mathcal{H}_{A,\bar{A}_{R}}^{*}C_{A,\bar{A}_{R}}^{\dagger}-C_{A,\bar{A}_{R}}\mathcal{H}_{A,\bar{A}_{R}}^T,\label{FDef} \\ \nonumber
F_{E} & =&\Gamma_{A}-\frac{1}{2}\left\{ \Gamma_{A}+\gamma_A,C_{A}\right\}.
\end{eqnarray}
From Eq.~\eqref{eq:C-equ}, we then obtain 
\begin{equation} \label{eq:CA-equ}
-i\dot{C}_{A} =\left[\mathcal{H}_{A}^{T},C_{A}\right] +
F_{\bar{A}_L} + F_{\bar{A}_R}-iF_E.
\end{equation}
The first term on the right-hand side describes the unitary dynamics within $A$, while the second
and third terms account for unitary evolution in $\bar{A}_{L}$ and $\bar{A}_{R}$, respectively. Finally, the $F_E$ term is due to the system-environment coupling within $A$. In a similar fashion, the time-dependent off-diagonal matrices $C_{{\bar A}_L,A}$ and $C_{A,\bar{A}_R}$ appearing in  Eqs.~\eqref{FDef} and \eqref{eq:CA-equ} follow from Eq.~\eqref{eq:C-equ}.
Inserting Eq.~\eqref{eq:CA-equ} into Eq.~\eqref{eq:IA-der-CA}, the contribution from $\left[\mathcal{H}_{A}^{T},C_{A}\right]$ vanishes because of the cyclic invariance of the trace and the relation $[C_A,f(C_A)]=0$ for arbitrary functions $f(C_A)$.  
The currents on the right-hand side of Eq.~(\ref{IAdeco}) are expressed through the subsystem's correlation matrix as follows: 
\begin{eqnarray}
\mathcal{I}_{{L/R}} & =&-i\,\mathrm{Tr}_A\left(F_{\bar{A}_{L/R}}\log_{2}\frac{C_{A}}{\mathbb{I}-C_{A}}\right),\nonumber \\
\label{eq:curr_block}
\mathcal{I}_{E} & =&-\mathrm{Tr}_A\left(F_{E}\log_{2}\frac{C_{A}}{\mathbb{I}-C_{A}}\right).
\end{eqnarray}

For the example of a chain of just $N=3$ sites, see Sec.~\ref{sec2c}, the unitary information currents in Eq.~(\ref{infocurN2}) follow from Eq.~\eqref{eq:curr_block} with 
\begin{equation}
    C(t)=\left(\begin{array}{cc} C_{11} & C_{12}\\
C_{12}^{*} & C_{22} \end{array}\right) 
\label{eq:matrix-Ct}
\end{equation}
in the form
\begin{eqnarray}\nonumber
\mathcal{I}_{(0,1),R} &=&  -iJ_{12}\left(C_{12}^{*}-C_{12}\right)\log_{2}\frac{C_{11}}{1-C_{11}},  \\ \label{unitarycurr2}
\mathcal{I}_{(0,2),L} &=& -iJ_{12}\left(C_{12}-C_{12}^{*}\right)\log_{2}\frac{C_{22}}{1-C_{22}}.
\end{eqnarray}
Similarly, the dissipative information currents are given by $(n=1,2)$ 
\begin{eqnarray}
\mathcal{I}_{(0,n),E}&=&\nonumber
-\bigl[\Gamma_{n}\left(1-C_{nn}\right)-\gamma_{n}C_{nn}\bigr]
\log_{2}\frac{C_{nn}}{1-C_{nn}},\\ 
\label{disscurr2}
\mathcal{I}_{(1,\frac{3}{2}),E}&=&-\mathrm{Tr}\left[\bigl(\Gamma-\frac12\{ \Gamma+\gamma,C\} \bigr)\log_{2}\frac{C}{\mathbb{I}-C}\right] .
\end{eqnarray}

\subsection{Particle currents} \label{sec22b}

For quasi-free fermions, the local particle currents \eqref{particlecurrentgeneral1} and \eqref{particlecurrentgeneral2} can be expressed in terms of the correlation matrix. For instance, Eq.~\eqref{particlecurrentgeneral1} can be rewritten as
\begin{equation}
\mathcal{I}_{j\to j+1}^{(p)} = i J_{j,j+1}(C_{j,j+1}-C_{j,j+1}^{*})= -\mathcal{I}_{j+1\to j}^{(p)},
\label{eq:I-particle}
\end{equation}
and Eq.~\eqref{particlecurrentgeneral2} takes the form
\begin{equation}
\mathcal{I}^{(p)}_{j\to E}=\gamma_j C_{jj}-\Gamma_j(1-C_{jj})=-\mathcal{I}^{(p)}_{E\to j}.
\label{eq:I-particle-E}
\end{equation}
From Eqs.~\eqref{unitarycurr2} and \eqref{disscurr2}, we can now read off general relations connecting the information currents flowing in the bottom layer $\ell=0$ of the information lattice to the above particle currents,
\begin{eqnarray} \nonumber
{\cal I}_{(0,n),R} &=& {\cal I}^{(p)}_{n\to n+1} \,\log_{2}\frac{C_{nn}}{1-C_{nn}}, \\ \label{connection}
{\cal I}_{(0,n),L} &=& - {\cal I}^{(p)}_{n-1\to n}\, \log_{2}\frac{C_{nn}}{1-C_{nn}}, \\ \nonumber
{\cal I}_{(0,n),E} &=& {\cal I}^{(p)}_{n\to E} \,\log_{2}\frac{C_{nn}}{1-C_{nn}}.
\end{eqnarray}
We note that if the environments only act on the chain ends, for the NESS, particle current conservation imposes
${\cal I}_{E\to 1}^{(p)}={\cal I}_{n\to n+1}^{(p)}={\cal I}_{N\to E}^{(p)}\equiv {\cal I}^{(p)}$.
Since the average occupancy $\bar n_j=\langle c_j^\dagger c_j^{}\rangle$ on site $j$ is given by $\bar n_j=C_{jj}$, see Eq.~\eqref{corrdef},
we observe from Eq.~\eqref{connection} that for half-filled sites, $\bar n_n=\frac12$, the corresponding $\ell=0$ information currents must vanish. 
Away from half-filling, particle and information currents propagate in the same (opposite) direction for $\bar n_n>\frac12$ ($\bar n_n<\frac12$).
We conclude that the information currents in Eq.~\eqref{connection} can be experimentally accessed by measuring standard particle currents and occupation numbers.

For higher layers of the information lattice, the local information currents  $\{\tilde{\cal I}_{(\ell,n)} \}$ could be measured by exploiting
the fact that they are fully determined by the  correlation matrix $C$, see Eqs.~\eqref{eq:curr_block} and \eqref{eq:effective_currents}.   For quasi-free fermions, 
$C$ can  be expressed by Wick's theorem in terms of particle number and/or particle current correlation functions, which can be measured  interferometrically 
\cite{Knap_2013,Pedernales_2014,DelRe_2024,Liu_2025,Yu_2011,Kastner_2020,Wang_2025}. A simpler alternative but approximate route to the measurement of local information currents is discussed in Sec.~\ref{sec22c}.

\subsection{Noise lattice}\label{sec22c}

Importantly, for open quasi-free fermionic systems, certain features of the information lattice can be captured by quantities that are more experimentally accessible than the von Neumann entropy. To that end, we introduce an additional triangular ``noise lattice" whose lattice sites are related to particle number fluctuations within subsystem $A=(\ell,n)$.
 
Using the variance of the particle number $\hat{n}_A$,   
\begin{equation}\label{varN}
\mathrm{Var}\left(\hat{n}_A\right)=\langle \hat{n}_A^2 \rangle - \langle \hat{n}_A^{} \rangle^2,  \quad \hat{n}_A=\sum_{j\in A} c^\dagger_j c_j^{},
\end{equation}
we note that for Gaussian states, ${\rm Var}(\hat{n}_A)$ can be fully expressed in terms of the 
correlation matrix $C_A$ by virtue of Wick's theorem. 
In the absence of anomalous correlations such as $\langle c_i c_j\rangle$, we find
\begin{equation}\label{eq:variance}
\mathrm{Var}\left(\hat{n}_A\right) = \mathrm{Tr}_A \left(C^{}_A-C_A^2\right) = 
\sum_{i=1}^{N_A} \mu_i (1-\mu_i),
\end{equation}
with the eigenvalues $\mu_i=\mu_i(t)$ of $C_A$. Evidently, both the variance \eqref{eq:variance} and the von Neumann information $I_A=N_A-S_A$ in  Eq.~\eqref{eq:von_neumann_entropy}, with the von Neumann entropy $S_A=S(\rho_A)$, are functions of the same spectral parameters $\{ \mu_i \}$.  
We next note that cumulant expansions of the von Neumann entropy (or related quantities such as R{\'e}nyi entropies) are highly useful for describing Gaussian states of fermionic many-body systems (or related spin models)~\cite{Eisert_2005, Klich_2009, Song_2010, Song_2011, Song_2012, Calabrese_2012, Petrescu2014, Thomas2015, Burmistrov2017},  including open (monitored) fermionic systems~\cite{Poboiko2023, Poboiko2023b}.
In Ref.~\cite{Klich_2009}, based on the model of a closed driven fermionic chain (with the impurity strength turned on and off periodically in time), it was shown that the subsystem von Neumann entropy $S_A$ for Gaussian fermionic states can be exactly represented through the full counting statistics of particle-number fluctuations (Klich-Levitov correspondence~\cite{Klich_2009}).
While this exact relation contains an infinite series of particle-number cumulants,  
the dominant contribution to $S_A$ is frequently given by the second cumulant~\cite{Klich_2009, Suesstrunk2012, Ivanov2013, Burmistrov2017, Poboiko2023, Zhang2024},\footnote{Note that here we use a convention for the normalization of $S_A$ which differs by a factor $\ln 2$ from the one in Ref.~\cite{Klich_2009}, where the coefficient relating $S_A$ and Var($\hat{n}_A$) is $\pi^2/3$.} 
\begin{equation}\label{klich}
    S_A \simeq \frac{\pi^2}{3 \ln{2}} \mathrm{Var}(\hat{n}_A).
\end{equation}
For details, see Appendix~\ref{app:FCS}.
In principle, it can happen that the cumulant expansion does not converge rapidly in some specific cases of unitary or Lindblad-type evolution \cite{Song_2011, Calabrese_2012, Znidaric_2014}. However, in our case, in which currents are injected from the environments at the boundaries, Eq.~\eqref{klich} is accurate, as checked numerically in Sec.~\ref{sec3}, especially away from boundary sites in large systems. 

In view of Eq.~\eqref{klich}, we define the noise lattice 
through the elements $i_{(\ell,n)}^{\mathrm{appr}}$, where
\begin{equation}
\label{info-noise}
i_{(\ell,n)}^{\mathrm{appr}}
= \delta_{\ell,0} - \frac{\pi^{2}}{3 \ln 2}\, \kappa_{(\ell,n)} \simeq i_{(\ell,n)} ,
\end{equation}
with
\begin{eqnarray}
\kappa_{(\ell,n)} &=& \mathrm{Var}\!\left(\hat{n}_{(\ell,n)}\right)
- \mathrm{Var}\!\left(\hat{n}_{(\ell-1,n-\frac12)}\right) \nonumber \\
& & -\, \mathrm{Var}\!\left(\hat{n}_{(\ell-1,n+\frac12)}\right)
+ \mathrm{Var}\!\left(\hat{n}_{(\ell-2,n)}\right).
\label{noiselatt}
\end{eqnarray}
For Gaussian states, using Wick’s theorem and recalling that subsystem $A=(\ell,n)$ contains all sites $j_L\le j\le j_R$ with $j_L=n-\frac{\ell}{2}$ and $j_R=j_L+\ell$, Eq.~\eqref{noiselatt} can be significantly simplified. For $\ell=0$, it reduces to
\begin{eqnarray}
    \kappa_{(0,n)} = C_{nn}-C_{nn}^2,
\end{eqnarray}
while for $\ell>0$, we have
\begin{eqnarray}
    \kappa_{(\ell,n)} = -2|C_{j_L j_R}|^2.
\end{eqnarray}
The noise lattice~\eqref{info-noise} is experimentally accessible via quantum noise measurements~\cite{Klich_2009, Clerk2010} for larger system sizes than the information lattice, which requires quantum state tomography~\cite{dariano2003quantum} or entropy measurements~\cite{islam2015measuring, vermersch2024many}. Hence, if the approximations discussed around Eq.~\eqref{klich} hold, the information lattice in the NESS can be experimentally estimated through the noise lattice.

In order to compute the approximated information currents $\{\tilde{\cal I}^{\rm appr}_{(\ell, n)}\}$ using the second-cumulant appoximation \eqref{klich} in the Klich--Levitov correspondence, we next consider the time derivative of particle-number variances.  From Eq.~\eqref{eq:variance}, we find
\begin{eqnarray}\label{ddtvar}
    \frac{d}{dt}{\rm Var}(\hat{n}_A) &=& \mathrm{Tr}_A \left[\left(\mathbb{I}-2C_A\right)\dot{C}_A\right] 
    \\ \nonumber &=& \mathrm{Tr}_A \left[\left(\mathbb{I}-2C_A\right) \left(iF_{\bar A_L}+iF_{\bar A_R}+F_E\right) \right],
\end{eqnarray}
where we used Eq.~\eqref{eq:CA-equ} with $F_{\bar A_L / \bar A_R / E}$ in Eq.~\eqref{FDef}. 
With the diagonal rate matrices \eqref{Gamdiag} and the Hamiltonian \eqref{HN3},  
$\{\tilde{\cal I}^{\rm appr}_{(\ell, n)}\}$ follows by combining Eqs.~\eqref{eq:curr_block}, \eqref{eq:effective_currents}, and \eqref{klich}, and subsequently using 
Wick's theorem.
For $\ell=0$ (i.e., $j_L=j_R=n$), we arrive at 
\begin{eqnarray}
 \nonumber
 \tilde{\cal I}_{(0,n),R}^{\rm appr} &=& \frac{\pi^2}{3\ln 2} {\cal I}^{(p)}_{n\to n+1} \left( 2C_{n n}-1\right), \\
\label{curappr_effective_ell0}
 \tilde{\cal I}_{(0,n),L}^{\rm appr} &=& -\frac{\pi^2}{3\ln 2} {\cal I}^{(p)}_{n-1\to n} \left( 2C_{n n}-1\right), \\
  \nonumber
 \tilde{\cal I}_{(0,n),E}^{\rm appr} &=& \frac{\pi^2}{3\ln 2} {\cal I}^{(p)}_{n\to E} \left( 2C_{n n}-1\right).
\end{eqnarray}
For $\ell>0$, with Eq.~\eqref{opcurr} and $\hat n_j=c_j^\dagger c_j^{}$, we find
\begin{eqnarray}\nonumber
  \tilde{\cal I}_{(\ell,n),R}^{\rm appr}\!\! &=&\!\!  -\frac{2\pi^2}{3\ln 2}\, {\rm Cov}\left( \, \hat{\cal I}_{j_R\to j_R+1}^{(p)} , \hat n_{j_L} \right),\\  \label{curappr_effective_ell}
 \tilde{\cal I}_{(\ell,n),L}^{\rm appr}\!\! &=& \!\! \frac{2\pi^2}{3\ln 2}\, {\rm Cov} \left(\, \hat{\cal I}_{j_L-1\to j_L}^{(p)} , \hat n_{j_R}\right),\\
\nonumber
 \tilde{\cal I}_{(\ell,n),E}^{\rm appr}\!\! &=&\!\! -\frac{\pi^2}{3\ln 2}\left(  \Gamma_{j_L}\!+\gamma_{j_L}\!+ \Gamma_{j_R}\!+\gamma_{j_R} \right){\rm Cov} \left(\hat n_{j_L},\hat n_{j_R}\right),
\end{eqnarray}
where the covariance of an operator pair $\hat a$ and $\hat b$ is
${\rm Cov}(\hat a,\hat b)=   \langle \hat a \hat b   \rangle  - \langle \hat a  \rangle \langle \hat b \rangle$.  As for the information lattice, the results for
$\{ \tilde{\cal I}_{(\ell,n)}^{\rm appr}\}$ in Eqs.~\eqref{curappr_effective_ell0} and
\eqref{curappr_effective_ell} are expected to be accurate away from the boundaries of the information lattice and for large system size.  
The above expressions show that experimental measurements of local information currents are possible. 
While the information currents pertaining to the bottom layer $\ell=0$ of the information lattice can already be obtained
without approximation by measuring particle currents and the onsite occupancies, see Eq.~\eqref{connection},  information currents for $\ell>0$ could be accessed 
by probing the covariances of particle currents and/or occupation numbers, see Eq.~\eqref{curappr_effective_ell}.

\section{Local information currents}\label{sec3}

In this section, we illustrate the theory implemented for quasi-free fermions in Sec.~\ref{sec22} with concrete examples.
By doing so, we reveal characteristic patterns for the 
information lattice, $\{ i_{(\ell,n)} \}$ in Eq.~\eqref{eq:triangles-to-sites}, and for the associated information currents, $\{ \tilde{\cal I}_{(\ell,n)}\}$, see
Eq.~\eqref{eq:effective_currents}, for open fermion chains under transport conditions.   
Through control parameters such as the bias voltage or gate voltages, one can then
change the information lattice and the corresponding information currents. In particular, gate voltages allow one to change the onsite energies $\epsilon_j$ and the hopping parameters $J_{j,j+1}$. As discussed in Sec.~\ref{sec22c}, the information lattice becomes easily experimentally accessible via the noise lattice.

For the Hamiltonian matrix ${\cal H}$ in Eq.~\eqref{HN3}, we assume homogeneous nearest-neighbor hopping amplitudes and onsite energies, $J_{j,j+1}=J$ and $\epsilon_j=\epsilon$. 
Short-range impurities are modeled by modulating the onsite energy $\epsilon_{j_0}\ne \epsilon$ at a single site $j=j_0$ (site defect).  As shown below, similar results emerge also for a bond defect with $J_{j_0,j_0+1}\ne J$.  
Below we focus on NESS configurations reached at asymptotically long times, where the choice of the initial state is irrelevant. In fact, our NESS results 
follow by using $\dot{C}=0$ in Eq.~\eqref{eq:C-equ}. Of course, for a generic Gaussian initial state, our theory also describes the transient dynamics toward the NESS by solving Eq.~\eqref{eq:C-equ}. 

In Sec.~\ref{sec3a}, we discuss the large-bias regime introduced in Sec.~\ref{sec2a}. While we mainly focus on this regime here, we also comment on the small-bias linear-response regime in Sec.~\ref{sec3b}.   Analytical arguments supporting the findings in this section can be found in Appendix \ref{appendix:exact for N=3}, where we study the case $N=3$.

\begin{figure}
\includegraphics*[width=0.75 \linewidth]{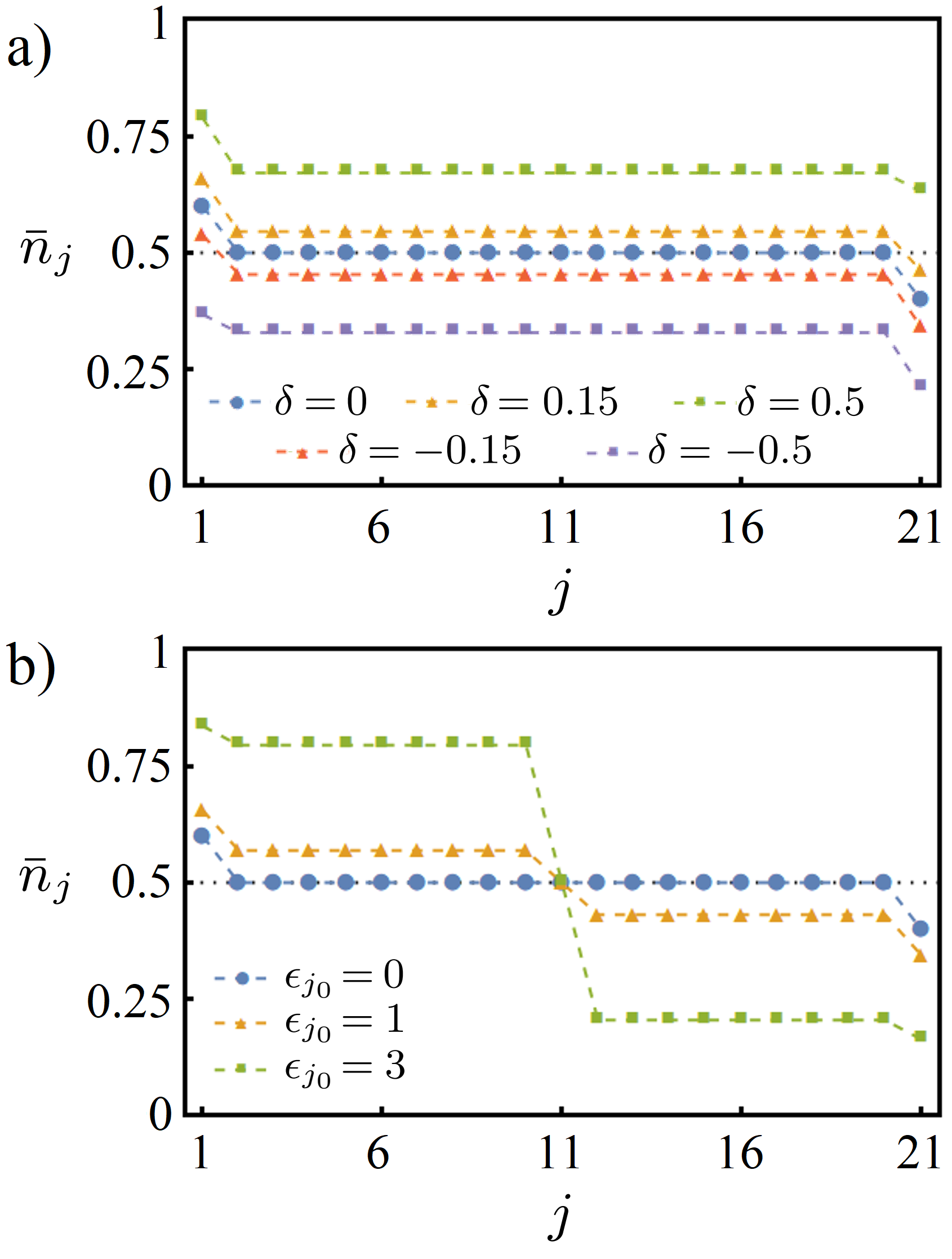}
\caption{Average occupation number $\bar n_j$ vs site index $j$ under large-bias conditions in the NESS for $N=21$, $g=J=1$, and $\epsilon=0$. Dashed lines are guides to the eye only.   (a) Impurity-free case for different particle-hole asymmetry parameters $\delta$. (b) Results for $\delta=0$ and a site defect at $j_0=11$ with different values of $\epsilon_{j_0}$. 
}
\label{fig5}
\end{figure}

\subsection{Large-bias regime}\label{sec3a}

We begin by studying the NESS under large-bias conditions as specified in Sec.~\ref{sec2a}. The injection and removal rates are then given by Eq.~\eqref{largebias}, where $g$ is the overall system-environment coupling and $\delta$ quantifies the particle-hole asymmetry. We consider the parameter choice $g=J=1$ and $\epsilon=0$, but allow for a site defect with energy $\epsilon_{j_0}\ne 0$ at site $j_0$. 

The resulting steady-state density profile $\bar n_j =C_{jj}$ for system size $N=21$ and $j_0=11$ is shown in Fig.~\ref{fig5}. These results are obtained from the stationary solution of Eq.~\eqref{eq:C-equ}, i.e., $\dot C=0$. Without impurity ($\epsilon_{j_0}=0$), see Fig.~\ref{fig5}(a), apart from the end sites, the occupation number is homogeneous and given by $\bar n_j=1/2$ for $\delta=0$.  With increasing $\delta>0$, this density profile shifts upward. For $\delta>\delta^*=\sqrt{5}-2\simeq 0.236$ (see Appendix \ref{appendix:exact for N=3} for a derivation of $\delta^\ast$), one finds that 
all sites have occupancy $\bar n_j>1/2$. Similarly, for $\delta<0$, the profile shifts downward, where for $\delta<-\delta^*$, we find $\bar n_j<1/2$ for all sites. 
Including a site defect,  Fig.~\ref{fig5}(b) reveals a step in the density profile at the location of the defect.  

\begin{figure}
\center
\includegraphics[width= \linewidth]{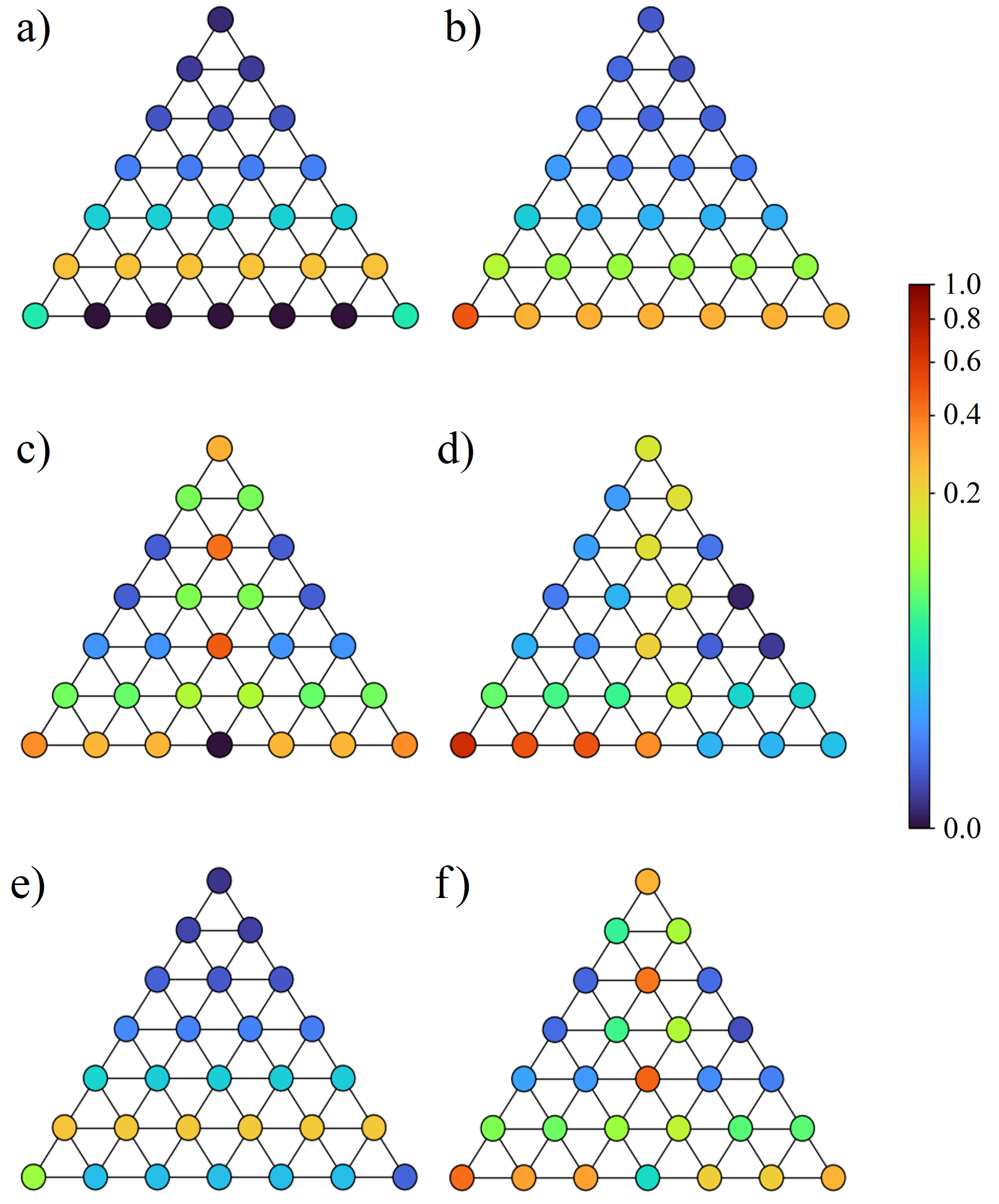}
\caption{Information lattice $\{ i_{(\ell,n)}\}$ in the NESS for $N=7, j_0=4, g=J=1,$ and $\epsilon=0$. Note the power-norm color scale \cite{power_norm} for $i_{(\ell,n)}$.  Results are shown for (a) $\delta=\epsilon_{j_0} = 0$, (b) $\delta=0.75$ and $\epsilon_{j_0} = 0$, (c)  $\delta = 0$ and $\epsilon_{j_0} = 3$, (d) $\delta = 0.75$ and $\epsilon_{j_0} = 3$, (e) $\delta=0.2$ and $\epsilon_{j_0} = 0$, and (f) $\delta=0.2$ and $\epsilon_{j_0} = 3$.}
\label{fig6}
\end{figure}

\begin{figure}
\includegraphics[width=\linewidth]{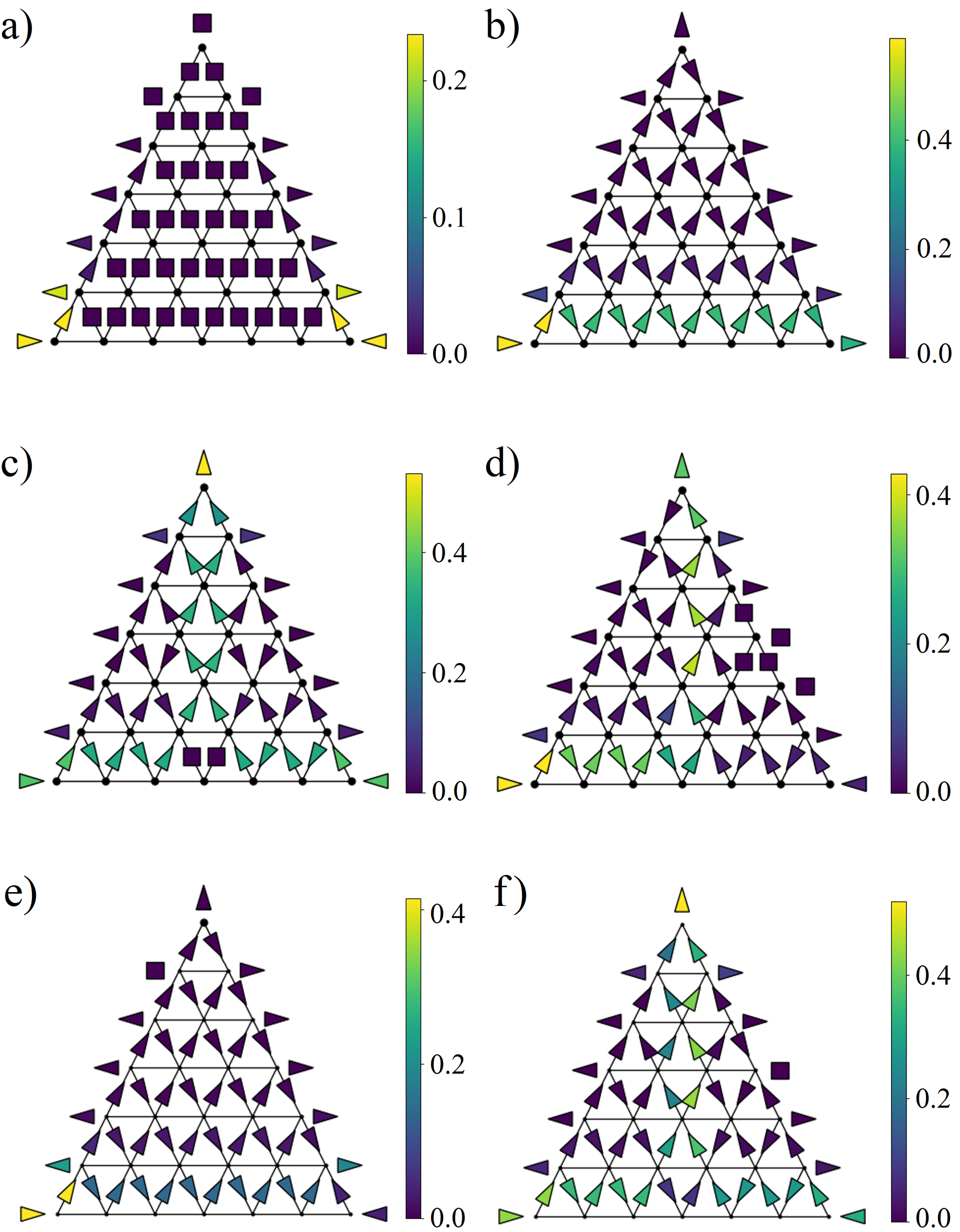}
\caption{
Information currents $\{\tilde{\cal I}_{(\ell,n)}\}$, see Eq.~\eqref{eq:effective_currents}, in the NESS for system size $N=7$, allowing for a site defect at $j_0=4$. We set $g=J=1$ and $\epsilon=0$ and consider the large-bias regime, see Eq.~\eqref{largebias}.  The local information current direction is indicated by arrows and the magnitude by the color scale. Note the different color scales in different panels. 
Information currents inside the lattice are due to unitary terms. Arrows pointing away from (or toward) the lattice represent dissipative information currents.
 Filled black squares indicate extremely small values, $|\tilde{\cal I}|<10^{-4}$.
(a) Particle-hole symmetric case for a clean chain, $\delta=\epsilon_{j_0} = 0$. (b) Broken particle-hole symmetry ($\delta=0.75$) with $\epsilon_{j_0} = 0$.
(c) Results for $\delta = 0$ with an impurity, $\epsilon_{j_0}=3$.
(d) Results for $\delta = 0.75$ and $\epsilon_{j_0}=3$. (e) Results for $\delta = 0.2$ and $\epsilon_{j_0}=0$. (f) Results for $\delta = 0.2$ and $\epsilon_{j_0}=3$.}
\label{fig7}
\end{figure}

We next show results for the information lattice and the information currents for system size $N=7$, with a site defect at $j_0=4$.
While an analytical solution for $\{ i_{(\ell,n)}\}$ and $\{ \tilde{\cal I}_{(\ell,n)}\}$ in the NESS can be given, the resulting expressions are lengthy and not particularly instructive.  Using color-scale plots, the information currents are shown in Fig.~\ref{fig6},
and the information lattice in Fig.~\ref{fig7}. In the four respective panels, we consider different combinations of $\epsilon_{j_0}$ and $\delta$. 
In all cases, the NESS particle current ${\cal I}^{(p)}$ is finite, with different values for each panel, and flows from left to right through the chain.

In Figs.~\ref{fig6}(a) and \ref{fig7}(a), we analyze the impurity-free and particle-hole-symmetric case, $\epsilon_{j_0}=\delta=0$, where the average occupancy of bulk sites ($1<j<N$) is $\bar n_j=1/2$, see Fig.~\ref{fig5}.
Figure \ref{fig6}(a) reveals that the information is mainly stored on layer $\ell=1$ of the information lattice, and becomes very small on longer scales ($\ell>1$) with an exponential decay~\cite{Artiaco2025}. 
Correlations are therefore short-ranged and mainly extend over neighboring sites.  In this case, dissipative information currents are not able to penetrate inside the information lattice, see Fig.~\ref{fig7}(a). Such information currents are injected from the environment at the end sites of layer $\ell=0$ (with the same intensity but in opposite directions), and then are almost completely reflected back into the environment on layer $\ell=1$. 
In the NESS, the interior part of the information lattice is thus effectively shielded from the environment. Consistent with this shielding effect, information currents inside the lattice are extremely small, see Fig.~\ref{fig7}(a).

This picture changes if particle-hole symmetry is broken $(\delta\ne 0)$. (For now, we still  assume a clean chain without impurity.)
For $0<\delta<\delta^*\simeq 0.236$, dissipative information 
currents are still exchanged between the environment and the information lattice 
sites $i_{(0,1)}$ and $i_{(0,N)}$. However, now a small information current can penetrate into the bulk of the information lattice from the left side, exiting at site $i_{(1,N-\frac12)}$, see  Figs.~\ref{fig6}(e) and \ref{fig7}(e).  Indeed, for $\delta> 0$,
the bulk occupancy is $\bar n_j>\frac12$, see Fig.~\ref{fig5}.  Since $C_{nn}=\bar n_n$ appears through logarithmic factors in the information 
currents ${\cal I}_{(0,n),R/L}$, see Eq.~\eqref{connection}, those currents vanish only if the system has particle-hole symmetry (i.e., for $\bar n_n=\frac12$). 
On the other hand, for $\delta>\delta^*$, the shielding effect described above breaks down completely, and information currents enter at $i_{(0,1)}$ and exit at $i_{(0,N)}$. This case is shown in Fig.~\ref{fig7}(b) for $\delta=0.75$. In fact, information is again mainly stored in layer $\ell=1$ but now is of significant magnitude also for the bottom layer $\ell=0$, see Fig.~\ref{fig6}(b).
The resulting information current flowing through the information lattice is unidirectional 
and alternates in a zig-zag manner between layers $\ell=0$ and $\ell=1$, see Fig.~\ref{fig6}(b). We note that information currents are (anti-)parallel to particle currents for positive (negative) $\delta$.  We emphasize that the information currents exhibiting the shielding effect and its breakdown are readily measurable as explained in Sec.~\ref{sec2f}.  
The physics behind this phenomenon could also be observed as a percolation‑like connectivity crossover at $\delta = \delta^\ast$ in the information lattice, inferred via its relation to the noise lattice discussed in Sec.~\ref{sec22c}. 
Indeed, the black dots in Fig.~\ref{fig6}(a) (with $\delta=0$) turn into colored dots in Fig.~\ref{fig6}(b) (with $\delta>\delta^*$), indicating that information starts to percolate through the bulk of the information lattice at $\delta=\delta^*$.

Another way how information currents can penetrate inside the information lattice is to include an impurity in the chain, see Figs.~\ref{fig6}(c) and \ref{fig7}(c), where we study a site defect with $\epsilon_{j_0}=3$ at the central position $j_0$ for $\delta=0$. 
The NESS information lattice in Fig.~\ref{fig6}(c) now features an ``information pillar'' centered above the impurity site, with pronounced values of the information $i_{(\ell,n)}$ on all layers $\ell>0$ for $n\approx j_0$.  The corresponding long-range quantum correlations generated by the impurity extend over the entire system. An intuitive interpretation for this information pillar, which is only found in the NESS, see Sec.~\ref{sec3b}, 
can be given in terms of scattering by a localized impurity as characterized by a transmission probability $0<\tau<1$ \cite{Fraenkel2023}. The wavefunction of an incoming fermion injected from the left reservoir then splits into a forward and a backward component. Both components propagate with the same velocity away from the impurity, i.e., they will have the same distance from the impurity.  The corresponding pair correlations encoded by the 
off-diagonal matrix elements of the correlation matrix $C$ are responsible for the information pillar observed in Fig.~\ref{fig6}(c), as also discussed around Fig.~\ref{fig13} and in Appendix \ref{appendix:exact for N=3}; see also Appendix~\ref{appendix:unraveling}, where the information lattice is discussed in terms of unraveling of the LME.

A similar pillar is also observed for the information currents, see Fig.~\ref{fig7}(c), where information currents injected from the leads propagate inside the lattice and then move all the way up to the top vertex along the pillar.  For $\delta\ne 0$, the information pillar generated by the defect acquires an asymmetric shape, see Figs.~\ref{fig6}(d) and \ref{fig7}(d), as well as Figs.~\ref{fig6}(f) and \ref{fig7}(f). In this case, information currents on the bottom layer $\ell=0$ propagate mainly on the left side of the pillar while they rise along the pillar mainly on the right side. 
We conclude that under NESS conditions, a local defect has a nontrivial effect on the information lattice and the corresponding information currents. In particular, it generates long-range correlations.  

\begin{figure}
\center
\includegraphics[width=0.8\linewidth]{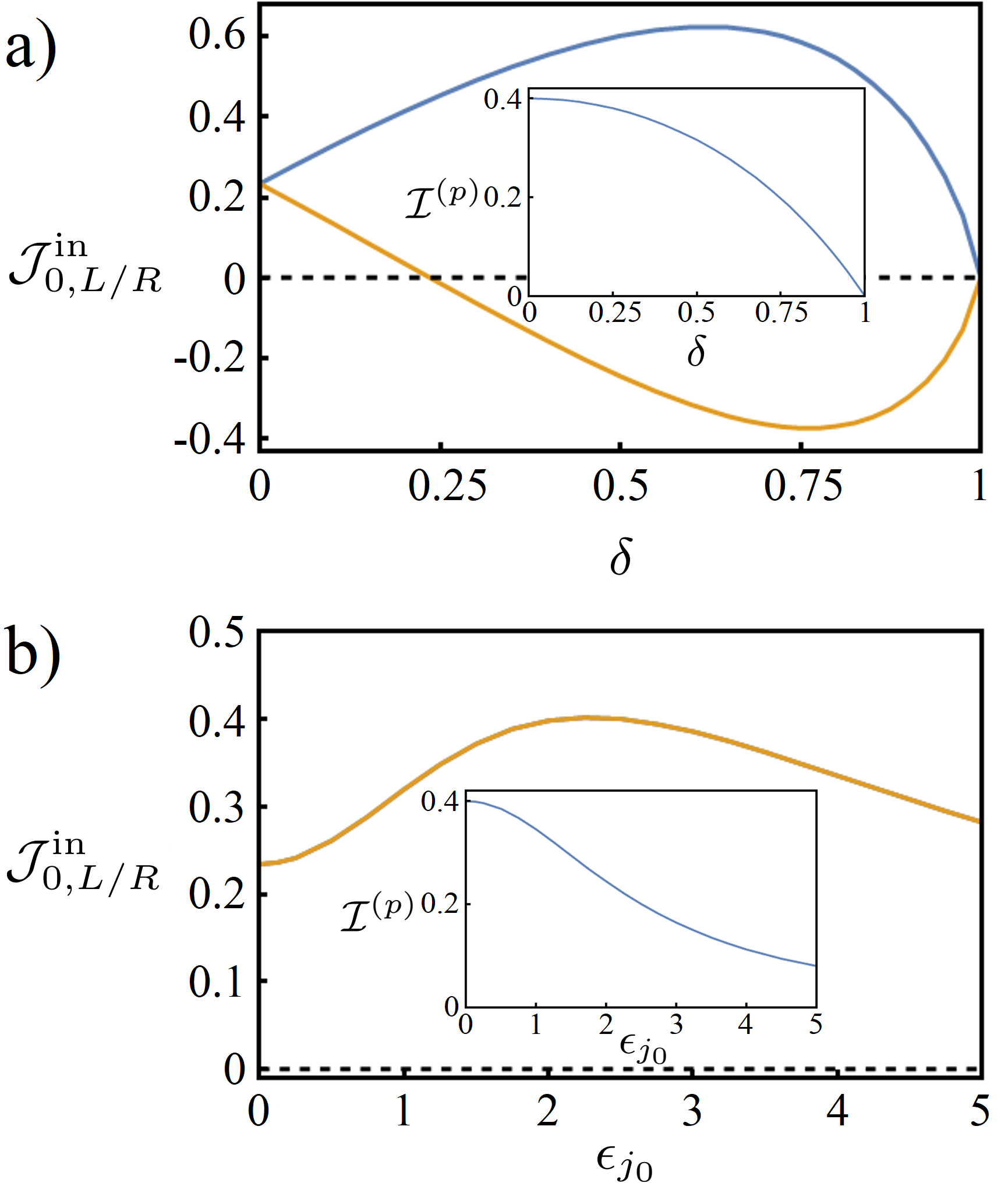} 
\caption{Information currents ${\cal J}_{0,L/R}^{\mathrm{in}}$ injected at the left ($j=1$, blue) and right end ($j=N$, orange curves) for $N=7, j_0=4, g=J=1,$ and $\epsilon=0$. Currents entering (flowing away from) the fermionic chain are positive (negative).
(a) Main panel: ${\cal J}_{0,L/R}^{\mathrm{in}}$ vs $\delta$ for $\epsilon_{j_0}=0$. Inset: Particle current ${\cal I}^{(p)}$ vs $\delta$ for the same parameters. (b) Main panel: ${\cal J}_{0,L/R}^{\mathrm{in}}$ vs $\epsilon_{j_0}$ for $\delta=0$, where both information currents coincide. Inset: ${\cal I}^{(p)}$ vs $\epsilon_{j_0}$ for the same parameters.
}
\label{fig8}
\end{figure}

We have focused on defects corresponding to a repulsive barrier with $\epsilon_{j_0}>0$.  For an attractive barrier with $\epsilon_{j_0}<0$, while 
scattering states are known to be insensitive to the sign of $\epsilon_{j_0}$, one expects an additional bound state localized at the defect \cite{Nazarov2009}.   However, as shown by our analytical results for $N=3$ in Appendix \ref{appendix:exact for N=3}, the sign of $\epsilon_{j_0}$ neither
affects the information lattice $\{ i_{(\ell,n)} \}$ nor the associated information currents $\{ \tilde{\cal I}_{(\ell,n)}\}$. We have also checked that this conclusion
remains valid for larger values of $N$.

In order to characterize information transport in terms of a few global currents instead of the full information current lattice, we next study the horizontal and vertical information currents introduced in Sec.~\ref{sec2d}. To begin, consider the information currents injected from the environment at the left ($L$) and right ($R$) ends of the chain, ${\cal J}_{0,L}^{\mathrm{in}}=-\mathcal{I}_{(0,1),E}$ and ${\cal J}_{0,R}^{\mathrm{in}}=-\mathcal{I}_{(0,N),E}$, for the same parameters as in Figs.~\ref{fig6} and \ref{fig7}.  They are shown 
as function of $\delta$ for a clean chain ($\epsilon_{j_0}=0$) in Fig.~\ref{fig8}(a), and as function of $\epsilon_{j_0}$ for $\delta=0$ in Fig.~\ref{fig8}(b). 
In the first case, we find ${\cal J}_{0,L}^{\mathrm{in}}>0$. This information current initially increases when increasing $\delta$, but decreases for larger $\delta$ and ultimately vanishes for $\delta\to 1$, where Eq.~\eqref{largebias} implies $\gamma_N=0$. At the same time, ${\cal J}_{0,R}^{\mathrm{in}}$ first decreases with increasing $\delta$. 
For $\delta>\delta^*$, we find ${\cal J}_{0,R}^{\rm in}<0$.  The value $\delta=\delta^*$ corresponds to the transition point where the shielding effect
observed in Fig.~\ref{fig7}(a) disappears completely and information current can freely flow through the information lattice.  Indeed, $\delta=\delta^*$ represents the point where the last site is half-filled, $C_{NN}=\frac12$, and thus both the onsite von Neumann information \eqref{eq:von_neumann_entropy} and the corresponding information current \eqref{eq:IA-der-CA}  vanish.   Eventually, for $\delta\to 1$, we note that also ${\cal J}_{0,R}^{\mathrm{in}}$ vanishes.
Next, for $\delta=0$ but $\epsilon_{j_0}\ne 0$, see Fig.~\ref{fig8}(b), both injected information currents are equal and exhibit a broad maximum as function of $\epsilon_{j_0}$ at $\epsilon_{j_0}\approx 2.5$. In particular, for the cut chain with $\epsilon_{j_0}\to \infty$, they vanish. The insets in Fig.~\ref{fig8}(a,b) show the respective dependence on $\delta$ or $\epsilon_{j_0}$ of the conventional particle current ${\cal I}^{(p)}$.  In contrast to the information currents, ${\cal I}^{(p)}$ monotonically decreases with increasing $\delta$ and/or $\epsilon_{j_0}$, with maximum value at $\delta=\epsilon_{j_0}=0$.

\begin{figure}
\center
\includegraphics[width=0.8 \linewidth]{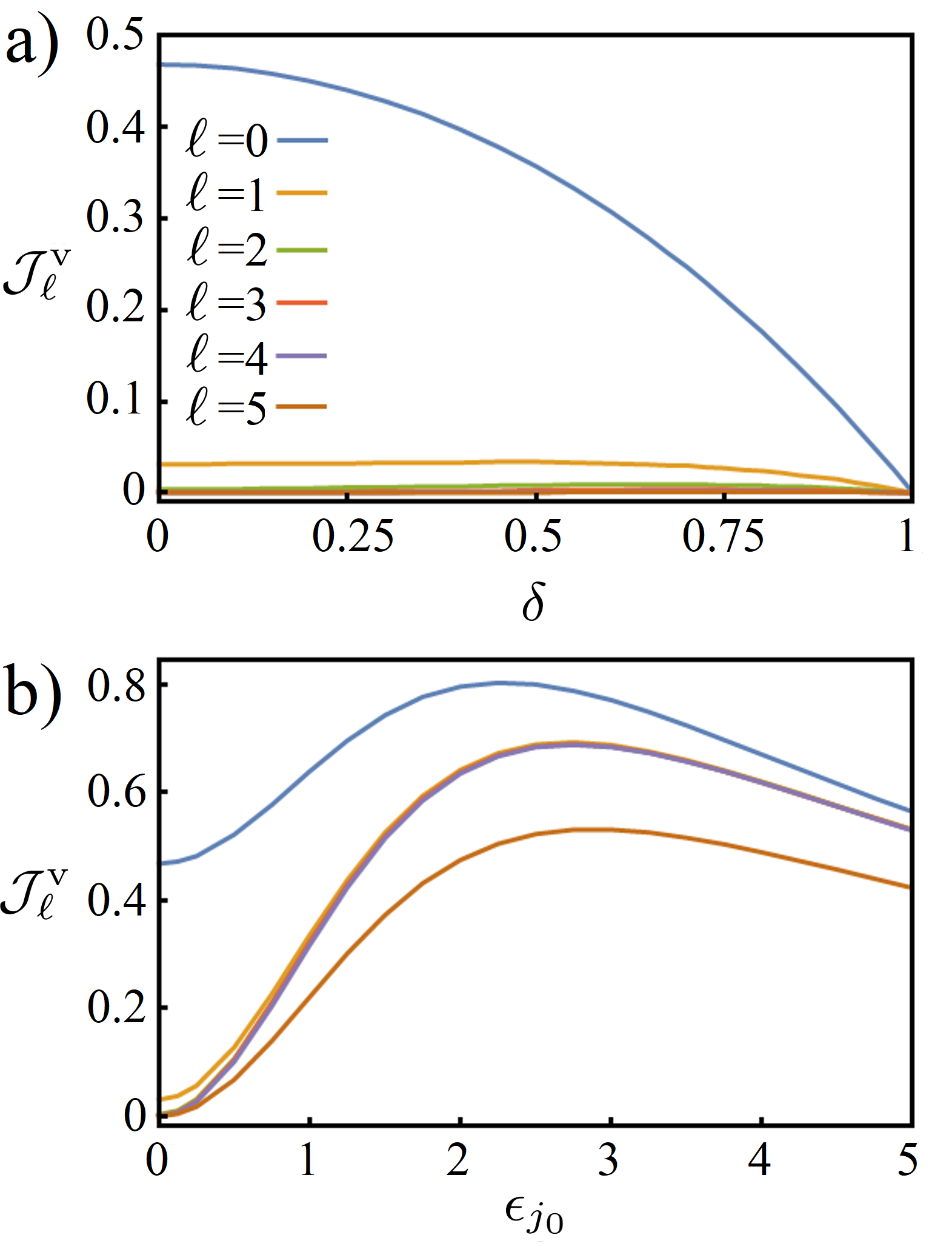}
\caption{Vertical currents $\mathcal{J}_{\ell}^{\mathrm{v}}$ in Eq.~\eqref{eq:summed_currents_vertical}, for $N=7, j_0=4, g=J=1,$ and $\epsilon=0$.  
(a) $\mathcal{J}_{\ell}^{\mathrm{v}}$ vs $\delta$ for different $\ell$ with $\epsilon_{j_0}=0$.
(b) $\mathcal{J}_{\ell}^{\mathrm{v}}$ vs $\epsilon_{j_0}$ for different $\ell$ with $\delta=0$.}
\label{fig9}
\end{figure}

\begin{figure}
\center
\includegraphics[width=0.8 \linewidth]{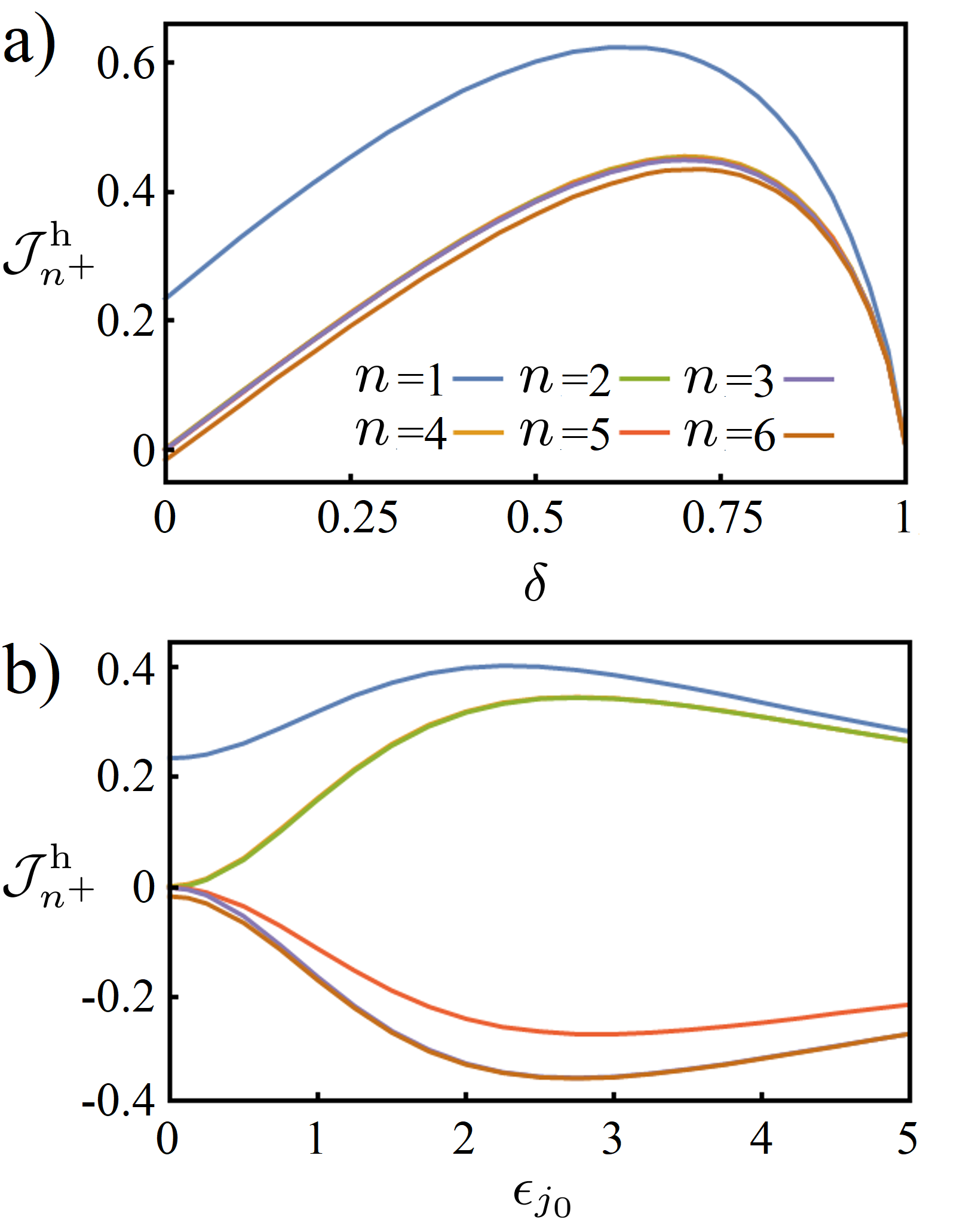}
\caption{Horizontal currents $\mathcal{J}_{n^+}^{\mathrm{h}}$ in Eq.~\eqref{eq:summed_currents_horizontal}, for $N=7, j_0=4, g=J=1,$ and $\epsilon=0$. 
(a) $\mathcal{J}_{n^+}^{\mathrm{h}}$ vs $\delta$ for different $n$ with $\epsilon_{j_0}=0$. 
(b) $\mathcal{J}_{n^+}^{\mathrm{h}}$  vs $\epsilon_{j_0}$ for different $n$ with $\delta=0$.}
\label{fig10}
\end{figure}

Let us next focus on the behavior of the unitary horizontal information currents  $\mathcal{J}_{n^\pm}^{\mathrm{h}}$ in Eq.~\eqref{eq:summed_currents_horizontal} and the
vertical  currents $\mathcal{J}_{\ell}^{\mathrm{v}}$ in Eq.~\eqref{eq:summed_currents_vertical}.
First, we consider the clean (impurity-free) case, see Figs.~\ref{fig9}(a) and \ref{fig10}(a).   Since information is then mainly stored in the information lattice layers $\ell=0$ and $\ell=1$, see Fig.~\ref{fig7}(a,b), $\mathcal{J}_{\ell}^{\mathrm{v}}$ is extremely small for $\ell>1$ and arbitrary $\delta$, as observed in Fig.~\ref{fig9}(a). On the other hand, the currents $\mathcal{J}_{n^+}^{\mathrm{h}}$ vanish for $1<n<N$ and $\delta=0$, see Fig.~\ref{fig10}(a), because of the shielding effect described above, 
see Fig.~\ref{fig6}(a).  For $0<\delta<1$, 
however, all currents $\mathcal{J}_{n^+}^{\mathrm{h}}$ are positive, signaling again a finite and unidirectional information current propagating along the chain, see Fig.~\ref{fig10}(a),
in accordance with Fig.~\ref{fig7}(b). 
 Figure~\ref{fig10}(a) also shows that all $\mathcal{J}_{n^+}^{\mathrm{h}}$ away from the boundary sites essentially coincide due to information current conservation. 
For $\delta\to 1$, where  $\gamma_N\to 0$ in Eq.~\eqref{ratesF}, 
we again find that all information currents vanish. 

Second, we consider the case $\delta=0$ in the presence of a central site defect, $\epsilon_{j_0}\ne 0$. The resulting information pillar, with vertical information current flowing from the bottom layer up to the vertex of the information lattice, see Fig.~\ref{fig7}(b,c), explains the observed behavior of $\mathcal{J}_{\ell}^{\mathrm{v}}$ in Fig.~\ref{fig9}(b). For $\epsilon_{j_0}\ne 0$, all vertical information currents become positive and equal to each other for $0<\ell<N-2$. Similarly, the behavior of $\mathcal{J}_{n^+}^{\mathrm{h}}$ shown in Fig.~\ref{fig10}(b) reveals the existence of the information pillar as well. Indeed, these information currents are positive (negative) on the left (right) side of the defect. 

\begin{figure}
\center
\includegraphics[width= \linewidth]{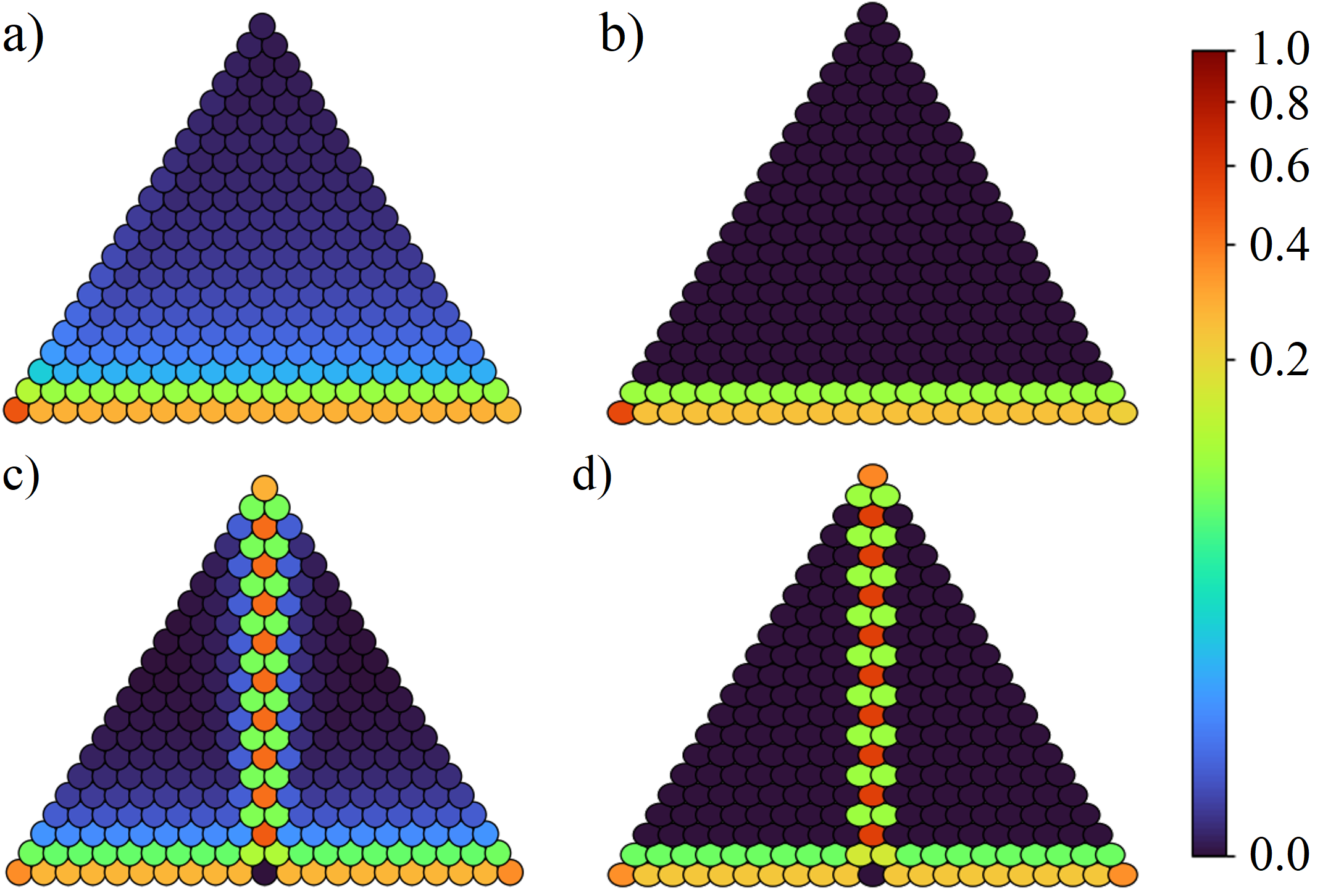}
\caption{Exact information lattice $\{ i_{(\ell,n)}\}$ and approximated information lattice 
$\{ i^{\rm appr}_{(\ell,n)}\}$ in Eq.~\eqref{info-noise} employing the noise lattice. We show results for $N=21, j_0=11, g=J=1$, and $\epsilon=0$, using power-norm color scales. 
Note that $i_{(\ell,n)}^{\rm appr}<0$ is possible (with small absolute value). Negative values are shown as zero in the plots for better comparison.
(a) $\{ i_{(\ell,n)}\}$ for $\delta=0.75$ and no impurity ($\epsilon_{j_0}=0$). 
(b) Same as (a) but for $\{ i^{\rm appr}_{(\ell,n)}\}$.
(c) $\{ i_{(\ell,n)}\}$ for $\delta=0$ and in the presence of a defect with 
$\epsilon_{j_0}=3$. (d) Same as (c) but for $\{ i^{\rm appr}_{(\ell,n)}\}$.}
\label{fig11}
\end{figure}

Next we compare results for the information lattice $\{ i_{(\ell,n)} \}$
to the approximated results $\{ i_{(\ell,n)}^{\rm appr} \}$ based on
the correspondence with the noise lattice in Eq.~\eqref{info-noise}.
Figure \ref{fig11} shows results for system size $N=21$.  In Fig.~\ref{fig11}(a,b), we 
consider the clean case with broken particle-hole symmetry, where good agreement between
the exact results and those based on the noise lattice is found.  Similarly, in 
Fig.~\ref{fig11}(c,d), qualitative agreement between both results is observed in the presence of a central defect for $\delta=0$. In particular, the information pillar centered at the defect site is clearly seen from the noise lattice.

\begin{figure}
\center
\includegraphics[width= \linewidth]{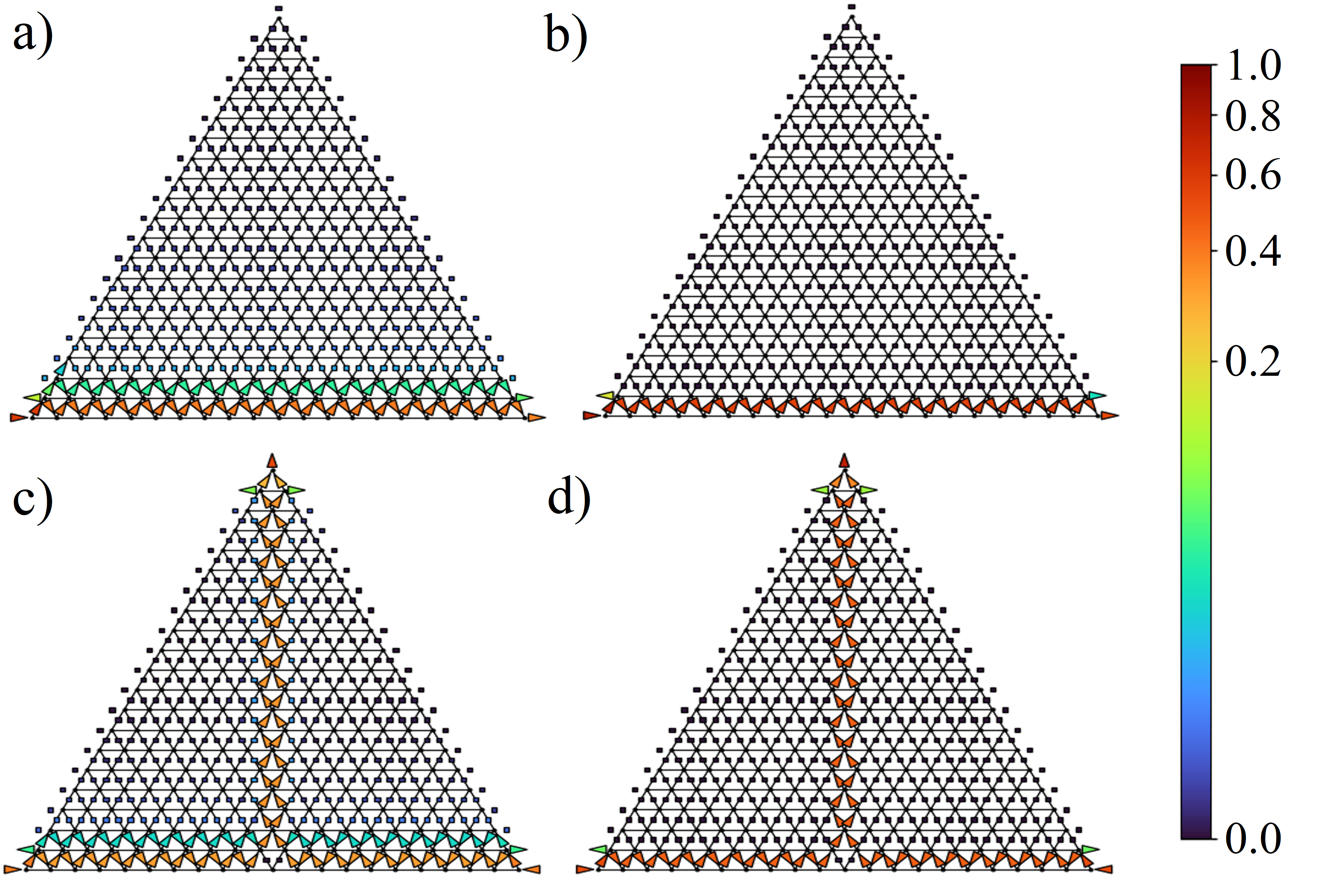}
\caption{Exact information currents $\{ \tilde{\cal I}_{(\ell,n)}\}$ and approximated information currents 
$\{ \tilde{\cal I}^{\rm appr}_{(\ell,n)}\}$ obtained from the second cumulant in the Klich--Levitov correspondence, see Eqs.~\eqref{curappr_effective_ell0} and \eqref{curappr_effective_ell}. 
As in Fig.~\ref{fig11}, results are shown for $N=21, j_0=11, g=J=1$, and $\epsilon=0$.
The local information current direction is indicated by arrows and the magnitude by the color scale, with filled black squares for $|\tilde{\cal I}|<10^{-2}$. Note the power-norm color scales.
(a) $\{ \tilde{\cal I}_{(\ell,n)}\}$ for $\delta=0.75$ and $\epsilon_{j_0}=0$. 
(b) Same as (a) but for $\{ \tilde{\cal I}^{\rm appr}_{(\ell,n)}\}$.
(c) $\{ \tilde{\cal I}_{(\ell,n)}\}$  for $\delta=0$ and $\epsilon_{j_0}=3$. 
(d) Same as (c) but for $\{ \tilde{\cal I}^{\rm appr}_{(\ell,n)}\}$.}
\label{fig12}
\end{figure}

Similarly, we compare the exact local information current pattern $\{\tilde{\cal I}_{(\ell,n)}\}$
to the approximated results $\{ \tilde{\cal I}^{\rm appr}_{(\ell,n)}\}$ obtained from the Klich--Levitov correspondence, see Eqs.~\eqref{curappr_effective_ell0} and
\eqref{curappr_effective_ell}.  As discussed in Sec.~\ref{sec22c}, the approximated information currents are 
easier to access experimentally since they can be obtained by measuring conventional particle currents, onsite occupations, and covariances involving particle numbers and/or particle currents. Especially away from the boundaries of the information lattice,
we find good agreement between exact and approximated information currents.   In particular, the information current pattern associated with the defect-induced information pillar is captured by the Klich--Levitov expressions.  While the Klich-Levitov correspondence always holds true for quasi-free fermions (both in open and closed systems), the second-cumulant approximation in Eq.~\eqref{klich} requires almost Gaussian fluctuations of particle numbers (not to be confused with state Gaussianity) such that higher cumulants are suppressed. This is typically justified for large subsystems in the bulk of the information lattice, where the central limit theorem works well. However, near the boundary, the counting statistics may change. In particular, for open systems, fluctuations and higher cumulants in the subsystems adjacent to the reservoirs tend to be enhanced by environment-induced random jumps.

\begin{figure}
\center
\includegraphics[width= \linewidth]{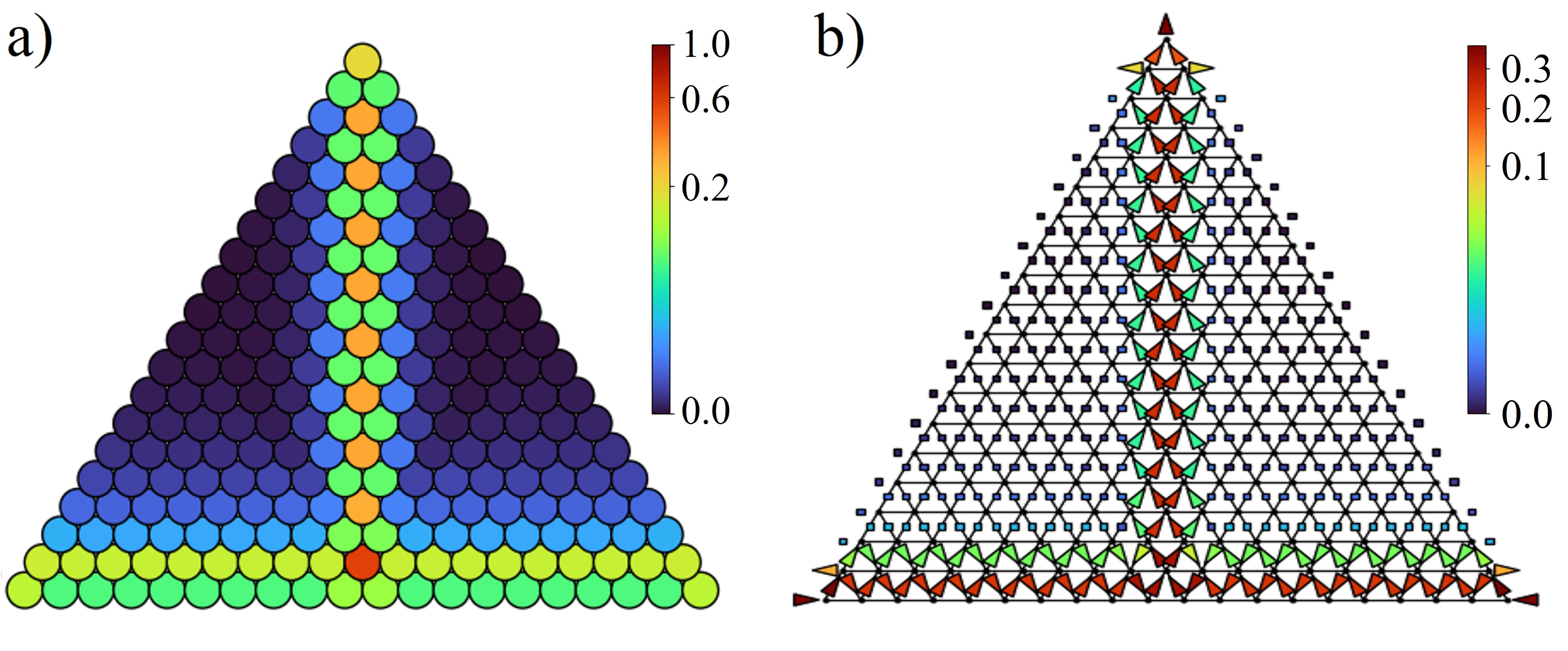}
\caption{Information lattice $\{ i_{(\ell,n)}\}$ and  information currents $\{\tilde{\cal I}_{(\ell,n)}\}$ in the presence of a bond defect for $N=20$, $g=J=1$, and $\epsilon=\delta=0$. The bond defect is modeled by $J_{j_0,j_0+1}=0.5$ with $j_0=10$. 
We use power-norm color scales.
(a) Information lattice. 
(b) Information currents indicated by arrows (direction) and colors (magnitude), with filled black squares for $|\tilde{\cal I}|<10^{-2}$. }
\label{fig13}
\end{figure}

In Fig.~\ref{fig13}, we illustrate the information lattice and the corresponding information currents for the case of a bond defect instead of the site defect.  Using a system of size 
$N=20$ with a central bond defect, we observe very similar features as for the case of a site defect, see Figs.~\ref{fig6}(c,d), \ref{fig7}(c,d), \ref{fig11}(c), and \ref{fig12}(c).  In particular, the information pillar is clearly visible both in the information lattice and in the information currents.

\begin{figure}
\center
\includegraphics[width= \linewidth]{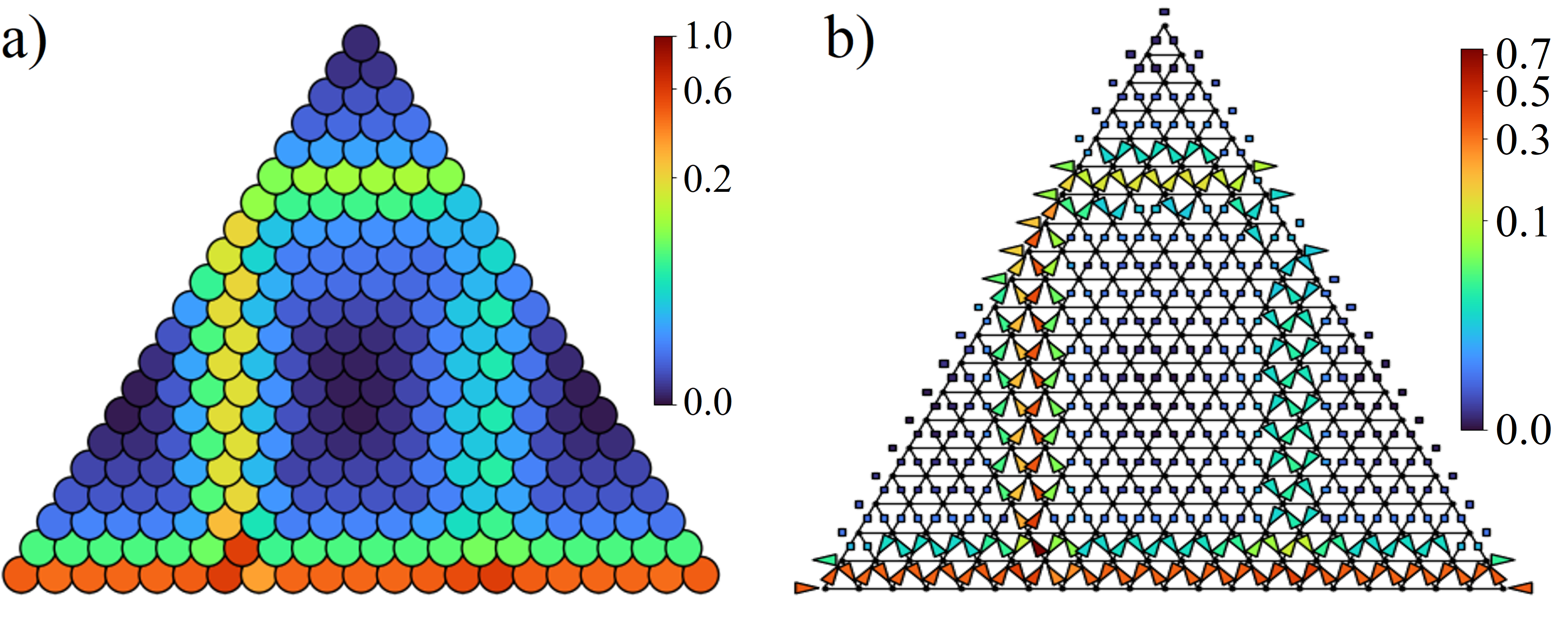}
\caption{Information lattice $\{ i_{(\ell,n)}\}$ and information currents 
$\{\tilde{\cal I}_{(\ell,n)}\}$ for $N=21, g=J=1, \epsilon=\delta=0$, for  
a site defect at $j_0=7$ with $\epsilon_{j_0}=3$. The defect position is away from the center of the chain.  We employ power-norm color scales. (a) Information lattice. (b) Information currents, with filled black squares for $|\tilde{\cal I}|<10^{-2}$.  } 
\label{fig14}
\end{figure}

In Fig.~\ref{fig14}, we next explore what happens if a site defect is present at an arbitrary site away from the center position.  For system size $N=21$ and putting the defect at site $j_0=7$, we observe that the information pillar still exists. However, it now does not extend all the way up to the top layer $\ell=N-1$ of the lattice anymore, but remains centered around the defect position $j_0$, in agreement with the interpretation given above for Fig.~\ref{fig6}(c).
In addition,  we observe a ``ghost'' pillar of smaller amplitude centered at the mirror position $j_1=N+1-j_0=15$.
Recalling the physical interpretation in terms of scattering by a localized impurity, the ghost pillar emerges due to multiple scattering processes between the initial backscattered component of the fermionic wavefunction and the local impurity. In particular, while the forwarded component travels to the right side, the backscattered component, after being reflected at the left endpoint of the chain, approaches again the impurity and a second splitting into a backward and a forward component takes place. The ghost pillar then encodes the pairwise correlations between the first and the secondary transmitted components of the wave function.  
The ``real'' pillar and the ghost pillar are also connected at the top by a horizontal strip in the information lattice with significant information amplitude, see Fig.~\ref{fig14}. We have checked that the position of the ghost pillar is not affected by changing the particle-hole asymmetry parameter $\delta$. 
Indeed, the Fermi velocities on both sides of the defect are identical---even though the value may depend on $\delta$---because of the uniform hopping strength $J$.
The multiple scattering processes causing a constructive interference pattern therefore imply a ghost pillar 
centered at the same location as for $\delta=0$. However, in accordance with the multiple scattering picture, we also
find that the ghost pillar becomes wider with increasing $|\delta|$.
By displacing the impurity from the center site, we therefore obtain surprisingly rich behavior in the information lattice and 
the associated information currents.  We note in passing that the 
ghost pillar can be turned into a real pillar by adding a second (and 
identical) site defect at the position $j_1$. 

\begin{figure}
\center
\includegraphics[width= \linewidth]{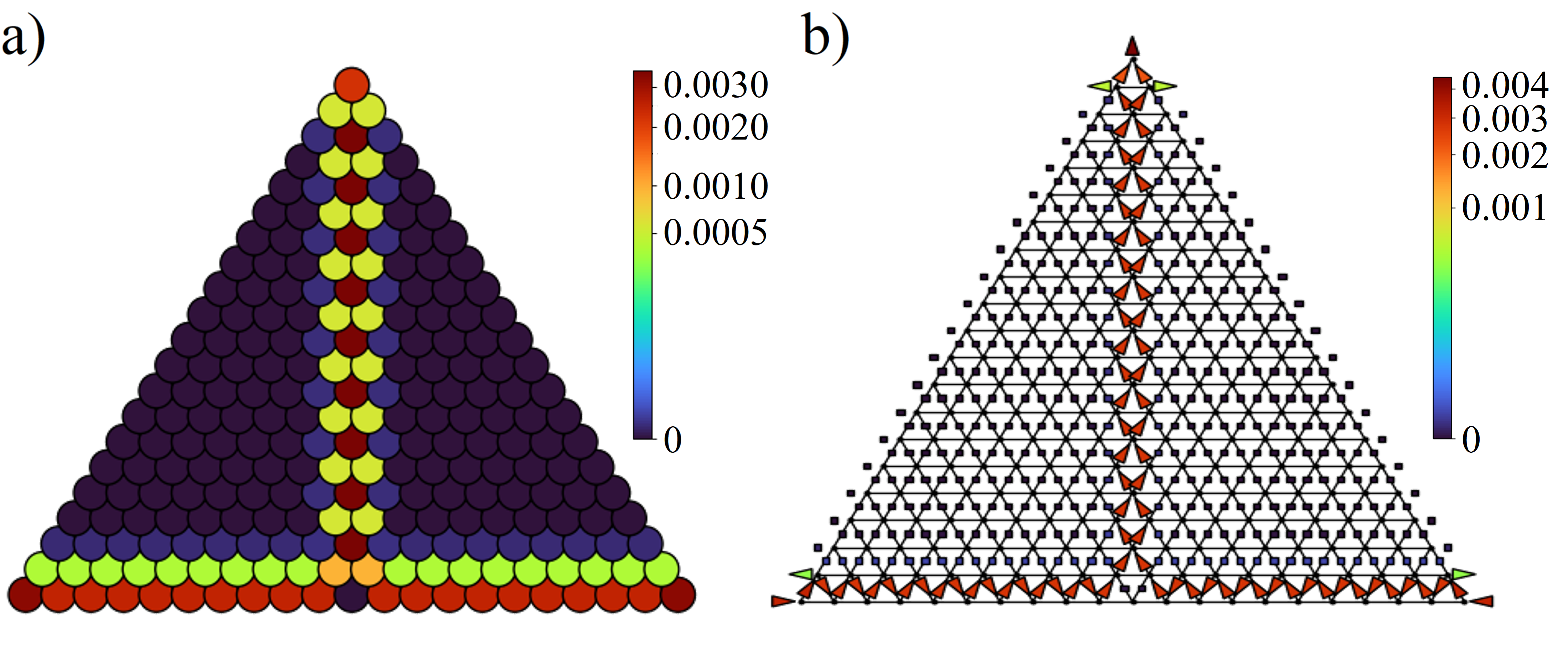}
\caption{Information lattice $\{ i_{(\ell,n)}\}$ and information currents $\{\tilde{\cal I}_{(\ell,n)}\}$ in the linear-response regime with the rates in Eq.~\eqref{smallbias}, using $N=21, j_0=11, g=J=1, \epsilon=\delta=0$, and drive amplitude $\phi=0.1$. The site defect at the central position $j_0$ has energy $\epsilon_{j_0}=3$. We use power-norm color scales. (a) Information lattice. (b) Information currents, with  
 filled black squares for $|\tilde{\cal I}|<10^{-4}$.} 
\label{fig15}
\end{figure}

\subsection{Linear-response regime}\label{sec3b}

We now turn to the linear-response regime introduced in Sec.~\ref{sec2a}, with the rates in Eq.~\eqref{smallbias}. The information lattice and the corresponding information currents in the NESS are shown in Fig.~\ref{fig15} for $N=21$ in the presence of a central site defect.
Evidently, the characteristic information pillar and the associated information current pattern are very similar to the results presented in Sec.~\ref{sec3a} for the large-bias limit. 
However, due to the much smaller drive amplitude, the magnitude of the information currents is greatly reduced, and also the information amplitudes $i_{(\ell,n)}$ are now much smaller.  For an experimental observation of the predicted information patterns, it is therefore advantageous to work in the large-bias limit.
We note in passing that the corresponding results in the absence of an impurity also closely resemble those reported in Sec.~\ref{sec3a}.
In particular, we have checked that the shielding effect observed in Fig.~\ref{fig7}(a) occurs in the linear-response regime as well,  including its lifting by particle-hole asymmetry $\delta\ne 0$.

It is important to note that the key features of the information lattice and the corresponding information currents discussed in this section are a consequence of the NESS induced by the environment. For instance, at equilibrium, achieved by sending the dimensionless drive amplitude $\phi\to 0$ in the linear-response rates \eqref{smallbias}, the scattering processes mentioned above will not result in an information pillar since the system is in a static configuration. (Note that the thermal equilibrium limit cannot be obtained from the large-bias rates \eqref{largebias}.) In the equilibrium case, all off-diagonal correlation matrix elements vanish, see Appendix \ref{appendix:exact for N=3} for details, and the information pillar is absent.  

Finally, as a topic for future studies, it will be highly interesting to understand the properties of information currents in relation to heat transport phenomena, particularly thermoelectric effects in the linear-response regime. 
Conventional thermoelectric phenomena describe the response of heat and charge currents to 
small applied bias voltages and/or temperature differences in terms of thermoelectric (e.g., Seebeck and Peltier) coefficients.  In particular, we have shown above that particle-hole asymmetry ($\delta\ne 0$) is needed for converting particle currents into information currents.  Similarly, it is well known that $\delta\ne 0$ is required for having non-zero
values of thermoelectric coefficients. This observation suggests the existence of generalized Onsager relations connecting the transport coefficients for particle currents and information currents in the 
linear-response regime.  Importantly, for such a formulation of an ``infoelectric'' transport theory and the corresponding kinetic coefficients, one needs to introduce the notion of ``information bias'' in analogy to the temperature bias in thermoelectric settings. 
It is natural to expect that quantum measurements, which directly address the information content of a quantum state, should be involved here.
We leave a detailed exploration of this perspective to future research, see  Refs.~\cite{Wichterich_2007,Barra_2015,Pereira_2018,Reis_2020,Cottet_2017,Oliveira_2025} for related works. 

\section{Transport-induced entanglement from fermionic negativity}\label{sec4}

In Sec.~\ref{sec3}, we have shown that a defect can stabilize NESS configurations featuring a finite local information 
for high layers of the information lattice.
However, the von Neumann entropy used in the definition of information in   Eq.~\eqref{eq:info_subsystem} quantifies bipartite entanglement only for pure states \cite{nielsen2010quantum,wilde2013quantum}. In order to address entanglement in the mixed states describing the NESSs encountered here, we employ the FN, which 
is a proper entanglement measure for mixed states of noninteracting fermion systems \cite{shapourian2017partial,shapourian2017many,Shapourian2019a,shapourian2019entanglement,eisert2018entanglement,Alba_2023}. 
In particular, for our open quasi-free fermion system, the negativity can be directly computed from the correlation matrix again. 
However, defining a ``negativity lattice" fully analogous to the information lattice, i.e., a decomposition of the FN into local contributions based solely on complementary bipartitions of the chain, is not straightforward. The FN is intrinsically defined for non-complementary subsystems, e.g., for two disjoint regions embedded in a larger system, and such correlations are not captured by the information lattice $\{ i_{(\ell,n)}\}$; see also Appendix \ref{appendix:exact for N=3}.  A complementary perspective on entanglement is offered in Appendix \ref{appendix:unraveling}, where we study pure-state trajectories obtained by unraveling of the LME.  For pure states, the von Neumann entropy quantifies bipartite entanglement.  

We first discuss how to quantify entanglement between two non-complementary regions ($A_1$ and $A_2$) of the total system in terms of the FN $\mathcal{E}$. We start from the general definition of the logarithmic negativity, 
\begin{equation}
    \mathcal{E}_{\rm ln} = \ln \mathrm{Tr}_A\left |\rho_{A}^{T_2} \right|,
\end{equation} 
for $A = A_1 \cup A_2$,
where the partially transposed density matrix $\rho_{A}^{T_2}$ follows from the reduced density matrix $\rho_A=\mathrm{Tr}_{\bar A} [\rho]$ by taking a partial transposition with respect to, say, region $A_2$. We note that the complementary region $\bar A$ neither has to be empty nor connected. The logarithmic negativity is, in general, difficult to compute. However, for open quasi-free fermion systems,  $\rho_{A}^{T_2} = e^{-i\pi/4}O_+ + e^{i\pi/4}O_-$
can be expressed by two Gaussian operators $O_{\pm}$, where the FN is given by
\begin{equation}
\mathcal{E} = \ln\mathrm{Tr}\sqrt{O_+ O_-}.
\end{equation}
This quantity can again be written in terms of the correlation matrix $C$. 
To this end, we introduce the matrices $G=2C-\mathbb{I}$ and $G_A$, where $G_A$ follows from $G$ by deleting all rows and columns related to $\bar A$. 
Expressing 
\begin{equation}
    G_{A}=\left(\begin{array}{cc} G_{A_{1},A_{1}} & G_{A_{1},A_{2}}\\
G_{A_{2},A_{1}} & G_{A_{2},A_{2}} \end{array}\right)
\end{equation} 
in block form with respect to $A_1$ and $A_2$, we define
\begin{eqnarray}
 \widetilde{G}_A &=& \frac12 \left[ \mathbb{I} - (\mathbb{I} + G_A^+ G_A^-)^{-1}(G_A^+ + G_A^-) \right ],\\ \nonumber
G_{A}^\pm&=&\left(\begin{array}{cc} G_{A_{1},A_{1}} & \pm i G_{A_{1},A_{2}}\\
\pm i G_{A_{2},A_{1}} & -G_{A_{2},A_{2}}\end{array}\right),
\end{eqnarray}
where $G_A^\pm$ corresponds to the covariance matrix of $O^\pm$. 
The FN is then expressed in terms of the eigenvalues $\{\mu_{i}\}$ of $C_A$ and $\{\lambda_{i}\}$ of $\widetilde{G}_A$ as \cite{Shapourian2019a}
\begin{equation}\label{negativity}
    \mathcal{E} = \sum_i \left[ \ln\left( \sqrt{\lambda_i}+\sqrt{1-\lambda_i}\right)  +  \ln\sqrt{ \mu_i^{2}+(1-\mu_i)^2 } \right].
\end{equation}

In Sec.~\ref{sec3}, we have reported that a single defect can generate an information pillar representing long-range correlations in the NESS of our open quasi-free fermion chain.
If this result can be mainly attributed to quantum entanglement instead of classical correlations, a finite FN is expected between 
chain segments located at the left and right sides of the defect. For the clean chain, where no long-range information correlations were found in Sec.~\ref{sec3}, we expect a vanishing FN.  Below, we study the large-bias regime as introduced in Sec.~\ref{sec2a} and discussed in Sec.~\ref{sec3a}.

\begin{figure}
\center
\includegraphics[width=0.95 \linewidth]{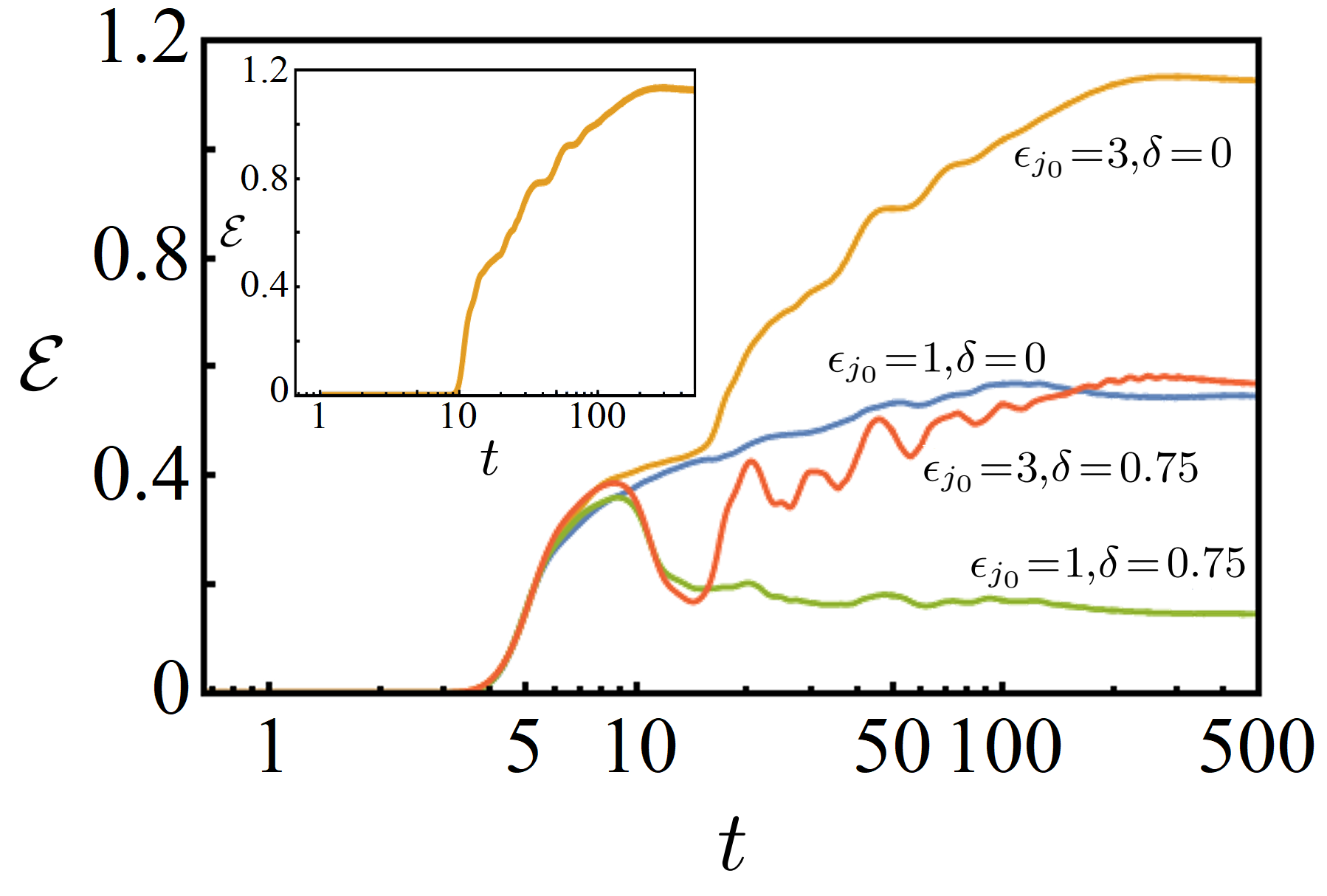}
\caption{Negativity $\mathcal{E}$ vs time $t$ between the first four ($A_1$) and the last four sites ($A_2$) of the chain,
see Eq.~\eqref{negativity}. We assume the large-bias limit with $N=21, g=J=1, \epsilon=0,$ and allow for a local defect at the central site $j_0=11$.  We consider different values of $\epsilon_{j_0}$ and $\delta$. Note the logarithmic time scale. Main panel: The initial state $\rho(0)$ is taken as steady state for $\epsilon_{j_0}=\delta=0$. 
At time $t=0^+$, $(\epsilon_{j_0}, \delta)$ are suddenly changed to the respective post-quench values specified for the four curves.  Inset: Results for $\epsilon_{j_0}=3$ and $\delta=0$ (see the top yellow curve in the main panel) but with $\rho(0)$ taken as the zero-particle state.}
\label{fig16}
\end{figure}

In order to observe how entanglement builds up in the system, we monitor the time-dependent FN \eqref{negativity} for system size $N=21$, allowing for a local defect in the center of the chain, $j_0=11$. We consider a sudden quench of the
parameters $(\epsilon_{j_0},\delta)$ to new post-quench values at time $t=0^+$.  Choosing the pre-quench values $\epsilon_{j_0}=\delta=0$ corresponding to a clean particle-hole symmetric chain, we study different post-quench parameters in Fig.~\ref{fig16}.
In the main panel of Fig.~\ref{fig16}, the initial state $\rho(0)$ is taken as the NESS for the pre-quench parameters, with the initial correlation matrix $C(0)$ in Eq.~\eqref{corrdef}.  The correlation matrix $C(t)$ then evolves according to Eq.~\eqref{eq:C-equ} toward the NESS correlation matrix for the post-quench parameters.
For a given choice of the segments $A_1$ and $A_2$, inserting the solution for $C(t)$ into Eq.~\eqref{negativity} yields the FN $\mathcal{E}(t)$. 
In Fig.~\ref{fig16}, we assume that $A_1$ ($A_2$) corresponds to the first (last) four sites of the chain. 
We then find $\mathcal{E}(t=0)=0$, as expected for $\epsilon_{j_0}=\delta=0$ since the information lattice then contains only short-range correlations. For $\epsilon_{j_0}=1$ and $\delta=0$, the FN increases toward a finite post-quench NESS value for $t\to \infty$. 
The corresponding value increases when further increasing the  impurity strength, e.g., for $\epsilon_{j_0}=3$, consistent with Fig.~\ref{fig8}(b).  
The finite negativity $\mathcal{E}(t\to \infty)$ shows that the long-range correlations associated with the defect-induced information pillar in the NESS are, at least partially, caused by quantum entanglement. (The FN is not obtainable from the von Neumann entropy \cite{Alba_2023}. It is thus not possible to directly extract the quantum component of the entanglement from the information lattice. However, the FN is a quantum entanglement quantifier.)

If also $\delta$ is quenched, see, e.g., the case $\epsilon_{j_0}=3$ and $\delta=0.75$ in Fig.~\ref{fig16}, $\mathcal{E}(t\to \infty)$ becomes smaller again. To rationalize this observation, we recall that $\delta\ne 0$ implies an asymmetric information pillar, see Fig.~\ref{fig6}(d), leading to a decrease of correlations between $A_1$ and $A_2$.   The inset of Fig.~\ref{fig16} shows that $\mathcal{E}(t\to \infty)$ does not depend on the initial state $\rho(0)$, which only affects the transient dynamics toward the NESS.  When starting from an empty chain for $t<0$, the FN remains zero for $t\alt 10$,  corresponding to the time needed for a particle injected at the left boundary to reach $A_2$ for the parameters considered here. 
 For $t\agt 10$, a fast increase of $\mathcal{E}(t)$ toward its post-quench NESS value is then observed.

\begin{figure}
\center
\includegraphics[width=0.95 \linewidth]{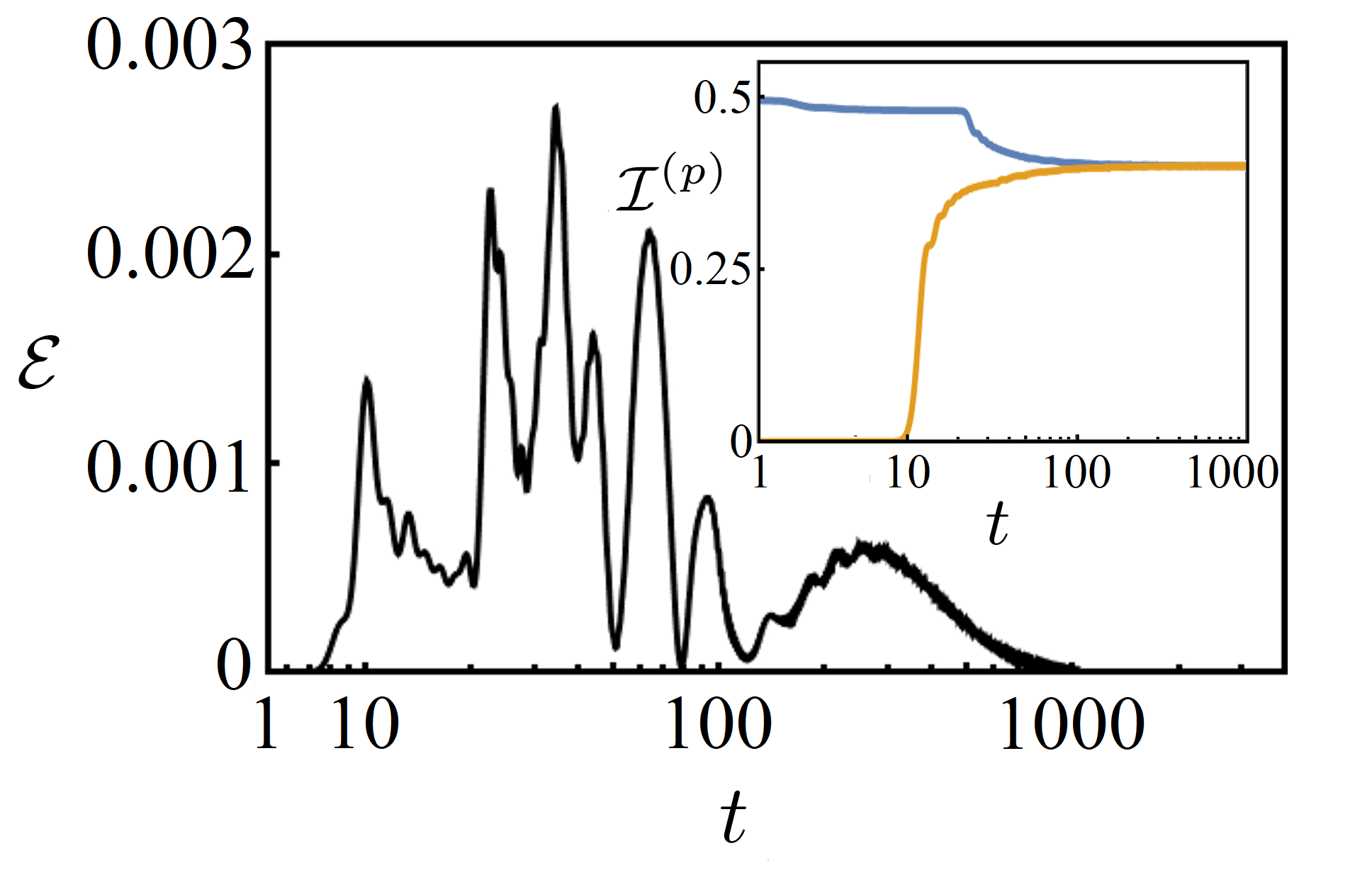}
\caption{Time-dependent FN $\mathcal{E}$ in Eq.~\eqref{negativity} between the first four ($A_1$) and the last four sites ($A_2$) of a clean particle-hole symmetric chain with $N=21$. 
Main panel: $\mathcal{E}$ vs time $t$ for $g=J=1$ and $\epsilon=\delta=0,$  where $\rho(0)$ is the zero-particle state.  Note the logarithmic time scale.
Inset: Corresponding particle currents injected at site $1$ (blue) and outgoing at site $N$ (yellow curve) vs $t$. }
\label{fig17}
\end{figure}

The main panel of Fig.~\ref{fig17} illustrates $\mathcal{E}(t)$ for a clean particle-hole symmetric chain with $N=21$ sites when starting from an initially empty state $\rho(0)$.
 We observe that both the initial and the final configuration are characterized by $\mathcal{E}=0$ as expected. 
 Indeed, for the $t\to\infty$ NESS, information correlations are short-ranged, see Fig.~\ref{fig6}(a). Nonetheless, 
 a small but finite FN is generated during the transient dynamics toward this NESS.  Again, we find that $\mathcal{E}(t)$ vanishes for $t\alt 10$ since only at later times, particles injected at the left end can reach region $A_2$.
 This effect is also visible in the corresponding particle currents ${\cal I}^{(p)}_{1}(t) = {\cal I}^{(p)}_{E \to 1}(t)$ and ${\cal I}^{(p)}_{N}(t) = {\cal I}^{(p)}_{N \to E}(t)$ shown in the inset of Fig.~\ref{fig17}.
 Indeed, the current injected at the left end ($j=1$) is maximal for $t=0$, since the chain is initially empty and thus there are no obstacles to the particle injection.
 However, an outgoing current at the right end ($j=N$) only appears for $t\agt 10$ because of the finite particle propagation time through the chain.
 For $10\alt t\alt 100$, both the FN $\mathcal{E}$ and the particle currents ${\cal I}^{(p)}_{1 (N)}$ evolve in time, with 
 $\mathcal{E}(t)$ displaying oscillatory behavior. One can rationalize the latter observation by noting that particles cannot leave the 
 chain by tunneling out from site $j=1$ under
 large-bias conditions, see Eq.~\eqref{largebias}. Since there is only a finite probability for particles to leave at the right end, particles can
 bounce back and forth inside the chain, resulting in the nontrivial behavior of $\mathcal{E}(t)$. For $t\agt 100$, the particle currents have already reached the 
 stationary value, while the true NESS configuration has not yet been reached as manifested by the finite FN observed for $100\alt t\alt 1000$ in Fig.~\ref{fig17}.
Only in the asymptotic limit $t\to\infty$, one reaches the expected NESS negativity $\mathcal{E}=0$. 

\begin{figure}[t]
\center
\includegraphics[width=0.85 \linewidth]{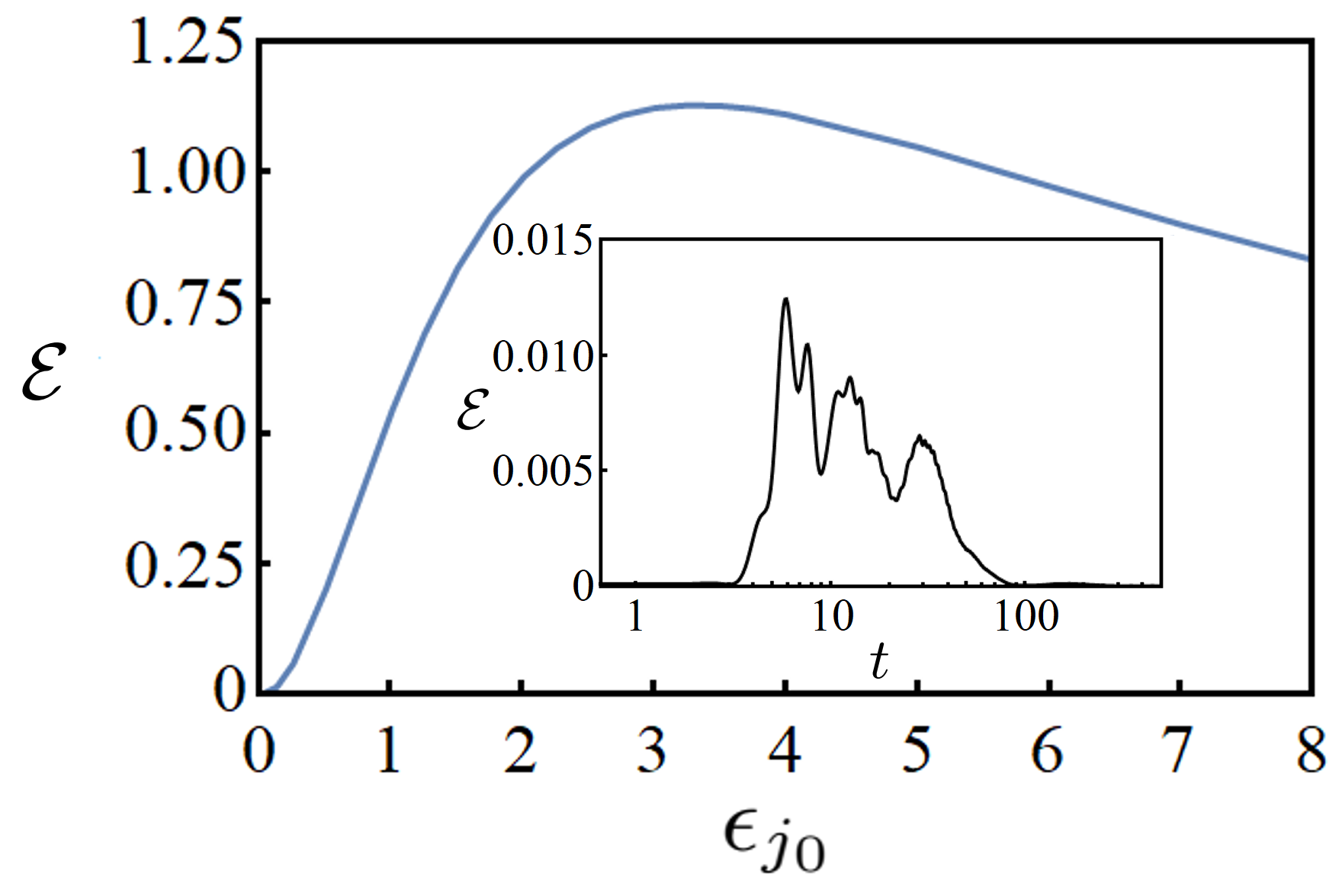}
\caption{FN $\mathcal{E}$ in Eq.~\eqref{negativity} for an onsite defect at site $j_0=11$. We study a fermionic chain in the large-bias limit with $N=21, g=J=1$, and $\epsilon=\delta=0$. Main panel: NESS value of $\mathcal{E}$ vs onsite energy $\epsilon_{j_0}$, where $A_1$ ($A_2$) contains the first (last) four sites of the chain. 
Inset: Time-dependent FN for $\epsilon_{j_0}=3$, where $A_1$ ($A_2$) now contains sites 1 and 2 (sites 6 and 7),  with $\rho(0)$ being the NESS configuration for $\epsilon_{j_0}=0$.
Note the logarithmic time scale.} 
\label{fig18}
\end{figure}

Similarly, in the presence of an impurity, we observe vanishing long-range correlations between chain segments both located at either the left ($j<j_0$) or the right ($j>j_0$) side of the impurity. This case is illustrated in the inset of Fig.~\ref{fig18}, 
where the non-complementary subsystems $A_1$ and $A_2$ used for computing the time-dependent FN $\mathcal{E}(t)$ refer to sites with $j<j_0$.  While the FN is finite at intermediate times, it vanishes in the asymptotic limit $t\to \infty$ as expected. 
In the main panel of Fig.~\ref{fig18}, we show the NESS value of $\mathcal{E}$, with $A_1$ ($A_2$) referring to the first (last) four sites of the chain, as a function of $\epsilon_{j_0}$ for $\delta=0$, again for the $N=21$ chain with large-bias conditions and a local defect at the central site $j_0=11$.  
Consistent with the corresponding observations for the information lattice in Sec.~\ref{sec3a}, the FN increases when 
increasing the energy $\epsilon_{j_0}$ up to $\epsilon_{j_0} \approx 3$. 
For even larger values of $\epsilon_{j_0}$, $\mathcal{E}$ decreases and is ultimately expected to vanish for a cut chain with $\epsilon_{j_0} \to \infty$.

\section{Conclusions}\label{sec5}

In this work, we have developed a detailed characterization of local information currents and transport-induced entanglement in nonequilibrium steady states (NESSs). Using the framework of the information lattice~\cite{Klein2022, Artiaco2025}, we have analyzed how von Neumann information is stored and redistributed across spatial scales. We have derived explicit expressions for local information currents in open (possibly interacting) fermion chains governed by Lindblad master equations, see Sec.~\ref{sec2}.   We have applied this theory to quasi-free fermions, see Sec.~\ref{sec22}, to illustrate how the flow of information is resolved both along the physical chain and within the layers of the information lattice, and to relate information transport to the underlying unitary dynamics and dissipative couplings at the chain boundaries, see Sec.~\ref{sec3}.

A central result of our analysis is that local impurities qualitatively modify the structure of information in current-carrying NESSs. While in the clean, particle-hole symmetric case, the information lattice displays predominantly short-range correlations and the NESS information currents are effectively shielded from the bulk of the information lattice, a single site or bond defect generates a pronounced ``information pillar'' extending across all scales of the lattice. This pillar reflects impurity-induced long-range correlations that span the entire chain and are absent in equilibrium. When particle-hole symmetry is broken, the information currents become unidirectional, and the pillar acquires a marked asymmetry, encoding how nonequilibrium driving and scattering conspire to shape the flow of quantum information. We have shown that these features are robust across different bias regimes (large-bias and linear-response) and impurity configurations, including off-center and multiple defects.

We have further established a practical connection between the information lattice and easily accessible experimental observables by introducing a ``noise lattice'' built from particle-number fluctuations in all subsystems. For quasi-free fermions, the von Neumann information and the variance of the particle number are both determined by the spectrum of the reduced correlation matrix. Exploiting this relation, guided by the Klich--Levitov correspondence between the von Neumann entropy and charge fluctuations~\cite{Klich_2009}, we have demonstrated that the information lattice can be accurately approximated from noise measurements alone, especially away from the boundaries.  Similarly, local information currents flowing on the information lattice can be measured by exploiting the Klich--Levitov correspondence as well.  For such measurements, one needs to determine standard particle currents, onsite occupation numbers, and covariances of standard particle-number currents and/or local occupation numbers.  An alternative measurement scheme, which avoids the approximations behind the Klich--Levitov correspondence, follows by measuring the correlation matrix $C$, e.g., by interferometric techniques \cite{Knap_2013,Pedernales_2014,DelRe_2024,Liu_2025,Yu_2011,Kastner_2020,Wang_2025}.
Indeed,  for quasi-free fermions, the local information currents are fully determined by $C$.
In particular, the information currents flowing between layers $\ell=0$ and $\ell=1$ of the information lattice can be  expressed 
in terms of particle currents and occupation numbers.
This opens up a promising experimental route toward probing information flow in driven open quantum many-body systems, using established tools for current- and density-noise detection  \cite{Clerk2010} in such platforms as ultracold atoms and arrays of quantum dots.

Finally, we have investigated entanglement transport in these NESSs using the fermionic negativity~\cite{shapourian2017partial,shapourian2017many, Shapourian2019a, shapourian2019entanglement, eisert2018entanglement, Alba_2023}. While the von Neumann information in the lattice accounts for the total amount of correlations, namely, both classical and quantum, and is strictly an entanglement measure only for pure states, the negativity provides a proper quantifier of mixed-state entanglement between spatially separated segments. We have shown that the impurity-induced information pillar is accompanied by a finite end-to-end negativity in the NESS, demonstrating that the associated long-range correlations have genuinely quantum-correlation contributions. By following the time evolution after parameter quenches, we have also highlighted the transient build-up and relaxation of entanglement, and clarified how it depends on impurity strength and particle-hole asymmetry. In contrast, clean particle-hole symmetric chains exhibit vanishing end-to-end negativity in the NESS, consistent with the short-range structure of the corresponding information lattice.

Our results suggest several directions for future work. A natural extension is to study stochastic unravelings of the Lindblad dynamics in terms of quantum trajectories, in particular if the reservoirs are realized as measurement devices \cite{Cao2019a, Fazio2025, Xing2025}. In such an unraveling, information currents in the lattice could be related directly to measurement records and information gain or loss along single trajectories, potentially clarifying the interplay between measurement backaction, classical randomness, and genuinely quantum information flow in open many-body systems.  We offer some thoughts along these lines in Appendix \ref{appendix:unraveling}. Since the unraveled trajectories refer to pure states, where the von Neumann entropy quantifies bipartite entanglement, such an approach can also directly probe local entanglement currents.

A related second direction is to develop an explicit ``negativity lattice''. Although we have used the FN to explore entanglement transport, a scale-resolved decomposition of mixed-state entanglement inspired by (but distinct from) the information lattice remains a challenge. Since the negativity is intrinsically sensitive to correlations between non-complementary regions, such a construction would necessarily go beyond bipartitions into contiguous segments and incorporate multipartite structures. If successful, this could provide the basis for a hydrodynamic description of entanglement transport in mixed states, where entanglement densities and currents obey effective continuity-like equations analogous to those we have derived for local information.

A third extension for future research concerns studies of heat transport in connection with information currents and entanglement transport, see Sec.~\ref{sec3b}.
This direction resonates with Landauer's principle that connects thermodynamic concepts to the time evolution of information in quantum computation protocols.
Moreover, in hydrodynamic approaches, the Navier-Stokes equation describing the evolution of particle currents is supplemented by a heat-balance equation
describing the evolution of entropy (and equations for other possible almost conserved modes, see, e.g., Ref.~\cite{Foster2009}).
Since heat is inherently related to the entropy transport, and entropy is linked to information
\cite{Wichterich_2007,Barra_2015,Pereira_2018,Reis_2020,Cottet_2017,Oliveira_2025},
a direct connection between heat transport and the local information currents studied here should
exist. In particular, information flows and the associated generation of entanglement could be driven by differences in reservoir temperatures rather than by an applied bias voltage. 

A fourth extension is to address the role of disorder.
In the presence of uncorrelated elastic disorder, one should compare the system length $L$ to the Anderson localization length $\xi.$  A qualitative discussion can be given as follows. For $\xi \ll L$, the system represents an insulator which can be divided into a bulk region and two boundary regions of length
$\approx \xi$, see Ref.~\cite{Berger2025}. The bulk is decoupled from the endpoint environments and acts like a closed system whose evolution depends on the initial conditions. In the large-bias regime, the left (right) boundary region will reach the fully filled (empty) steady state. In both cases (corresponding to maximum or zero occupancy), the information lattice is trivially zero for $\ell>0$ while $I_{(\ell=0)}=1$.  
On the other hand, for $\xi > L$, the system is still able to conduct but with a reduced charge current. At the same time, the average occupation number profile is not flat anymore in the bulk as in Fig.~\ref{fig5}, see Ref.~\cite{Nava_2021}. Similarly to the single-impurity case in Sec.~\ref{sec3a}, disorder should therefore break the shielding effect observed in the clean case. Disorder is also expected to affect the information pillar since each disorder site acts as a scattering impurity, giving rise to a proliferation of information pillars interfering with each other. Finally, different types of disorder will have different effects. For instance, binary disorder can be used to manipulate the density of pillars in the bulk.

In addition, the present study could be extended by investigating the role of local dephasing and/or dissipation in the bulk of the chain. For instance, by allowing for local onsite Hermitian jump operators $L_k \sim c_j^\dagger c_j^{}$ in Eq.~\eqref{eq:rho-equ}, one can model dephasing effects without spoiling the solvability of the Lindblad equation in terms of the correlation matrix \cite{Fazio2025}.
In a similar manner, one may investigate the impact of monitoring the onsite occupation numbers of the open fermion chain by performing (weak) measurements. 
For the monitoring of boundary-driven systems, see, e.g., Refs.~\cite{Turkeshi_2021_2,Turkeshi_2022,Turkeshi_2022_2,Turkeshi_2024}.
Including quantum measurements within our framework could also allow one to define the notion of ``information bias'' (or ``information gradient''), in analogy to the thermal bias (temperature gradient) in heat transport, see our discussion in Sec.~\ref{sec3b}. Thereby, a direct link to measurement-induced transport based on the ``steering'' of quantum states~\cite{Roy2020, Herasymenko2023, Poepperl2023, Morales2024} could be established.

On the experimental side, our noise-lattice approach suggests concrete protocols to access information transport in realistic platforms. It would be interesting to combine full counting statistics techniques
(including statistics of measurement outcomes \cite{Turkeshi_2021}) with time- and space-resolved detection to go beyond second-order fluctuations and test the range of validity of the noise-information correspondence in the presence of interactions or non-Gaussian reservoirs. 
 While the general formalism in Sec.~\ref{sec2} remains valid, the Gaussianity of fermion states is broken, and the system dynamics is not described by only the correlation matrix anymore. 
Further, for non-Gaussian states, the Klich--Levitov correspondence between full counting statistics and entanglement \cite{Klich_2009} breaks down. 
In the context of measurement-induced transitions, this amounts to ``information-charge separation'' \cite{Poboiko2025a}, see also Ref.~\cite{Foster2025}, which implies a splitting between the ``charge sharpening'' (or learnability) transition~\cite{Agrawal2022, Barratt2022, Vasseur2022} and the entanglement transition \cite{Fisher2023}.

Extending the present analysis from quasi-free to interacting systems, e.g., using approximate numerical schemes such as Local-Information Time Evolution~\cite{Klein2022, Artiaco2024, Harkins2025}, can also clarify whether the impurity-stabilized information pillars and the transport-induced entanglement reported above are generic features of driven open quantum matter, and how they manifest in different universality classes and phases. In this context, one may also employ quantum-field-theory approaches similar to those in Refs.~\cite{Agrawal2022, Barratt2022, Poboiko2025a, Foster2025} in order to tackle non-Gaussianity effects on information currents and the negativity, especially under the conditions of monitored dynamics. A related exciting direction concerns the development of a hydrodynamic theory, which seems in reach in view of the locality offered by the 
information lattice.  Hydrodynamic theories are formulated in terms of coupled transport equations that govern information currents, particle currents, and heat currents,  
and thus may allow for analytical studies of interacting models.
Concerning the specific physical phenomena predicted for noninteracting fermion chains in this work, we expect them to be qualitatively robust against the inclusion of weak interactions. For instance, the information pillar discussed in Sec.~\ref{sec3a} 
should still be present but is likely broadened by interactions.
However, strongly repulsive interactions, e.g., of Hubbard type, can significantly affect our noninteracting results, in particular the shielding effect and the information pillar. 
Indeed, strong interactions could significantly modify the average occupation numbers from flat to site-dependent profiles, which in turn may break the information current shielding reported in Sec.~\ref{sec3a}. Strong interactions are also known to enhance the backscattering of low-energy electrons off impurities, which affects the information pillar at large scales, $\ell \gg 1$, by diluting the corresponding information currents. We note in passing that integrability is not expected to play a key role
for our findings. The concepts and ideas put forward in this paper 
apply both to integrable and non-integrable models.

From a broader perspective, recasting entanglement dynamics through the local information transport paves the way for a genuinely hydrodynamic description of entanglement in driven many-body systems. This suggests a unifying language, in which emergent conservation laws for information currents coexist with---and constrain---the buildup of multipartite entanglement across different length scales. In the long term, such a hydrodynamic description could provide a deeper understanding of the entanglement dynamics in quantum systems beyond equilibrium, effectively connecting microscopic scrambling, dephasing, and decoherence to macroscopic transport in a unified way. In this sense, our findings mark a step toward reconciling the nonlocal character of entanglement with the local structure of physical laws---a direction that promises to reshape how we think about dynamics and correlations in complex quantum environments far from equilibrium.

\begin{acknowledgments} 
We thank J. H. Bardarson, M. Buchhold, S. Diehl, D. Giuliano, A. Mirlin, I. Poboiko, and A. Rosch for discussions.
We acknowledge funding by the Deutsche Forschungsgemeinschaft (DFG, German Research Foundation) under Projektnummer 277101999 -- TRR 183 (projects A01 and B02) and projects EG 96/14-1 and GO 1405/7-1,
and under Germany’s Excellence Strategy – Cluster of Excellence
Matter and Light for Quantum Computing (ML4Q) EXC 2004/2 – 390534769. C.~A.\ acknowledges funding from the European Research Council (ERC) under the European Union’s Horizon 2020 research and innovation program (Grant Agreement No.~101001902). Y.~G.\ acknowledges support from the National Science Foundation (NSF-2338819)–-Binational Science Foundation (BSF-2023666).
\end{acknowledgments}

\section*{Data availability}

The data underlying the figures in this work are available at Zenodo \cite{Zenodo}.


\appendix

\section*{APPENDIX}

Below we provide analytical results for the NESS configuration in the large-bias limit for the case $N=3$, see Appendix \ref{appendix:exact for N=3}. Further, in Appendix \ref{appendix:unraveling}, we discuss the unraveling of the Lindblad equation and the associated information currents.

\section{Exact solution of $N=3$ NESS} 
\label{appendix:exact for N=3}

We here provide the analytical solution for the $N=3$ NESS of quasi-free fermions. Let us start with the large-bias limit, see Eq.~\eqref{largebias}, where the only nonvanishing rates are 
$\Gamma_1=g(1+\delta)$ and $\gamma_3=g(1-\delta)$.  We assume $J_{12}=J_{23}=J$ and $\epsilon_1=\epsilon_3=0$, but allow for a finite energy $\epsilon_2$ at site $j_0=2$. 
Equation \eqref{eq:C-equ} is then solved for the steady-state correlation matrix with $\dot C=0$.  
With $C_{ji}=C_{ij}^\ast$, we find
\begin{eqnarray}
C_{11} &=& 1 - \frac{4 J^2 \gamma_3 (4 J^2 + \Gamma_1 \gamma_3)}
{(\Gamma_1 + \gamma_3) \left[ (4 J^2 + \Gamma_1 \gamma_3)^2 + 4 \Gamma_1 \gamma_3 \epsilon_2^2 \right]}, \nonumber \\
C_{12} &=&  
-\frac{2 i J \Gamma_1 \gamma_3 \left[ 4 J^2 + \gamma_3 (\Gamma_1 - 2i \epsilon_2) \right]}
{(\Gamma_1 + \gamma_3) \left[ (4 J^2 + \Gamma_1 \gamma_3)^2 + 4 \Gamma_1 \gamma_3 \epsilon_2^2 \right]}, \nonumber \\
C_{13} &=& \frac{8 i J^2 \Gamma_1 \gamma_3 \epsilon_2}
{(\Gamma_1 + \gamma_3) \left[ (4 J^2 + \Gamma_1 \gamma_3)^2 + 4 \Gamma_1 \gamma_3 \epsilon_2^2 \right]}, \nonumber \\
C_{22} &=& 
\frac{\Gamma_1 \left[ (4 J^2 + \Gamma_1 \gamma_3)(4 J^2 + \gamma_3^2) + 4 \Gamma_1 \gamma_3 \epsilon_2^2 \right]}
{(\Gamma_1 + \gamma_3) \left[ (4 J^2 + \Gamma_1 \gamma_3)^2 + 4 \Gamma_1 \gamma_3 \epsilon_2^2 \right]}, \nonumber \\
C_{23} &=&  
-\frac{2 i J \Gamma_1 \gamma_3 \left[ 4 J^2 + \Gamma_1 (\gamma_3 + 2 i \epsilon_2) \right]}
{(\Gamma_1 + \gamma_3) \left[ (4 J^2 + \Gamma_1 \gamma_3)^2 + 4 \Gamma_1 \gamma_3 \epsilon_2^2 \right]}, \nonumber \\ \label{C3}
C_{33} &=& \frac{\Gamma_1}{\gamma_3}\left(1-C_{11}\right).
\end{eqnarray}
The particle current in the NESS follows as
\begin{equation}\label{ip3}
\mathcal{I}^{(p)}= \Gamma_{1}\left(1-C_{11}\right)=-2J\, {\rm Im}\left(C_{12}\right).
\end{equation}
It becomes maximal for $\Gamma_1=\gamma_3=g=2J$ (where $\delta=0$) and $\epsilon_2=0$, which defines the optimal working point. By further increasing $\Gamma_1$ and $\gamma_3$, one encounters a negative differential conductance regime due to the quantum Zeno effect \cite{Benenti_2009,Nava_2021}. 
We note in passing that for the clean case $\epsilon_2=0$, Eq.~\eqref{C3} also determines the NESS correlation matrix for arbitrary $N$. To this end, one has to replace $C_{33}\to C_{NN}$, $C_{22}\to C_{jj}$ and $C_{12}\to C_{j,j+1}=C^*_{j+1,j}$ for $1<j<N$, where $C_{i,j}\to 0$ for all $|i-j|>1$ \cite{Nava_2021}.

With increasing $\epsilon_2$, Eq.~\eqref{C3} yields a decrease in $C_{12}$ and $C_{23}$, while $C_{13}\propto \epsilon_2$ vanishes for the clean case. 
This observation provides an important clue that impurities can create long-range correlations in the NESS.  For larger system sizes, long-range correlations centered around the impurity site are similarly induced by the presence of the impurity \cite{Fraenkel2023}.  These correlations are responsible for the information pillar discussed in Sec.~\ref{sec3}, which emerges in the NESS around a defect.

Let us next explore the relation between such correlations and the information lattice. Using Eq.~\eqref{C3}, the information lattice follows by means of Eqs.~\eqref{eq:info_subsystem} and \eqref{eq:triangles-to-sites}. We note that in the information lattice, there is no element $i_{(\ell,n)}$ containing direct correlations between disconnected sites or regions.  Let us then apply  Eqs.~\eqref{eq:info_subsystem} and \eqref{eq:triangles-to-sites} to the subsystem $A$ composed of sites 1 and 3, where
\begin{eqnarray}
    i_{(1,A)}&=&I_{(1,A)}-I_{(0,1)}-I_{(0,3)} ,\\ \nonumber
    I_{(1,A)}&=&2+\mathrm{Tr}_A\left[C_{A}\log_{2}C_{A}+(\mathbb{I}-C_{A})\log_{2}(\mathbb{I}-C_{A})\right].
\end{eqnarray}
To make progress, we expand to first order in the particle-hole asymmetry $\delta$. The NESS correlation matrix $C_{ij}=C_{ij}^\delta$, with $C_{ij}^0$ for $\delta=0$, then has the form 
\begin{eqnarray}
C^\delta_{11} & =& 1- C^0_{33} ( 1- \delta),\quad\ C^\delta_{33}  = C^0_{33} ( 1+ \delta),\nonumber \\
C^\delta_{22}  &=&\frac{1}{2} + \frac{ 16J^4-g^4 + 4g^2 \epsilon_2^2}{2|d|^2} \delta, \quad\
C^\delta_{13}  =  C^0_{13} ,\label{C3d}\\
C^\delta_{12} &=& C^0_{12}+ \frac{2Jg^2\epsilon_2}{|d|^2} \delta , \nonumber \quad\
C^\delta_{23}  =-C^{0,\ast}_{12}+\frac{2Jg^2\epsilon_2}{|d|^2} \delta, 
\end{eqnarray}
where we define the complex-valued quantity
\begin{equation}\label{ddef}
d=g^2+4J^2+2ig\epsilon_2.
\end{equation}
To lowest order in $\delta$, we thus obtain an increase (for $\delta>0$) or decrease (for $\delta<0$) of the occupancy factors $\bar n_j=C_{jj}$ on all sites, without affecting the current \eqref{ip3} since ${\rm Im}\left(C_{j,j+1}\right)$ is unchanged. Assuming $\delta>0$, the increase in $C_{nn}$ also increases the information stored in $i_{(0,n)}$. 
At the same time, the dissipative information currents for $\delta\ll 1$ are given by
\begin{equation}
\mathcal{I}_{(0,n=1,3),E}  =-A_1 \log_{2} \left(1+A_2\right) \mp A_3\delta,
\end{equation}
with the positive constants 
\begin{align}
A_1  &= \frac{2gJ^2(g^2+4J^2)}{|d|^2},\quad  A_2 = g^2\, \frac{g^2+4J^2+4\epsilon_2^2}{2J^2(g^2+4J^2)},\notag
\\
A_3 &=\frac{2 g J^2 (g^2+4J^2)}{\ln 2\ [|d|^2-2J^2(g^2+4J^2)]}.
\end{align}
For $\delta=0$, both information currents $\mathcal{I}_{(0,1),E}$ and $\mathcal{I}_{(0,3),E}$ are negative, indicating information flow from the environment into the system. However, $|\mathcal{I}_{(0,1),E}|$ increases with increasing $\delta$, while $|\mathcal{I}_{(0,3),E}|$ is reduced. For sufficiently large $\delta$, $\mathcal{I}_{(0,3),E}$ changes sign, and a transition between the shielding regime in Sec.~\ref{sec3a} and a regime with information current flowing from the left to the right side occurs.  The transition happens when $\log_{2}(C_{33}/(1-C_{33}))=0$, i.e., for $C_{33}=\frac12$. 
For $\epsilon_2=0$ and $g=J$, we then find from Eq.~\eqref{C3} that the transition occurs at $\delta^*=\sqrt5-2$.  In fact, this result holds for arbitrary $N$ in the absence of impurities and can easily be generalized to the case $g\ne J$ for which we find
\begin{equation}\label{A8}
    \delta^* = \frac{1}{g^2} \left( \sqrt{g^4 + 4 J^4} - 2J^2 \right).
\end{equation}
A similar argument applies to the case $\delta<0$, where the transition happens for $C_{11}=\frac12$, leading to $\delta=-\delta^*$.
Let us briefly comment on different limiting behaviors implied by Eq.~\eqref{A8}. First, in the strong-coupling limit $g\gg J$, we encounter a quantum Zeno 
regime with $\delta^\ast \to 1$, with an associated ``information blockade". In the opposite weak-coupling limit $g \ll J$, 
one instead finds $\delta^\ast \ll 1$, indicating that the shielding regime becomes very small.

Finally, we provide the analytical solution for the NESS correlations in the linear-response small-bias regime, see Sec.~\ref{sec2a}. We focus on the case $\delta=0$. With the rates in Eq.~\eqref{smallbias} and $d$ in Eq.~\eqref{ddef}, we find 
\begin{eqnarray}
C^0_{11} &=& \frac12 +  \left( 1- \frac{4J^2(g^2+4J^2)}{ |d|^2} \right) \frac{\phi}{2}, \nonumber \\
C^0_{22} &=& \frac12,\quad C^0_{33} = 1-C^0_{11},\quad
C^0_{12} = 
-\frac{i  g J}{d}\, \phi,\nonumber\\ C^0_{23} &=&  
-\frac{i g J}{d^\ast} \, \phi, \quad
C^0_{13} = \frac{4i g J^2 \epsilon_2}{|d|^2}\, \phi.
\label{C3b}
\end{eqnarray}
For vanishing drive amplitude $\phi\rightarrow 0$, all off-diagonal matrix elements vanish.

\section{Second-cumulant approximation for the entropy}
\label{app:FCS}

The accuracy of the truncation \eqref{klich} admits a simple
model-independent criterion. A useful way to understand its often encountered
accuracy is to formulate the problem directly in terms of the spectrum of the
correlation matrix restricted to subsystem $A$. Writing its eigenvalues as
$\mu_i=1/(1+e^{E_i})$,
we introduce the corresponding spectral parameters
\begin{equation}
 E_i=\ln\frac{1-\mu_i}{\mu_i}
\end{equation}
and their density $\rho(E)$. The effective energies $E_i$ form the
single-particle spectrum entering the quadratic representation of
$-\ln\rho_A$ (the entanglement Hamiltonian, also called the modular
Hamiltonian); they should not be confused with the physical single-particle
energies of the Hamiltonian.

The two sides of Eq.~\eqref{klich} can then be written as
\begin{equation}
 S_A
 =\!
 \frac{1}{\ln2}\!\int\! dE\,\rho(E)\,h(E),
 \quad
 \mathrm{Var}(\hat n_A)
 =
 \!\int\! dE\,\rho(E)\,v(E),
\end{equation}
where
\begin{equation}
 h(E)=-\mu\ln\mu-(1-\mu)\ln(1-\mu),
 \quad
 v(E)=\mu(1-\mu).
\end{equation}
The integration extends over the full real axis, $-\infty<E<\infty$,
since $E=\ln[(1-\mu)/\mu]$ maps $0<\mu<1$ onto $\mathbb{R}$.
Both kernels are even functions of $E$ and are concentrated at
$|E|\lesssim {\cal O}(1)$. Indeed,
\begin{equation}
 h(E)\sim (|E|+1)e^{-|E|},
 \quad
 v(E)=\frac{1}{4\cosh^2(E/2)}
 \sim e^{-|E|}
\end{equation}
for $|E|\gg1$. Thus, spectral parameters with $|E|\gg1$ give
exponentially small contributions to both the entropy and the variance. Moreover,
\begin{equation}
 \int_{-\infty}^{\infty}\!dE\,h(E)
 =
 \frac{\pi^2}{3}
 \int_{-\infty}^{\infty}\!dE\,v(E)
 =
 \frac{\pi^2}{3}.
 \label{eq:FCS_kernel_integrals}
\end{equation}
Hence, Eq.~\eqref{klich} is exact for a flat spectral density $\rho(E)$.
Here and below, $\rho(E)$ denotes a coarse-grained density, with the
coarse-graining scale chosen large compared to the local spacing between $E_i$, but small compared to the ${\cal O}(1)$ width of
the kernels. Such a smooth-density description is meaningful when the
window $|E|\lesssim{\cal O}(1)$ contains many values of $E_i$. In the
approximately flat regime, this corresponds to $\rho(0)\gg1$ and, hence,
$\mathrm{Var}(\hat n_A)\simeq\rho(0)\gg1$. If only a few isolated modes
contribute, the smooth-density expansion is not controlled; however, the general
one-sided bound given below remains applicable.

The origin of the approximation can be stated directly by defining
\begin{equation}
 d(E)
 =
 h(E)-\frac{\pi^2}{3}v(E).
\end{equation}
Equation~\eqref{eq:FCS_kernel_integrals} gives
\begin{equation}
 \int_{-\infty}^{\infty}\!dE\,d(E)=0,
\end{equation}
and therefore
\begin{align}
 S_A-S_A^{\rm var}
 &=
 \frac{1}{\ln2}
 \int\!dE\,\rho(E)d(E),
 \label{eq:FCS_error_kernel}
 \\
 S_A^{\rm var}
 &=
 \frac{\pi^2}{3\ln2}\,
 \mathrm{Var}(\hat n_A).
 \label{eq:S2}
\end{align}
Thus, the constant part of $\rho(E)$ cancels from the error exactly. The
accuracy of Eq.~\eqref{klich} is controlled only by the nonconstant part of
the spectral density in the ${\cal O}(1)$ window selected by the kernels.
There are two useful ways in which this nonconstant part can become
parametrically unimportant.

First, suppose that $\rho(E)$ itself varies only on a broad scale. Expanding
within the relevant window,
\begin{equation}
 \rho(E)
 =
 \rho(0)
 +
 \rho'(0)E
 +
 \frac12\rho''(0)E^2
 +\ldots ,
 \label{eq:FCS_rho_expand}
\end{equation}
the linear term drops out because both kernels are even. Using
\begin{equation}
 \int\!dE\,E^2v(E)=\frac{\pi^2}{3},
 \qquad
 \int\!dE\,E^2h(E)=\frac{7\pi^4}{45},
\end{equation}
one finds
\begin{equation}
 \frac{S_A^{\rm var}}{S_A}
 \simeq
 1-\frac{\pi^2}{15}
 \frac{\rho''(0)}{\rho(0)}.
 \label{curv}
\end{equation}
If the derivatives satisfy
\begin{equation}
 \frac{\rho^{(2m)}(E)}{\rho(E)}
 =
 {\cal O}\!\left(\Lambda^{-2m}\right),
 \qquad
 |E|\lesssim{\cal O}(1),
 \label{eq:FCS_smoothness}
\end{equation}
with $\Lambda\gg1$, Eq.~\eqref{klich} is the leading term of a
Sommerfeld-type gradient expansion, whose first relative correction is of
order $\Lambda^{-2}$.

The same curvature can be diagnosed directly from the full counting statistics (FCS). For a single mode with occupation probability
$\mu(E)$, its contribution to the FCS is Bernoulli
\cite{Klich_2009,Song_2010}, with cumulant-generating function
\begin{align}
 K_E(t)
& =
 \ln\!\left[1-\mu(E)+\mu(E)e^t\right]
 =
 F(E-t)-F(E),
 \notag
 \\
 F(E)&=\ln(1+e^{-E}).
\end{align}
The corresponding single-mode cumulants are therefore
\begin{equation}
 c_n(E)
 =
 \left.\partial_t^n K_E(t)\right|_{t=0}
 =
 (-1)^n F^{(n)}(E),
\end{equation}
which gives
\begin{equation}
 c_{n+1}(E)=-\partial_E c_n(E).
\end{equation}
Since $c_2(E)=\mu(1-\mu)=v(E)$, one has
$c_4(E)=\partial_E^2v(E)$. Hence, after two integrations by parts,
\begin{equation}
 \mathcal{C}_4
 =
 \int\!dE\,\rho''(E)v(E).
\end{equation}
Thus, in the smooth regime,
\begin{equation}
 \frac{\mathcal{C}_4}{\mathcal{C}_2}
 \simeq
 \frac{\rho''(0)}{\rho(0)},
\end{equation}
so that Eq.~\eqref{curv} becomes
\begin{equation}
 \frac{S_A^{\rm var}}{S_A}
 \simeq
 1-\frac{\pi^2}{15}
 \frac{\mathcal{C}_4}{\mathcal{C}_2},
 \qquad
 \frac{\pi^2}{15}\simeq0.658.
 \label{eq:FCS_C4_diagnostic}
\end{equation}
Thus, $\mathcal{C}_4/\mathcal{C}_2$ provides a practical leading diagnostic of
the accuracy of Eq.~\eqref{klich}. More generally,
\begin{equation}
 \mathcal{C}_{2m}
 =
 \int\!dE\,\rho^{(2m-2)}(E)v(E),
 \label{eq:FCS_higher_derivatives}
\end{equation}
so that, under the stronger smoothness assumption
\eqref{eq:FCS_smoothness}, successive even cumulants are suppressed by
successive powers of $\Lambda^{-2}$.

This also identifies a particularly simple limiting case. A genuinely
``Gaussian FCS'',
\begin{equation}
 \ln\chi_A(\lambda)
 =
 i\mathcal C_1\lambda-\frac{\mathcal C_2}{2}\lambda^2 ,
\end{equation}
has $\mathcal C_{n>2}=0$ and therefore gives the second-cumulant result
exactly. Here, ``Gaussian'' refers to the functional form of the FCS, not
to the Gaussianity of the fermionic state. In the continuum-spectrum
representation, this case is precisely associated with a flat density
$\rho(E)$. Indeed, a constant $\rho(E)$ makes all derivatives entering
Eq.~\eqref{eq:FCS_higher_derivatives} vanish. Conversely,
\begin{equation}
 \partial_t^2 K(t)
 =
 \int dE\,\rho(E)\,v(E-t),
\end{equation}
and the Fourier transform
$\widetilde v(k)=\pi k/\sinh(\pi k)$ has no zeros on the real axis.
Thus, a strictly $t$-independent $\partial_t^2K$, i.e., Gaussian FCS,
requires a flat $\rho(E)$ within this continuum representation. The
perfectly transmitting equilibrium quantum point contact provides the standard regularized
example for this Gaussian limit \cite{Klich_2009}.

A second mechanism does not require such a hierarchy of derivatives. Suppose
that
\begin{equation}
 \rho(E)=A+\delta\rho(E),
 \qquad
 A\gg |\delta\rho(E)|
 \quad
 \text{for } |E|\lesssim{\cal O}(1).
 \label{eq:FCS_flat_background}
\end{equation}
The large constant part $A$ obeys Eq.~\eqref{klich} exactly, while all
corrections originate from $\delta\rho(E)$. Their relative magnitude is
therefore controlled by $\delta\rho/A$, even if $\delta\rho(E)$ varies on an
energy scale of order unity. In this case the higher cumulants need not form
the hierarchy
$\mathcal{C}_{2m}/\mathcal{C}_2\sim\Lambda^{-2m+2}$; rather, they can all be
smaller than $\mathcal{C}_2$ by the same large factor. This large-flat-background
mechanism is the one realized by the controlled Fermi-sea examples discussed
below.
An extreme version is a density that is constant throughout a broad interval
$|E|\lesssim\Lambda$ and varies only near its edges. Then all local
algebraic gradient corrections around $E=0$ vanish, and the remaining
relative error is controlled by the exponentially small tails of the
kernels,
\begin{equation}
 \frac{S_A-S_A^{\rm var}}{S_A}
 =
 {\cal O}\!\left(\Lambda e^{-\Lambda}\right).
\end{equation}
This provides a simple mechanism by which Eq.~\eqref{klich} can be much more
accurate than a generic curvature estimate.

To connect the spectral criterion above with the cumulant expansion
underlying Eq.~\eqref{klich}, it is sufficient to retain the first
correction beyond the variance.  The Klich--Levitov expansion gives
\cite{Klich_2009,Song_2010}
\begin{equation}
 S_A\ln2
 =
 \frac{\pi^2}{3}\mathcal{C}_2
 +
 \frac{\pi^4}{45}\mathcal{C}_4
 +\ldots .
 \label{eq:FCS_formal_KL}
\end{equation}
The $\mathcal{C}_4$ correction is consistent with
Eq.~\eqref{eq:FCS_C4_diagnostic}. 
Using $\mathcal{C}_4/\mathcal{C}_2\simeq\rho''(0)/\rho(0)$ in the smooth-density
regime reproduces Eq.~\eqref{curv}.  Thus, the curvature expansion provides
a spectral interpretation of the first correction to the second-cumulant
approximation.
No separate assumption about convergence
of the infinite cumulant series is needed for the spectral argument above.
Indeed, for generic FCS, the fixed-coefficient Klich--Levitov series is only
formal because sufficiently high cumulants can grow factorially; convergent
cutoff-dependent representations were constructed in
Refs.~\cite{Song_2010,Song_2011,Song_2012}.
The spectral criterion used here does not require convergence of
the formal cumulant series. In the smooth-density regime,
$\mathcal C_4/\mathcal C_2$ gives the leading diagnostic, whereas in the
large-flat-background regime, the small parameter is directly the
nonconstant part of $\rho(E)$ relative to its flat background.

It is equally important that Gaussianity in the central-limit sense does
not provide such a criterion. For a voltage-biased quantum point contact with transmission
probability $D$ and
$N=eVt/h$ transmission attempts, the FCS is binomial,
\begin{equation}
 \ln\chi(\lambda) =
 N\ln\!\left(1-D+De^{i\lambda}\right),
 \qquad \mathcal C_2=ND(1-D),
\end{equation}
whereas the exact entropy reads as~\cite{Klich_2009}
\begin{equation}
 S_A = -\frac{N}{\ln2}
 \left[D\ln D+(1-D)\ln(1-D)\right].
\end{equation}
At fixed $D$, all generically nonzero cumulants are of order $N$.
Thus, although the central part of the counting distribution approaches a
Gaussian as $N\to\infty$, the higher cumulants remain of the same extensive
order as $\mathcal C_2$. Correspondingly,
\begin{equation}
 \frac{S_A^{\rm var}}{S_A} =
 -\frac{(\pi^2/3)D(1-D)}
 {D\ln D+(1-D)\ln(1-D)}
\end{equation}
is independent of $N$. Thus, neither a large number of particles nor a
Gaussian central-limit form of the probability distribution implies
second-cumulant dominance. What matters is the unnormalized cumulant
structure, or, equivalently in the spectral formulation above, the
structure of $\rho(E)$ in the entropy-sensitive window.

For the ground-state Fermi sea in a clean 1D fermion chain
\cite{Calabrese_2012, Suesstrunk2012, Ivanov2013}, with $L_A$ the subsystem length, the
spectral density contains a large approximately constant contribution
proportional to $\ln(k_F L_A)$, while its $E$-dependent part remains
subleading. This is an example of Eq.~\eqref{eq:FCS_flat_background}: the
second cumulant contains the leading logarithm, whereas the higher cumulants
are not logarithmically enhanced. Their contribution to the entropy is thus suppressed
relative to $\mathcal{C}_2$ by the large logarithmic background.

The same leading structure survives the introduction of weak disorder. In a
metal with mean free path $l_{\rm dis}$ satisfying
\begin{equation}
 \lambda_F\ll l_{\rm dis}\ll L,
\end{equation}
the variance and entropy contain the logarithmic enhancement
\begin{equation}
 \mathcal{C}_2
 \propto
 (k_FL)^{d-1}\ln(k_Fl_{\rm dis}),
\end{equation}
whereas the higher cumulants obey the plain area law
\begin{equation}
 \mathcal{C}_{2m}
 \propto
 (k_FL)^{d-1},
 \qquad m>1,
\end{equation}
without the logarithmic factor \cite{Burmistrov2017}. 
The logarithm arises from ballistic momentum scales
$l_{\rm dis}^{-1}\ll q\ll k_F$, with the mean free path replacing the
subsystem size as the infrared cutoff of the clean logarithm, while the
genuinely diffusive contribution obeys the unenhanced area law. At the level
of the leading logarithmic contribution, disorder therefore realizes the
same mechanism of large flat background with a shifted infrared cutoff, rather
than a distinct Sommerfeld expansion in successive powers of a new
parameter.

For the present boundary-driven problem, there is no comparably obvious
microscopic large parameter as long as the system is far from the
ground-state regime. Nevertheless, Eq.~\eqref{klich} remains accurate
numerically both near equilibrium and far from equilibrium, with the
agreement improving away from the boundaries and in larger systems, as
shown in Sec.~\ref{sec3}. In the spectral language developed above, this
behavior is consistent with a density $\rho(E)$ whose nonconstant part in
the window $|E|\lesssim{\cal O}(1)$ is sufficiently small. Thus, the
second-cumulant approximation can remain accurate in quasi-free-fermion open or
monitored systems even when its microscopic control parameter is not
apparent. Direct inspection of $\rho(E)$ and of
$\mathcal{C}_4/\mathcal{C}_2$ provides a quantitative test of this explanation.

Finally, there is a useful one-sided bound that does not rely on any
smooth-density assumption. The binary-entropy inequality
\begin{equation}
 h(\mu)\ge 4\mu(1-\mu)\ln(2)
\end{equation}
implies \cite{Gioev2006}
\begin{equation}
 \frac{S_A^{(2)}}{S_A}
 \le
 \frac{\pi^2}{12\ln2}
 \simeq1.1866,
 \qquad
 S_A^{(2)}
 =
 \frac{\pi^2}{3\ln2}\,
 \mathrm{Var}(\hat n_A).
\end{equation}
Thus, the second-cumulant approximation can overestimate the exact entropy
by at most about $18.7\%$. The bound is saturated when all nontrivial
single-particle modes have $\mu_i=1/2$. There is no analogous small bound on
its possible underestimate.

\begin{figure}[t!]
\center
\includegraphics[width=\linewidth]{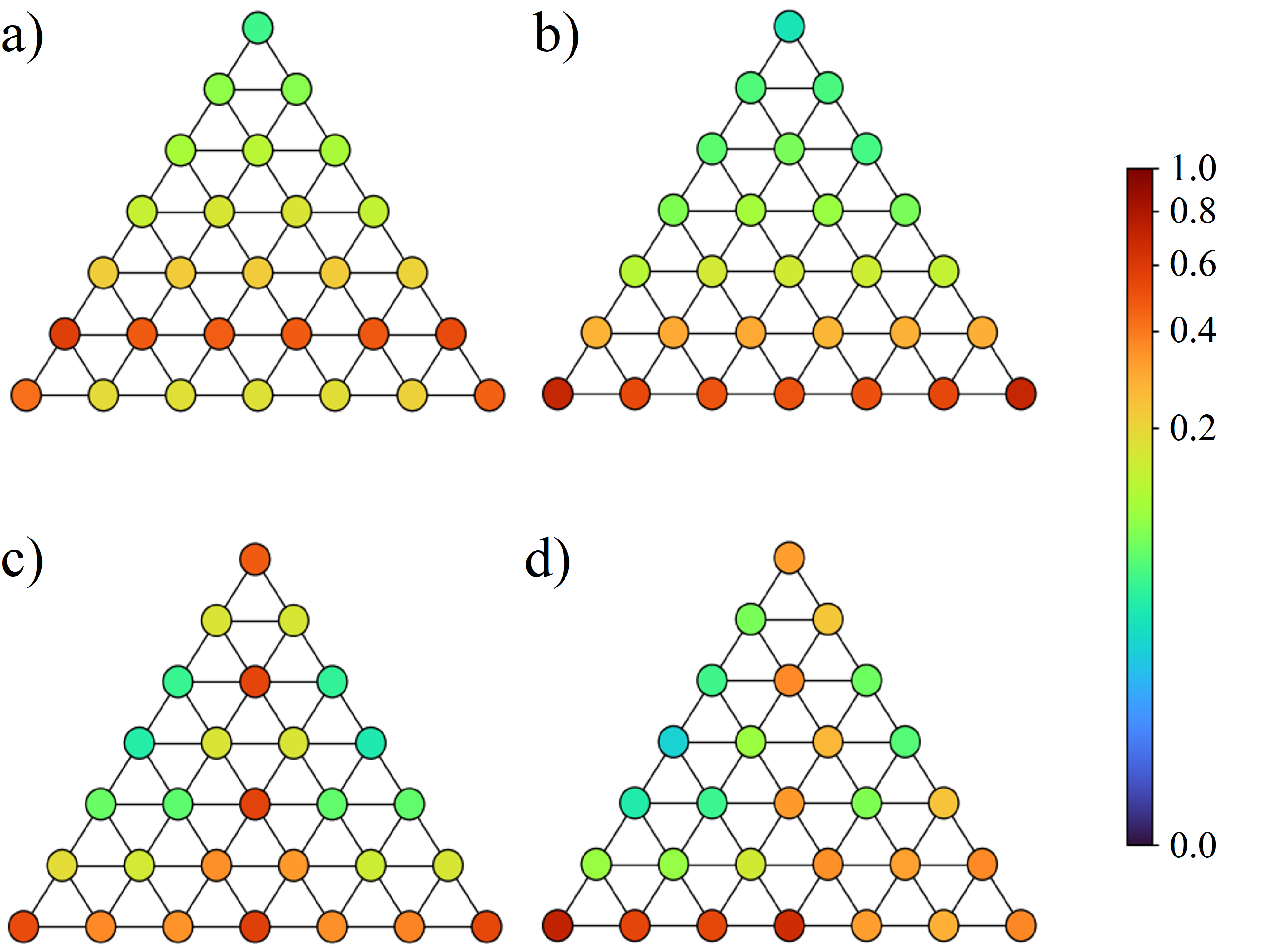}
\caption{Trajectory-averaged information lattice $\{\,\widetilde{i}_{(\ell,n)}\}$ for the same parameters as in Fig.~\ref{fig7}, i.e., $N=7, j_0=4, g=J=1,$ and $\epsilon=0$, with (a) $\delta=\epsilon_{j_0} = 0$, (b) $\delta=0.75$ and $\epsilon_{j_0} = 0$, (c)  $\delta = 0$ and $\epsilon_{j_0} = 3$, and (d) $\delta = 0.75$ and $\epsilon_{j_0} = 3$.  We use $N_{\rm traj}=1000$ stochastic pure-state trajectories for unraveling the LME, see Eq.~\eqref{eq:unraveled_info_subsystem}. With the shortest decay time $\tau^{-1} = \rm{max}\{\Gamma_{1},\Gamma_{N},\gamma_{1},\gamma_{N}\}$, which gives $\tau \approx 0.57$ for $\delta = 0.75$, we use the time step size $\Delta t=0.1$ (such that $\Delta t<\tau$) in numerical simulations for $\delta = 0$ and $\delta=0.75$. The trajectory average is performed after $35000$ time steps, such that the NESS configuration has safely been reached.}
\label{fig19}
\end{figure}

\section{Trajectory-averaged information lattice} 
\label{appendix:unraveling}

Here, we discuss some implications of unraveling the LME, Eq.~\eqref{eq:rho-equ}, into $N_{\rm traj}\gg 1$ stochastic pure-state trajectories $|\psi^{(k)}(t)\rangle$ \cite{Breuer2007, Cao2019a, Alberton_2021, Fazio2025}, corresponding to the individual events of particle exchanges between the fermion chain and the reservoirs.
Since the von Neumann entropy is a nonlinear function of $\rho$, one may distinguish the subsystem von Neumann information \eqref{eq:info_subsystem} obtained from the trajectory-averaged density matrix from its unraveled variant, 
\begin{equation}\label{eq:unraveled_info_subsystem}
\widetilde{I}_{A}(t)=\frac{1}{N_{\rm traj}}\sum^{N_{\rm {traj}}}_{k=1}I^{(k)}_{A}(t),
\end{equation}
where $I^{(k)}_{A}(t)$ is the von Neumann information of the subsystem for the density matrix $\rho^{(k)} = |\psi^{(k)}(t)\rangle \langle \psi^{(k)}(t)|$. In Eq.~(\ref{eq:unraveled_info_subsystem}), the averaging is performed after calculating this information for individual trajectories. After a transient regime that depends on the initial condition (similar to Fig.~\ref{fig15} for the FN), $I^{(k)}_{A}(t)$ reaches a quasi-stationary regime interspersed by stochastic fluctuations due to quantum jumps. 
By combining Eqs.~\eqref{eq:unraveled_info_subsystem} and \eqref{eq:triangles-to-sites}, we then compute the trajectory-averaged NESS 
information lattice $\{\,\widetilde{i}_{(\ell,n)}\}$.   

In general, trajectory-averaged values of nonlinear functionals of $\rho$ depend on the unraveling method itself \cite{Cao2019a,Piccitto_2022}. For simplicity, here we only discuss a single unraveling type based on the so-called Monte-Carlo wave-function method \cite{Dalibard1993}. We use the QuTiP library \cite{qutip5} to simulate the evolution of the $2^N\times 2^N$ density matrix along independent stochastic trajectories starting from an initially empty chain.  
Since the trajectory-averaged information \eqref{eq:unraveled_info_subsystem} is computed for pure-state trajectories, where the von Neumann entropy quantifies bipartite entanglement, $\{ \,\widetilde{i}_{(\ell,n)}\}$ and the associated information currents provide direct access to entanglement transport in the system.  

In  Fig.~\ref{fig19}, the corresponding results are shown for the same parameters as in Fig.~\ref{fig7}. Comparing Figs.~\ref{fig6}(c) and \ref{fig19}(c), we observe the defect-induced information pillar also for the trajectory-averaged information lattice, although slightly less sharp than the one found from the LME in Fig.~\ref{fig6}(c). Similarly, in Fig.~\ref{fig19}(d), we again encounter the pillar asymmetry caused by particle-hole asymmetry. Finally, in the clean case, both the LME approach, see Fig.~\ref{fig6}(a,b), and the unraveled version, see Fig.~\ref{fig19}(a,b), predict the von Neumann information to be localized on the two bottom layers of the information lattice.  Since for the trajectory-averaged information lattice, information is directly connected to quantum entanglement, the similarity between Figs.~\ref{fig7} and \ref{fig19} indicates that the information currents $\{\tilde{\cal I}_{(\ell,n)}\}$ obtained from the LME also approximately describe entanglement transport in this quasi-free fermion system.

\bibliography{biblio}

\end{document}